\documentclass[11pt,reqno,twoside]{atmp}

\usepackage{amsmath,amssymb}

\usepackage{ifpdf}
\usepackage{graphicx}
\usepackage{epsfig}
\usepackage{subfigure}
\usepackage{exscale}
\usepackage[mathscr]{eucal}
\usepackage{mathrsfs}
\usepackage{dsfont}

\usepackage[all]{xy}
\usepackage{color}
\usepackage{fancyhdr}

\begin{document}

\title[Quantum Instability of Attractive Bose Systems]
{On the Quantum Instability of Attractive Bose Systems}

%\arxurl{<hep_reference_#>}

\author[George E. Cragg]{$^1$George E. Cragg}
\author[Arthur K. Kerman]{$^2$Arthur K. Kerman} 
\address{$^1$Los Alamos National Laboratory, Los Alamos, New Mexico 87545, USA\\
cragg@lanl.gov}
%\addressemail{cragg@lanl.gov}
\address{$^2$Center for Theoretical Physics, Laboratory for Nuclear Science and Department of Physics, Massachusetts Institute of Technology,\\ Cambridge, Massachusetts 02139, USA\\
kerman@mitlns.mit.edu}
%\address{}  %lines should be separated with double backslashes: \\
%\addressemail{cragg@lanl.gov, }
%\addressemail{kerman@mitlns.mit.edu}

\begin{abstract}
We explore the zero-temperature behavior of an assembly of bosons interacting through a zero-range, attractive potential.  Because the two-body interaction admits a bound state, the many-body model is best described by a Hamiltonian that includes the coupling between atomic and molecular components.  Due to the positive scattering length, the low-density collection is expected to remain stable against collapse despite the attraction between particles.  Although a variational many-body analysis indicates a collapsing solution with only a molecular component to its condensate at low density, the expected atomic condensate solution can be obtained if the chemical potential is allowed to be complex valued.  In addition to revealing two discrete eigenfrequencies associated with the molecular case, an expansion in small oscillations quantifies the imaginary part of the chemical potential as proportional to a coherent decay rate of the atomic condensate into a continuum of collective phonon excitations about the collapsing lower state.
\end{abstract}

\maketitle

\tableofcontents

\newpage

% Section: Introduction
\section{Introduction}

This paper is intended to provide a complete quantum mechanical description of the instability inherent in zero-temperature atom-molecule Bose systems with attractive interparticle interactions, but with positive scattering length.  At low-density it is well-known that the pressure in these systems is proportional to $a \rho^2$, where $a$ is the s-wave scattering length and $\rho$ is the number density.  This suggests that despite being innately attractive, the ensemble can remain stable against collapse if there is a nonnegative scattering length \cite{Rb85BEC}.  Nevertheless, we find a ground state that tends toward collapse, with the expected, positive-pressure case realized only at the cost of including an inherent quantum instability \cite{Complexmu, IndepPair, CoherentDecay}.  Elucidation of this result is provided by a series of derivations presented in the following sections.  \\
\indent Because many-body interactions are built up from the pairwise sum of those between individual particles, Section \ref{TwoBodyScattering} begins with a review of the low-energy two-body scattering.  After relating the scattering length to the interaction strength, it is shown that a separable interaction not only captures the low-energy physics, but yields analytically tractable results.  In addition, we include coupling to a molecular Feshbach state, thus enabling a tuning of the effective scattering length through a resonance.  Solving the two-body problem provides a convenient context to model the many-body atom-molecule system, which is the subject of Section \ref{HamiltonianofCoupledSystem}.  As discussed in Section \ref{GaussianVariationalPrinciple}, a Gaussian variational procedure is used to find the lowest energy states of the system.  Upon deriving general variational expressions for these states, the static, uniform case is considered in Section \ref{StaticUniformSolution} where it is found that a collapsing two-piece ground state persists even in the case of a positive scattering length.  \\
\indent Notwithstanding, the expected solution can be obtained by allowing the chemical potential to assume complex values, where the imaginary part quantifies the decay of the condensate.  Section \ref{ComplexMu} demonstrates this result for the special case of a uniform system whereas Section \ref{DecayinNonuniformCase} proves the persistence of this feature even when the uniformity constraint is removed.  In the final section, we expand in small oscillations, obtaining a pair of discrete excitation frequencies corresponding to the collapsing solution.  Additionally, the expected energy per particle resides within a continuum of excitations above the collapsing ground state.  Energy conservation indicates that the decay represents a transition into the continuum, thus revealing that the expected solution evolves into states that tend toward collapse as well.  This analysis completes the result as it reveals the imaginary part of the chemical potential to be associated with a decay rate into collective modes of the collapsing lower state.   
\cutpage %move this line so that the first page breaks at the appropriate place.

\setcounter{page}{5}

\noindent
% Section: Two-Body Scattering
\section{Two-Body Scattering} \label{TwoBodyScattering}
Because microscopic models of many-particle assemblies are built from all pairwise interactions between the constituents, a reasonable place to begin is by considering the interaction just between two particles.  After identifying the s-wave scattering length as the relevant low-energy scattering parameter, this quantity is then related to the interaction strength.  In doing so, we recognize relationships that any model potential must satisfy in order to accurately describe the interparticle interaction.  From this analysis, we find that the separable potential is simple enough to be of calculational advantage, yet nonetheless retains sufficient flexibility to depict low-energy scattering.\\
\indent With the separable interaction, the scattering description is extended to include the possibility of molecular binding, a crucial element in the loss mechanisms at work in atomic condensates \cite{Dalibard,Shlya}.  Due to the presence of the hyperfine-induced molecular state, there exists a Feshbach resonance that occurs when the effective scattering length becomes unbounded upon tuning an externally applied magnetic field.  In completing the analysis, an effective range equation is derived for this coupled system.  From these two-body results, a many-body Hamiltonian may be postulated in which subsequent investigation reveals the properties of the collective.
% Subsection
\subsection{The Scattering Length}
\begin{figure} [h]
\begin{center}
\epsfxsize=4in\epsfbox{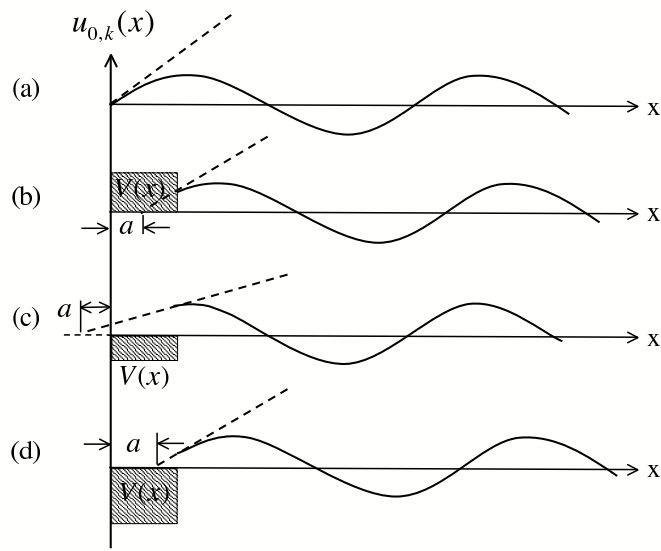}
%\vspace{-.5in}
\end{center}
\caption{ At points outside the influence of the potential, $V(x)$, this schematic depicts the s-wave ($\ell = 0$) radial wave function (solid line) and the corresponding ``zero energy'' linear asymptote (dashed line).  Part (a) shows the free forms of both.  For a repulsive potential, (b) indicates a negative phase shift [compared with (a)], as the wave function is ``repelled'' by the potential.  The point where the zero energy line intersects the radial axis is the scattering length, $a$, which must be non-negative for repulsion.  Note that as the range of $V(x)$ shrinks to zero, so does the scattering length.  For an attractive potential, (c) shows that the intercept can land on the other side of the axis, indicating a negative scattering length.  As the well depth increases, (d) shows that the wave function can get ``pulled in'' far enough to give a positive value for $a$.  Thus, for attractive potentials, the scattering length may have either sign. \label{Radial_WaveFn}}
\end{figure}

In the center of mass, two-body scattering processes are equivalent to reduced masses, $m/2$, impinging on a force center that is identical to the interparticle interaction.  Within the influence of this potential, $V$, the reduced mass obeys the full time independent Schr\"{o}dinger equation,
% Eq. 1
\setlength{\jot}{.1in}
\begin{equation} \label{TBSE}
	H | \psi \rangle = E | \psi \rangle ,
\end{equation}
\noindent where the Hamiltonian is a sum of a free kinetic term plus the potential, $H = H_0 + V$.  Long before collision, the incoming particles lie far outside the potential's range, thus allowing the incident flux to be represented as plane waves obeying the free Schr\"{o}dinger equation,
% Eq. 2
\begin{equation} \label{TBFSE}
	H_0 | \mathbf{k} \rangle = E | \mathbf{k} \rangle .
\end{equation}
\noindent It is convenient to relate the plane waves to the full scattering state, $| \psi \rangle$, by introducing the ${\mathcal T}$ operator, 
% Eq. 3
\begin{equation} \label{TBTOp}
	V | \psi \rangle = {\mathcal T} (E_k) | \mathbf{k} \rangle .
\end{equation}
\noindent From this definition, the solution to the full Schr\"{o}dinger equation (\ref{TBSE}) may be written
% Eq. 4
\begin{equation} \label{TBSESol}
	| \psi \rangle = | \mathbf{k} \rangle + G_0(2 k^2 + i \varepsilon) 
	{\mathcal T} (2 k^2) | \mathbf{k} \rangle ,
\end{equation}
\noindent where the free-particle Green's function\footnote{To connect with the many-body analysis, all energies are on the scale of $\hbar^2 / 2m$, where $m$ is the actual particle mass, not the reduced mass.  Hence, the energy eigenvalue is denoted as $2 k^2$.}has the coordinate space representation
% Eq. 5
\begin{align} \label{TBGreen}
	\begin{split}
		\langle \mathbf{x} | G_0(2 k^2 + i \varepsilon) | \mathbf{x'} \rangle 
		&= \int \limits_\mathbf{k'}
		\frac{e^{i \mathbf{k'} \cdot (\mathbf{x} - \mathbf{x'})}}{2 k^2 + i \varepsilon - 2 k'^2} \\
		&= - \frac{1}{8 \pi} \frac{e^{i k | \mathbf{x} - \mathbf{x'} |}}{| \mathbf{x} - \mathbf{x'} |} .
	\end{split}
\end{align}
\noindent In this expression, the integral is evaluated with complex contour integration in the standard way.\footnote{We use the notation $\int \limits_\mathbf{k} = (2 \pi)^{-3} \int d^3k.$} \\
\indent Obtaining the asymptotic behavior for large $| \mathbf{x} |$ requires the usual expansion $| \mathbf{x} - \mathbf{x'} | = x ( 1 - \mathbf{x} \cdot \mathbf{x'} / x^2 + \cdots )$, where the leading term is kept in the denominator, but the first two terms are retained in the exponential in (\ref{TBGreen}).  Substitution of this expanded Green's function into (\ref{TBSESol}) leads to the asymptotic scattering wave function,
% Eq. 6
\begin{equation} \label{TBWveFun}
	\psi_\mathbf{k} (\mathbf{x}) \xrightarrow[x \to \infty] {} e^{i \mathbf{k} \cdot \mathbf{x}} - \frac{1}{8 \pi}
	\langle k \mathbf{\hat{x}} | {\mathcal T} (2 k^2) | \mathbf{k} \rangle \frac{e^{i k x}}{x} .
\end{equation}
\noindent Comparing this with the usual form $\psi_\mathbf{k} (\mathbf{x}) \rightarrow e^{i \mathbf{k} \cdot \mathbf{x}} - \mathcal{A}(\theta, \varphi) e^{i k x}/x$, it is seen that, up to an overall constant, the matrix element is simply the scattering amplitude:
% Eq. 7
\begin{equation} \label{TBScattAmp}
	\mathcal{A}(\theta, \varphi) =  \frac{1}{8 \pi} \langle k \mathbf{\hat{x}} | {\mathcal T} (2 k^2) | \mathbf{k} \rangle,
\end{equation}
\noindent where the polar angles are those between $\mathbf{\hat{x}}$ and $\mathbf{k}$. \\
\indent Since we will ultimately be concerned with zero-temperature bosonic systems, it is useful to examine the low-energy form of the scattering amplitude.  Figure \ref{Radial_WaveFn} shows the radial wave function, which at ``zero energy'' has a linear form where the s-wave scattering length is the axial intercept:
% Eq. 8
\begin{equation} \label{TBRadFunc}
	u_{0, k}(x) \rightarrow x - a .
\end{equation}
\noindent Comparing $u_{0, k} / x$ with the asymptotic form of the wave function, the scattering length is identified as the zero energy ${\mathcal T}$ matrix element,
% Eq. 9
\begin{equation} \label{TBTmatrix_a}
	8 \pi a = \langle \mathbf{k} | {\mathcal T} (2 k^2) | \mathbf{k'} \rangle 
	\bigl|_{\mathbf{k} = \mathbf{k'} = 0} .
\end{equation}
%Figure ******************
%Figure*******************
\subsection{Separable Potential} \label{Sec:SepPot}
\indent A two-body model interaction is sought that is not only consistent with the low-energy scattering physics summarized by Fig.\:\ref{Radial_WaveFn}, but that leads to analytically tractable results as well.  In particular, it is required that there exist the possibility for either positive or negative scattering lengths in the case of attractive interactions.  To address these considerations, we use a separable potential characterized by the center of mass matrix element depending on the strength, $\lambda$, in addition to the form factor, $f(\mathbf{k})$:\footnote{The zero momentum form factor is arbitrary, since its value does not alter any of the calculations.  For simplicity, we therefore take $f(0) = 1$.}
% Eq. 10
\begin{equation} \label{TBSeparablePotential}
	\langle \mathbf{k} | V | \mathbf{k'} \rangle = \lambda f(\mathbf{k}) f(\mathbf{k'}).
\end{equation}
\noindent Multiplication of (\ref{TBSESol}) by $V$, then applying definition (\ref{TBTOp}) results in the Lippmann-Schwinger equation, ${\mathcal T} = V + V G_0 {\mathcal T}$, from which it is possible to relate the strength with the scattering length.  Using the momentum space representation of the Green's function, $\langle \mathbf{k''} | G_0 (2 k^2 + i \varepsilon) | \mathbf{k'''} \rangle = \delta(\mathbf{k''} - \mathbf{k'''}) / (2 k^2 + i \varepsilon - 2 k''^2)$, obtains the ${\mathcal T}$ matrix as
% Eq. 11
\begin{equation} \label{TBTMatrix}
	\langle \mathbf{k} | {\mathcal T} (2 k^2) | \mathbf{k'} \rangle = \frac{f(\mathbf{k}) \lambda f(\mathbf{k'})}
	{\displaystyle 1 + \lambda 
	\int \limits_{\mathbf{k''}} \frac{f(\mathbf{k''})^2}{2 k''^2 - 2 k^2 - i \varepsilon}}.
\end{equation}
\noindent Equation (\ref{TBTmatrix_a}) then relates the scattering length to the strength by
% Eq. 12
\begin{equation} \label{TBa_lambda}
	\frac{1}{8 \pi a_{bg}} = \frac{1}{\lambda} + \frac{1}{b},
\end{equation}
\noindent in which $b$ is identified as the range through $1/b = \int \limits_\mathbf{k} f(\mathbf{k})^2/ 2 k^2$.  Here, the scattering length carries a subscript to distinguish the background value from the case where a Feshbach state is included. \\
\indent Equation (\ref{TBa_lambda}) is consistent with the intuitive results depicted in Fig.\:\ref{Radial_WaveFn}, thus demonstrating that the separable form captures the low-energy scattering physics.  According to (\ref{TBa_lambda}), an attractive interaction $( \lambda < 0 )$ can yield either positive or negative scattering lengths, which is qualitatively consistent with Fig.\:\ref{Radial_WaveFn} (c) and (d).  Alternatively, a repulsive interaction $(\lambda > 0)$ only gives rise to non-negative scattering lengths, which is as required.  Henceforth, we confine our attention to attractive interactions in which the range is allowed to vanish for simplicity.  In this limit, we take $f(\mathbf{k}) \to 1$, with $\lambda$ approaching zero from below in such a way that $a_{bg}$ remains finite. \\
% subsection
\subsection{Coupled Channels Scattering}
\begin{figure} [h]
\begin{center}
\epsfxsize=2.5in\epsfbox{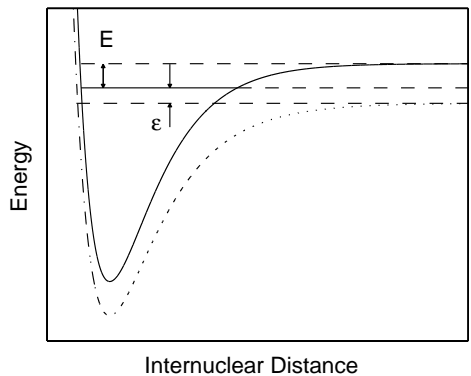}
\end{center}
\caption{ Schematic showing the potential of the closed channel (solid curve) with that of the open channel (dash-dot curve).  The binding energy $E$ and the detuning $\epsilon$ are shown relative to the free level of each case. \label{Channel_Potentials}}
\end{figure}

\begin{figure}[h]
\centering
\subfigure[]{{\label{Rba_vs_B}}\epsfxsize=2.25in\epsfbox{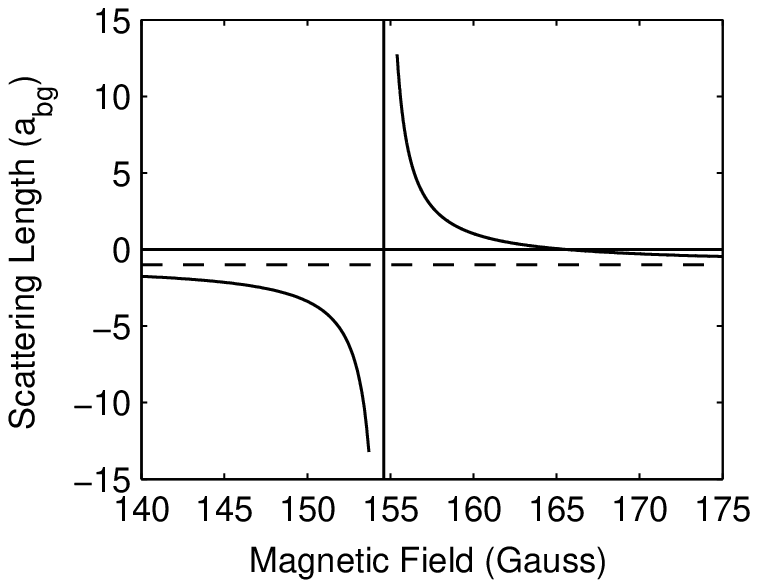}} \hspace{.25in}
\subfigure[]{{\label{RbEb_vs_B}}\epsfxsize=2.25in\epsfbox{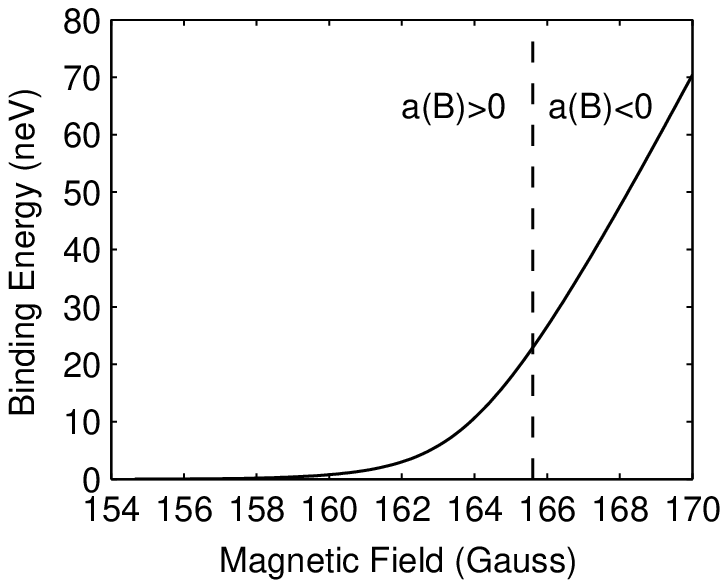}} \\ [0 in] 
\caption{For the case of $^{85}$Rb, (a) shows the full scattering length of Eq. \!(\ref{TBa_B}), with the horizontal dashed line as the asymptotic background value and the solid vertical indicating the location of the resonance at 154.6 G.  (b)  With a vertical, dashed boundary line separating the region of positive from that of negative scattering length, a plot is shown of the $^{85}$Rb binding energy corresponding to Eq. \!(\ref{TBBE}).  Approaching the resonance, both the scattering length and molecular size become unbounded, indicating an increasingly weakly bound molecular state.  Confirming this intuition, the binding energy is seen to increase from zero on resonance. \label{Eba_vs_B}}
\end{figure}

\indent Up to this point, all that has been considered pertains to the case of a single channel outcome in which two particles enter in a particular state or channel, scatter, then reemerge in the same channel.  Since there also exists the possibility for bound states to arise, an analysis is undertaken in which the energetically favored channel of the scattering state is coupled to the energetically unfavorable molecular state.  This coupled channels analysis forms the basis of the description of the Feshbach resonance, a phenomenon in which the effective scattering length becomes unbounded. \\
\indent As shown in Fig.\:\ref{Channel_Potentials}, the difference in electronic spin configuration gives rise to a distinct potential for each of the scattered and bound states \cite{EddyPhysRep}.  Furthermore, the schematic illustrates the molecular binding energy, $E$, as the energy difference between the bound state and two free atoms, relative to the molecular (closed channel) potential.  Analogously, the detuning, $\epsilon$, is defined as the same, but relative to the scattering (open channel) potential.  Since both potentials arise due to a difference in spin states, the detuning, and hence the binding energy, can be adjusted using the Zeeman interaction in the presence of an applied magnetic field, $B$.  Therefore, the detuning varies linearly with the field,
% Eq. 13
\begin{equation} \label{TBdetuning}
	\epsilon \to \epsilon + \Delta g B ,
\end{equation}
\noindent where $\Delta g$ is the difference in the $g$ factors between atoms and molecules. \\
%Figure ******************
%Figure******************* 
To include the coupling between the two states, it is most convenient to define projection operators $P$ and $Q$ onto the respective scattering and bound Hilbert spaces, $\mathscr{H}_S$ and $\mathscr{H}_B$ \cite{HFeshbach}.  For notational convenience, the appropriate projections of the full wave function, $| \psi \rangle$, are given by
% Eq. 14
\begin{align} \label{TBWFProjections}
	\begin{split}
		P | \psi \rangle &= | \psi^P \rangle \\
		Q | \psi \rangle &= | \psi^Q \rangle.
	\end{split}
\end{align}
\noindent Likewise, the various projections of the Hamiltonian are given by
% Eq. 15
\begin{align} \label{TBHProjections}
	\begin{split}
		P H P &= H_{PP} \qquad
		Q H Q = H_{QQ} \\
		P H Q &= H_{PQ}  \qquad
		Q H P = H_{QP}.
	\end{split}
\end{align}
\noindent Using the standard relationships, $P = P^\dag$, $Q = Q^\dag$ and $Q^\dag Q + P^\dag P = 1$, the projection of the Schr\"{o}dinger equation, $H | \psi \rangle = E | \psi \rangle$, onto the scattering space gives
% Eq. 16
\begin{equation} \label{TBScat_Schrodinger}
	H_{PP} | \psi^P \rangle + H_{PQ} | \psi^Q \rangle = E | \psi^P \rangle .
\end{equation}
\noindent Similarly, the bound space counterpart is
% Eq. 17
\begin{equation} \label{TBBound_Schrodinger}
	H_{QQ} | \psi^Q \rangle + H_{QP} | \psi^P \rangle = E | \psi^Q \rangle.
\end{equation}
\indent Consider first the momentum space representation of the scattering space projection of the Schr\"{o}dinger equation (\ref{TBScat_Schrodinger}).  In this case, the Hamiltonian, $H_{PP}$, contains both kinetic and potential pieces, which in momentum space are
% Eq. 18
\begin{equation} \label{TBHPP_momentumsp}
	\langle \mathbf{k} | H_{PP} | \psi^P \rangle = 2 k^2 \psi(\mathbf{k}) + \int \limits_\mathbf{k'}
	\langle \mathbf{k} | V | \mathbf{k'} \rangle \psi(\mathbf{k'}),
\end{equation}
\noindent where $\psi(\mathbf{k}) = \langle \mathbf{k} | \psi^P \rangle$.  Consisting of a linear combination of the orthonormal basis functions, $\{ | \phi_n \rangle \in \mathscr{H}_B \}$, the bound state projection is written
% Eq. 19
\begin{equation} \label{TBPsiQ}
	| \psi^Q \rangle = \sum_n c_n | \phi_n \rangle.
\end{equation}
Thus, upon substitution of the separable potential, the momentum space representation of Eq. \!(\ref{TBScat_Schrodinger}) is
% Eq. 20
\begin{equation} \label{TBScat_Schro_momentumspce}
	(2 k^2 - E) \psi (\mathbf{k}) + \lambda f(\mathbf{k}) \int \limits_\mathbf{k'} f(\mathbf{k'}) 	
	\psi(\mathbf{k'}) + \sum_n c_n \langle \mathbf{k} | H_{PQ} | \phi_n \rangle = 0 .
\end{equation}
\noindent Elimination of the coupling term requires an examination of the bound state projection.  \\
\indent Since the $\{ | \phi_n \rangle \}$ are eigenstates of $H_{QQ}$, their eigenvalues are identified with detunings, $\epsilon_n$, as in (\ref{TBdetuning}) and Fig.\:\ref{Channel_Potentials}.  Projecting the molecular equation (\ref{TBBound_Schrodinger}) onto one of the bound states, $|\phi_n \rangle$, results in a solution for the coefficients,
% Eq. 21
\begin{equation} \label{TBcoeff}
	c_n = - \frac{\lambda \alpha_n}{\epsilon_n - E} \int \limits_\mathbf{k'} f(\mathbf{k'}) \psi(\mathbf{k'}),
\end{equation}
\noindent with coupling constants, $\alpha_n$, defined through
% Eq. 22
\begin{equation} \label{TBcoupling}
	H_{QP} | \mathbf{k'} \rangle = \sum_j \lambda \alpha_j f(\mathbf{k'}) | \phi_j \rangle .
\end{equation}
\noindent Because it must vanish as the two-body strength is taken to zero, the coupling is represented by the product $\lambda \alpha_n$.  Substitution of the coefficients into Eq. \!(\ref{TBScat_Schro_momentumspce}) gives
% Eq. 23
\begin{equation} \label{TBEffective_Schrodinger}
	\left(2 k^2 - E \right) \psi(\mathbf{k}) + \left( \lambda - \lambda^2 \sum_n 
	\frac{\alpha_n^2}{\epsilon_n - E} \right) f(\mathbf{k}) \int \limits_\mathbf{k'} f(\mathbf{k'}) 
	\psi(\mathbf{k'}) = 0 ,
\end{equation}
\noindent thus illustrating the coupling to several bound states.  Assuming the effect of the highest lying molecular state to be dominant over the ones below, a single bound state (denoted with the subscript $1$) suffices for the description.  Comparison with the usual single channel result (obtained when $\alpha_1 = 0$) reveals that the molecular state can simply be included by a shift in the strength,
% Eq. 24
\begin{equation} \label{TBlambda_shift}
	\lambda \rightarrow \lambda - \frac{\lambda^2 \alpha_1^2}{\epsilon - E} .
\end{equation}
\indent After setting the energy to zero, a combination of (\ref{TBlambda_shift}) and (\ref{TBa_lambda}) arrives at an expression for the effective scattering length due to the presence of the molecular state,
% Eq. 25
\begin{equation} \label{TBa_eff}
	\frac{1}{a(\epsilon)} = \frac{1}{a_{bg}} + \frac{8 \pi \alpha_1^2}{\epsilon - \lambda \alpha_1^2} .
\end{equation}
From the detuning (\ref{TBdetuning}), the scattering length can be put into the standard form, explicitly dependent upon the magnetic field,
% Eq. 26
\begin{equation} \label{TBa_B}
	a(B) = a_{bg} \left( 1 - \frac{\Delta B}{B - B_0} \right)\!,
\end{equation}
\noindent where the resonance width, $\Delta B$, and the resonant field, $B_0$, are defined by
% Eq. 27
\begin{align} \label{TBRes_width_B0}
	\begin{split}
		\Delta B &= 8 \pi a_{bg} \alpha_1^2 / \Delta g \\
		B_0 &= (\lambda \alpha_1^2 - \epsilon )/ \Delta g - \Delta B .
	\end{split}
\end{align}
\noindent In addition, the full $T$ matrix is obtained by the substitution of (\ref{TBlambda_shift}) into (\ref{TBTMatrix}), thus giving
% Eq. 28 
\begin{equation} \label{TBFull_TMatrix}
	\langle \mathbf{k} | {\mathcal T} (2 k^2) | \mathbf{k'} \rangle = \frac{\displaystyle f(\mathbf{k}) 
	\left( \lambda 
	- \frac{\lambda^2 \alpha_1^2}{\epsilon - 2 k^2} \right) f(\mathbf{k'})}
	{\displaystyle 1 + \left( \lambda - \frac{\lambda^2 \alpha_1^2}{\epsilon - 2 k^2} \right)
	 \int \limits_\mathbf{k''} 
	\frac{f(\mathbf{k''})^2}{2k''^{\, 2} - 2 k^2 - i \varepsilon}}.
\end{equation}
\indent Since a pole in the ${\mathcal T}$ matrix indicates a bound state, the binding energies are given by the values of $k^2$ where the denominator vanishes.  After some algebraic manipulations, there emerges the following equation for the zeros:
% Eq. 29
\begin{equation} \label{TBTMatrix_Poles}
	\frac{1}{8 \pi a_{bg}} - \frac{\gamma_0^2}{2} \int \limits_\mathbf{k''} \frac{1}{k''^{\, 2}} 
	\frac{f(\mathbf{k''})^2}{k''^{\, 2} + \gamma_0^2} + \frac{\alpha_1^2}{\epsilon + 2 \gamma_0^2
	- \lambda \alpha_1^2} = 0 ,
\end{equation}
\noindent with the binding energy defined as $-k^2 = \gamma_0^2$.  For negative background scattering lengths, the zero range ($b \to 0$)  limit is taken in which $\lambda \to 0^-$ and $f(\mathbf{k}) \to 1$.  In this limit, (\ref{TBTMatrix_Poles}) reduces to a cubic equation in the square root of the binding energy:
% Eq. 30
\begin{equation} \label{TBBE}
	\left( \epsilon + 2 \gamma_0^2 \right) \left( \frac{1}{a_{bg}} - \gamma_0 \right) = -8 \pi \alpha_1^2 .
\end{equation}
\noindent By the integral in (\ref{TBTMatrix_Poles}), only roots with $\gamma_0 > 0$ are permissible.  Figure \ref{Eba_vs_B} shows the zero-range cases of the scattering length resonance along with the binding energy for the $^{85}$Rb system.
% Figure ***************
% Figure ***************
% Subsection: Effective Range
\subsection{Effective Range}
Frequently, the partial wave analysis is employed where the determination of the phase shifts allows subsequent derivation of any other quantity of physical interest.  In particular, the scattering length, $a$, and the effective range, $r_{eff}$, are extracted from the expansion
% Eq. 31
\begin{equation} \label{TBER}
	k \cot \delta_0(k) = -a^{-1} + \frac{1}{2} r_{eff} k^2 + \ldots \, ,
\end{equation}
\noindent where $\delta_0(k)$ is the s-wave phase shift for particles of energy $2 k^2$.  Recall that the scattering amplitude is given by a sum over all partial waves, $-\mathcal{A}(\theta, \varphi) = k^{-1} \sum_\ell (2 \ell + 1) e^{i \delta_\ell(k)} \sin \delta_\ell(k) P_\ell (\cos \theta)$.  If only the s-waves contribute, this reduces to $\mathcal{A} = (8 \pi)^{-1} {\mathcal T} = - k^{-1} e^{i \delta_0(k)} \sin \delta_0(k)$, thus suggesting that the effective range may be found by an expansion of the ratio of real to imaginary parts of ${\mathcal T}$:
% Eq. 32
\begin{equation} \label{TBkcot_deltaT}
	\cot \delta_0(k) = \frac{{ \rm Re}\{{\mathcal T} \} }{{\rm Im} \{ {\mathcal T} \} }.
\end{equation}
These parts are obtained from (\ref{TBFull_TMatrix}) after the integral is evaluated with $\int dk' / (k' \, ^2 - k^2) = (2 \, k)^{-1} [\ln |(kb - 4 \pi^2)/(kb + 4 \pi^2)| + i \pi]$.  Expanding the logarithms for small $k$, then taking the ratio in (\ref{TBkcot_deltaT}) obtains
% Eq. 33
\begin{equation} \label{TBkcot_delta_effrng}
	k \cot \delta_0(k) = - \frac{1}{a} + \left[ \frac{b}{2 \pi^3}  
	- 16 \pi \alpha_1^2 \left( \epsilon - \lambda \alpha_1^2 \right)^{-2} 
	\right] k^2 + \ldots \, ,
\end{equation}
\noindent giving the effective range
% Eq. 34
\begin{equation} \label{1SepPot_EffRng}
	r_{eff} = \pi^{-3} b - 32 \pi \alpha_1^2 
	\left( \epsilon - \lambda \alpha_1^2 \right)^{-2}.
\end{equation}
Note that in the absence of the molecular state ($\alpha_1 = 0$), the effective range is simply proportional to the range, $b$, which justifies the identification.  In the zero range limit, the effective range becomes negative, $r_{eff} \to - 32 \pi \alpha_1^2 / \epsilon^2$, due to the presence of the molecular state \cite{BruunJackson}.
% Section: Many-Body Hamiltonian of the Coupled System
\section{Many-Body Hamiltonian of the Coupled System}\label{HamiltonianofCoupledSystem}
Having reviewed the two-body scattering physics, we now discuss a corresponding Hamiltonian for the many-body case.  In field operator language, this Hamiltonian is
% Eq. 35
\begin{align} \label{H_Hamilt}
	\begin{split}
		\hat{H}  = \sum_{\alpha \beta} \hat{\Psi}_\alpha^\dag & T_{\alpha \beta} 
		 \hat{\Psi}_\beta + \frac{1}{2} \sum_{\alpha \beta \gamma 
		\delta} \hat{\Psi}_\alpha^\dag \hat{\Psi}_\beta^\dag V_{\alpha \beta; \delta \gamma} 
		\hat{\Psi}_\gamma \hat{\Psi}_\delta \\
		& + \epsilon \sum_\alpha \hat{\Phi}_\alpha^\dag \hat{\Phi}_\alpha
		+ \frac{\lambda \alpha_1}{\sqrt{2}} \sum_{\alpha \beta \gamma} \hat{\Phi}_\alpha^\dag
		F_{\alpha; \beta \gamma} \hat{\Psi}_\gamma \hat{\Psi}_\beta + \mbox{H. c.} .
	\end{split}
\end{align}
\noindent Here, the first term is the kinetic energy whereas the second is the sum over all pairwise contributions from the two-body interaction.  Given by the molecule number times the detuning, the next term accounts for the energy of the molecular condensate component.  With a proportionality constant $\lambda \alpha_1$ and a molecular form factor $F$, the final term is the interaction energy due to coupling of atoms to molecules and vice versa \cite{Eddy}. \\
\indent Before embarking on the analysis, it is worthwhile to attempt a simplification of the above model.  Suppose there were no two-body potential present.  Having only a single Feshbach state with a coupling parameter $g$ and detuning $\varepsilon$, Eq. \!(\ref{TBEffective_Schrodinger}) becomes
% Eq. 36
\begin{equation} \label{H_Effective_Pot}
	\left(2 k^2 - E \right) \psi(\mathbf{k}) - \frac{g^2}{\varepsilon - E} f(\mathbf{k}) \int \limits_\mathbf{k'}
	f(\mathbf{k'}) \psi(\mathbf{k'}) = 0 .
\end{equation}
\noindent For a detuning far enough away from the binding energy, $| \varepsilon | \gg | E |$, the effective interaction strength is $\approx - g^2/ \varepsilon$ which can be chosen to be equal to the original separable potential strength, $\lambda$.  Accordingly, an appropriate detuning, $\varepsilon = 2 / \lambda^2$, and a coupling, $g = \sqrt{-2 / \lambda}$, are defined such that the requirement $| \varepsilon| \gg | E |$ is self-consistently fulfilled in the limit of zero-range attraction, $\lambda \to 0^-$.  Replacing the two-body potential with a term that couples the atoms to an auxiliary molecular field, $\hat{X}$, gives an effective Hamiltonian \cite{Altland},
% Eq. 37
\begin{align} \label{H_2Res_Hamilt}
	\begin{split}
		\hat{H} = & \sum_{\alpha \beta} \hat{\Psi}_\alpha^\dag T_{\alpha \beta} 
		\hat{\Psi}_\beta + \epsilon \sum_\alpha \hat{\Phi}_\alpha^\dag \hat{\Phi}_\alpha 
		+  \varepsilon \sum_\alpha \hat{X}_\alpha^\dag \hat{X}_\alpha \\
		& + \frac{\lambda \alpha_1}{\sqrt{2}} \sum_{\alpha \beta \gamma} \hat{\Phi}_\alpha^\dag
		F_{\alpha; \beta \gamma} \hat{\Psi}_\gamma \hat{\Psi}_\beta 
		+ \frac{g}{\sqrt{2}} \sum_{\alpha \beta \gamma} \hat{X}_\alpha^\dag 
		F_{\alpha; \beta \gamma} \hat{\Psi}_\gamma \hat{\Psi}_\beta + \mbox{H. c.} .
	\end{split}
\end{align}
\noindent Although we have replaced the two-body interaction by a single Feshbach state, it must be emphasized that the two Hamiltonians in (\ref{H_Hamilt}) and (\ref{H_2Res_Hamilt}) are in general different for nonzero range.\footnote{It turns out that that the two models are equivalent in the zero-range limit.}  
% Section: Gaussian Variational Principle
\section{Gaussian Variational Principle}\label{GaussianVariationalPrinciple}
To obtain an equation of state for the coupled system, we employ a variational procedure in which a Gaussian trial functional is used in calculating the expectation value of the Hamiltonian \cite{GVP}.  An extremization then determines the various solutions arising from the model.  As a first step in this procedure, we decompose the field operators in terms of their corresponding coordinates, $\hat{\psi}$, $\hat{\phi}$, $\hat{\chi}$, and their conjugate momenta, $\hat{\pi}$, $\hat{\omega}$, $\hat{\nu}$:
% Eq. 38
\begin{subequations} \label{GVP_OpDecomp}
	{\allowdisplaybreaks \begin{align}
		\hat{\Psi}_\alpha &= \frac{1}{\sqrt{2}} \left( \hat{\psi}_\alpha + i \hat{\pi}_\alpha \right); 
		\quad \hat{\Psi}_\alpha^\dag = \frac{1}{\sqrt{2}} \left( \hat{\psi}_\alpha - i \hat{\pi}_\alpha \right) 
		\\
		\hat{\Phi}_\alpha &= \frac{1}{\sqrt{2}} \left( \hat{\phi}_\alpha + i \hat{\omega}_\alpha \right); 
		\quad \hat{\Phi}_\alpha^\dag = \frac{1}{\sqrt{2}} \left( \hat{\phi}_\alpha - 
		i \hat{\omega}_\alpha \right) \\
		 \hat{X}_\alpha &= \frac{1}{\sqrt{2}} \left( \hat{\chi}_\alpha + i \hat{\nu}_\alpha \right); 
		\quad \hat{X}_\alpha^\dag = \frac{1}{\sqrt{2}} \left( \hat{\chi}_\alpha - i \hat{\nu}_\alpha \right) 
		.
	\end{align}} \\
\end{subequations}
\noindent Where appropriate, only the atomic field is used in illustrating the required operator relationships since the molecular counterparts follow by a straightforward comparison.  As usual, the field operators' commutation relations are
% Eq. 39
\begin{subequations} \label{GVP_Commutators}
	\begin{align} \label{GVP_Commutators1}
		\begin{split}
			\left[ \hat{\Psi}_\alpha, \hat{\Psi}_\beta \right] 
			= \left[ \hat{\Psi}_\alpha^\dag, \hat{\Psi}_\beta^\dag \right] &= 0 \\
			\left[ \hat{\Psi}_\alpha, \hat{\Psi}_\beta^\dag \right] &= \delta_{\alpha \beta}
		\end{split}
	\end{align}
	\parbox{\textwidth}{which imply}
	\begin{align} \label{GVP_Commutators2}
		\begin{split}
			\left[ \hat{\psi}_\alpha , \hat{\psi}_\beta \right] = \left[ \hat{\pi}_\alpha , \hat{\pi}_\beta \right]
			&= 0 \\
			\left[ i \hat{\pi}_\alpha , \hat{\psi}_\beta \right] &= \delta_{\alpha \beta} .
		\end{split}
	\end{align}
\end{subequations}
\indent In the functional Schr\"{o}dinger picture, the state vector, $| \Psi , t \rangle$, depends on the field $\psi'$,
% Eq. 40
\begin{equation} \label{GVP_WaveFn}
	| \Psi, t \rangle \rightarrow \Psi[ \psi', t] .
\end{equation}
\noindent Analogous to the coordinate space representation of single-particle operators, the action of $\hat{\psi}$ and $\hat{\pi}$ on the many-body state is
% Eq. 41
\begin{subequations} \label{GVP_psipi_Action}
	\begin{align}
		\hat{\psi}_\alpha | \Psi, t \rangle &\rightarrow \psi'_\alpha \Psi[ \psi',t]  \label{GVP_psi_Action} \\
		\hat{\pi}_\alpha | \Psi, t \rangle &\rightarrow -i \frac{\delta}{\delta \psi'_\alpha}
		\Psi [ \psi',t ] . \label{GVP_pi_Action}
	\end{align}
\end{subequations}
Explicitly, the coherent many-body state is an ansatz consisting of a product of Gaussian functionals, with one for each of the three fields:
% Eq. 42
{\allowdisplaybreaks \begin{align} \label{GVP_TrialFunc}
	\begin{split}
		\Psi [ \psi', \phi', \chi', t] &= N_\psi \exp \left\{ - \sum_{\alpha \beta} \delta \psi'_\alpha (t) 
		\left[ \frac{1}{4} G^{-1}_{\alpha \beta}(t) - i \Sigma_{\alpha \beta}(t) \right] \delta \psi'_\beta (t)
		\right. \\
		& \left. \qquad \qquad \qquad \qquad \qquad \qquad \qquad 
		+ \, i \sum_\alpha \pi_\alpha(t) \delta \psi'_\alpha (t) \right\} \\
		& \quad \cdot N_\phi \exp \left\{ - \sum_\alpha \left[ \frac{1}{2} \delta \phi'_\alpha (t)^2 
		- i \omega_\alpha(t) \delta \phi'_\alpha(t) \right] \right\} \\
		& \quad \cdot N_\chi \exp \left\{ - \sum_\alpha \left[ \frac{1}{2} \delta \chi'_\alpha(t)^2
		- i \nu_\alpha(t) \delta \chi'_\alpha(t) \right] \right\} ,
	\end{split}
\end{align}} \\
\noindent where $N_\psi$, $N_\phi$ and $N_\chi$ are the normalization constants and the fluctuations are given by the fields minus their mean values:
% Eq. 43
\begin{subequations} \label{GVP_Fluctuations}
	\begin{align}
		\delta \psi'_\alpha(t) &= \psi'_\alpha - \psi_\alpha(t) \label{GVP_psiFluct} \\
		\delta \phi'_\alpha(t) &= \phi'_\alpha - \phi_\alpha(t) \label{GVP_phiFluct} \\
		\delta \chi'_\alpha(t) &= \chi'_\alpha - \chi_\alpha(t) . \label{GVP_chiFluct}
	\end{align}
\end{subequations}
\indent Unlike the molecular functionals, the atomic Gaussian has extra degrees of freedom as it is parametrized by a symmetric width composed of a real part, $G^{-1}_{\alpha \beta}/4= G^{-1}_{\beta \alpha}/4$, along with its canonical conjugate, $\Sigma_{\alpha \beta} = \Sigma_{\beta \alpha}$.  With their conjugates set to zero, the corresponding molecular widths are taken to be $1/2$.  This simplification is justified due to the absence of any $\hat{\Phi} \hat{\Phi}$ and $\hat{X} \hat{X}$ terms in the Hamiltonian.  Under a variational analysis, general molecular widths simply reduce to the aforementioned values as anticipated in the ansatz (\ref{GVP_TrialFunc}).  Because each term contains a balance of creation-destruction operator pairs, the Hamiltonian remains invariant under any phase transformation of the trial functional:
% Eq. 44
\begin{equation} \label{GVP_WaveFunc_Phase}
	| \Psi , t \rangle \rightarrow e^{-i \hat{N} \theta (t)} | \Psi , t \rangle.
\end{equation}
Put another way, the invariance with respect to the phase angle, $\theta(t)$, must be connected with a fixed total particle number, ${\mathcal N}$.  Due to the continuous symmetry in $\theta$, there must exist a mode of zero energy, otherwise known as the Goldstone mode.  Upon conducting a small oscillation analysis in the random phase approximation (RPA), the presence of this zero frequency is explicitly demonstrated. \\
\indent In this formalism, the mean of any operator, ${\mathcal O}$, is calculated from the functional integral
% Eq. 45
\begin{equation} \label{GVP_OpMean}
	\langle \Psi, t | \hat{{\mathcal O}} | \Psi, t \rangle = \int {\mathcal D} \psi'  \, \Psi^\ast[ \psi', t] 
	\, \hat{{\mathcal O}} \, \Psi[\psi', t] .
\end{equation}
For calculating the mean, it is useful to construct a basis of creation and destruction operators for the atomic Gaussian in (\ref{GVP_TrialFunc}).  As linear combinations of $\hat{\varphi}$ and $\hat{\pi}$, these operators are found to be
% Eq. 46
\begin{subequations}  \label{GVP_cdagc}
	\begin{align}
		\hat{c}^\dag_\alpha &= \frac{1}{\sqrt{2}} \sum_\beta G_{\alpha \beta} \left[ 2 \sum_\gamma
		\left( \frac{1}{4} G^{-1}_{\beta \gamma} + i \Sigma_{\beta \gamma} \right) 
		\left( \hat{\psi}_\gamma - \psi_\gamma \right) - i\left( \hat{\pi}_\beta - \pi_\beta \right)
		\right] \label{GVP_cdag} \\ 
		\hat{c}_\alpha &= \frac{1}{\sqrt{2}} \sum_\beta G_{\alpha \beta} \left[ 2 \sum_\gamma
		\left( \frac{1}{4} G^{-1}_{\beta \gamma} - i \Sigma_{\beta \gamma} \right) 
		\left( \hat{\psi}_\gamma - \psi_\gamma \right) + i\left( \hat{\pi}_\beta - \pi_\beta \right)
		\right] \!, \label{GVP_c}
	\end{align}
\end{subequations}
\noindent omitting the explicit time dependence for convenience.  A direct application on the many-body state reveals that
% Eq. 47
\begin{subequations} \label{GVP_cdagc_App}
	\begin{align}
		\hat{c}^\dag_\alpha (t) | \Psi, t \rangle & \rightarrow \frac{1}{\sqrt{2}} 
		\delta \psi'_\alpha (t) \Psi[ \psi',t] \label{GVP_cdag_App} \\
		\hat{c}_\alpha(t) | \Psi, t \rangle &= 0 , \label{GVP_c_App}
	\end{align}
\end{subequations}
\noindent thus verifying the construction.  Using the commutators (\ref{GVP_Commutators2}) along with the symmetry of $G$ and $\Sigma$ we have
% Eq. 48
\begin{equation} \label{GVP_cdagc_Commutators}
	\left[ \hat{c}^\dag_\alpha (t), \hat{c}_\beta(t) \right] = - \frac{1}{2} G_{\alpha \beta}(t) .
\end{equation}
\indent Transforming to the $\{ \hat{c}^\dag, \hat{c} \}$ basis, it is possible to easily calculate all quantities of interest. %through the application of Eqs. (\ref{GVP_cdagc_App}) and (\ref{GVP_cdagc_Commutators}).  
Inversion of (\ref{GVP_cdagc}) leads to the desired expression for the field operators
% Eq. 49
\begin{align} \label{GVP_FieldOP_cdagc}
	\begin{split}
		\hat{\Psi}_\alpha(t) &= 2 \sum_\beta \left[ \frac{1}{4} G^{-1}_{\alpha \beta} (t) 
		+ i \Sigma_{\alpha \beta}(t) + \frac{1}{2} \delta_{\alpha \beta} \right] \hat{c}_\beta (t) \\
		&-2 \sum_\beta \left[ \frac{1}{4} G^{-1}_{\alpha \beta}(t) - i\Sigma_{\alpha \beta}(t) 
		-\frac{1}{2} \delta_{\alpha \beta} \right] \hat{c}^\dag_\beta(t) 
		+ \frac{1}{\sqrt{2}} \left[ \psi_\alpha(t) + i \pi_\alpha(t) \right] .
	\end{split}
\end{align}
Employing Eqs. \!\!(\ref{GVP_cdagc_App})-(\ref{GVP_FieldOP_cdagc}), all required mean values are calculated in the following:
% Eqs. 50
\setlength{\jot}{.2in}
\begin{subequations} \label{GVP_MeanValues}
	{\allowdisplaybreaks \begin{align}
	%\begin{equation}
		\langle \Psi , t | \hat{\Psi}_\alpha (t)| \Psi , t \rangle &= \frac{1}{\sqrt{2}} \left[ \psi_\alpha (t) + i 
		\pi_\alpha (t) \right] = \Psi_\alpha (t)  \label{GVP_Meana} \\
	%\end{equation}
	%\begin{equation} 
		\langle \Psi , t | \hat{\Psi}_\alpha (t) \hat{\Psi}_\beta (t) | \Psi , t \rangle &= -D_{\alpha \beta} (t) +
		\Psi_\alpha (t) \Psi_\beta (t) \label{GVP_Meanb} \\
	%\end{equation}
	%\begin{equation} 
		\langle \Psi , t | \hat{\Psi}^\dag_\alpha (t) \hat{\Psi}_\beta (t) | \Psi , t \rangle  
		&= R_{\alpha \beta} (t) + \Psi^\ast_\alpha (t) \Psi_\beta (t) \label{GVP_Meanc} \\
	%\end{equation}
	%\begin{align} 
		\langle \Psi , t | \hat{\Psi}^\dag_\alpha (t) \hat{\Psi}^\dag_\beta (t) \hat{\Psi}_\gamma (t)
		\hat{\Psi}_\delta (t) | \Psi , t \rangle &= D^\ast_{\alpha \beta} (t) D_{\gamma \delta} (t) -
		D^\ast_{\alpha \beta} (t) \Psi_\gamma (t) \Psi_{\delta} (t) \qquad \qquad 
		\nonumber \\ &  \negthickspace \negthickspace \negthickspace \negthickspace
		\negthickspace \negthickspace \negthickspace \negthickspace \negthickspace
		\negthickspace \negthickspace \negthickspace \negthickspace \negthickspace
		+ R_{\beta \gamma} (t)
		R_{\alpha \delta} (t)  + R_{\beta \delta} (t) R_{\alpha \gamma} (t) \nonumber \\ 
		& \negthickspace \negthickspace \negthickspace \negthickspace 
		\negthickspace \negthickspace \negthickspace \negthickspace \negthickspace
		\negthickspace \negthickspace \negthickspace \negthickspace \negthickspace
		+ \Psi^\ast_\beta (t) \Psi_\delta (t) R_{\alpha \gamma} (t) 
		+ \Psi^\ast_\beta (t) \Psi_\gamma (t) R_{\alpha \delta} (t) \label{GVP_Meand}
		\\ & \negthickspace \negthickspace \negthickspace \negthickspace 
		\negthickspace \negthickspace \negthickspace \negthickspace \negthickspace
		\negthickspace \negthickspace \negthickspace \negthickspace \negthickspace
		+ \Psi^\ast_\alpha (t) \Psi_\delta (t) R_{\beta \gamma} (t) + \Psi^\ast_\alpha (t) 
		\Psi_\gamma (t) R_{\beta \delta} (t) \nonumber \\ 
		& \negthickspace \negthickspace \negthickspace \negthickspace 
		\negthickspace \negthickspace \negthickspace \negthickspace \negthickspace
		\negthickspace \negthickspace \negthickspace \negthickspace \negthickspace
		- \Psi^\ast_\alpha (t) \Psi^\ast_\beta (t)D_{\gamma \delta}(t)  
		+ \Psi^\ast_\alpha (t) \Psi^\ast_\beta (t) \Psi_\gamma (t) \Psi_\delta(t) 
		\nonumber
	\end{align}}
	\begin{align} 
		\begin{split} \label{GVP_Meane}
			\langle \Psi , t | i \frac{\delta}{\delta t} | \Psi , t & \rangle =  \sum_\alpha 
			\left[ \pi_\alpha (t) \dot{\psi}_\alpha (t) 
			+ \omega_\alpha(t) \dot{\phi}_\alpha(t) + \nu_\alpha(t) \dot{\chi}_\alpha(t) \right]
			\\ & + \sum_{\alpha \beta} \Sigma_{\alpha \beta} (t) 
			\dot{G}_{\beta \alpha} (t)+ {\mathcal N} \dot{\theta} (t) + \mbox{total time derivatives}.
		\end{split}
	\end{align}
\end{subequations}
In the expressions above, we have introduced $R_{\alpha \beta}(t)$ and $D_{\alpha \beta}(t)$ as the respective fluctuations of $\langle \hat{\Psi}^\dag \hat{\Psi} \rangle$ and $\langle \hat{\Psi} \hat{\Psi} \rangle$ about their mean field values of $| \Psi |^2 /2$ and $\Psi^2/2$.  Explicitly, these fluctuations are parametrized by $G^{-1}_{\alpha \beta}(t)$ and its conjugate, $\Sigma_{\alpha \beta}(t)$:
% Eqs. 51
\begin{subequations} \label{GVP_RD}
	{\allowdisplaybreaks \begin{align}
		R_{\alpha \beta} (t) & = \frac{1}{2} \left[ \frac{1}{4} G^{-1}_{\alpha \beta} (t) + 
		G_{\alpha \beta} (t) - \delta_{\alpha \beta} \right] + 2 \sum_{\gamma \delta} 
		\Sigma_{\alpha \gamma} (t) G_{\gamma \delta} (t) \Sigma_{\delta \beta} (t) \label{GVP_R} , \\
		\begin{split} \label{GVP_D}
			D_{\alpha \beta} (t) & = \frac{1}{2} \left[ \frac{1}{4} G^{-1}_{\alpha \beta} (t) 
			- G_{\alpha \beta} (t) \right] + 2\sum_{\gamma \delta} \Sigma_{\alpha \gamma} (t) 
			G_{\gamma \delta} (t) \Sigma_{\delta \beta} (t) \\ & -i \sum_\gamma 
			\left[ \Sigma_{\alpha \gamma} (t) G_{\gamma \beta} (t) + G_{\alpha \gamma} (t) 
			\Sigma_{\gamma \beta} (t) \right] . 
		\end{split}
	\end{align}}
\end{subequations}
\noindent \negthickspace Note that Eq. \!(\ref{GVP_Meane}) takes account of the wave functional phase, as given in the transformation (\ref{GVP_WaveFunc_Phase}).  \\
\indent Lastly, Hamilton's equations of motion follow from the effective action which is defined as
% Eq. 52
\begin{equation} \label{GVP_Action}
	S = \int L(t)dt = \int dt \, \langle \Psi ,t | i \partial_t - \hat{H} | \Psi ,t \rangle .
\end{equation} 
Substitution of the mean values (\ref{GVP_MeanValues}) into the action obtains
% Eq. 53
\begin{align} \label{GVP_Action2}
	\begin{split}
		S &= \int dt \,\Biggl\{ \sum_\alpha \left[ \pi_\alpha(t) \dot{\psi}_\alpha(t) 
		+ \omega_\alpha(t) \dot{\phi}_\alpha(t) + \nu_\alpha(t) \dot{\chi}_\alpha(t) \right] \Biggr. \\
		&\qquad  \quad \;
		\Biggl. + \sum_{\alpha \beta} \Sigma_{\alpha \beta}(t) \dot{G}_{\beta \alpha}(t)
		+ {\mathcal N} \dot{\theta}(t) - {\mathcal H} \Biggr\} ,
	\end{split}
\end{align}
where
% Eq. 54
{\allowdisplaybreaks \begin{align} 
	{\mathcal H} &= \langle \Psi,t | \hat{H} | \Psi, t \rangle  \label{GVP_<H>} \\
%\end{equation}
%\noindent and
% Eq. 55
%\begin{equation} 
	%\begin{split}  
		{\mathcal N} &= \langle \Psi, t | \hat{N} | \Psi, t \rangle; \label{GVP_<N>} \\
		\hat{N} &= \sum_\alpha \left( \hat{\Psi}^\dag_\alpha \hat{\Psi}_\alpha 
		+ 2 \hat{\Phi}_\alpha^\dag \hat{\Phi}_\alpha + 2 \hat{X}_\alpha^\dag \hat{X}_\alpha \right) .
		\nonumber
	%\end{split}
\end{align}} \\
\indent Since ${\mathcal N}$ is constant, $\dot{\theta}$ must be time independent, thus giving $\dot{\theta} = 0 \Rightarrow \theta = \mbox{const.} = \mu$, which we identify as the chemical potential at zero temperature.  By stationarizing the action, the equations of motion emerge as
% Eqs. 56
\begin{subequations} \label{GVP_Hamilteom}
	\begin{align}
	%\begin{equation} 
		\dot{G}_{\alpha \beta} (t)  &= \frac{\delta ({\mathcal H} - \mu {\mathcal N})}
		{\delta \Sigma_{\alpha \beta} (t)} 
		\label{GVP_Hamiltoneom_a} \\
	%\end{equation}
	% \begin{equation}  
		\dot{\Sigma}_{\alpha \beta} (t)  &= - \frac{\delta ({\mathcal H} - \mu {\mathcal N})}
		{\delta G_{\alpha \beta} (t)}
		\label{GVP_Hamiltoneom_b} \\ 
	%\end{equation}
	%\begin{equation} 
		\dot{\psi}_\alpha (t) &= \frac{\delta ({\mathcal H} - \mu {\mathcal N})}{\delta \pi_\alpha (t)} 
		\label{GVP_Hamiltoneom_c} \\
	%\end{equation}
	%\begin{equation} 
		\dot{\pi}_\alpha (t) &= -\frac{\delta ({\mathcal H} - \mu {\mathcal N})}{\delta \psi_\alpha (t)} .
		\label{GVP_Hamiltoneom_d}
	%\end{equation}
	\end{align}
\end{subequations}
Not shown are the two molecular fields since their derivatives are straightforward analogs of (\ref{GVP_Hamiltoneom_c}) and (\ref{GVP_Hamiltoneom_d}).  From these it is evident that (${\mathcal N}, \theta$), ($\pi, \psi$), ($\omega, \phi$), ($\nu, \chi$) and ($\Sigma, G$) are canonical pairs. \\
\indent Employing Eqs. \!(\ref{GVP_MeanValues}) and (\ref{GVP_<N>}), the grand canonical Hamiltonian corresponding to (\ref{H_2Res_Hamilt}) has the momentum space expectation value of 
% Eq. 57
{\allowdisplaybreaks \begin{align} \label{GVP_HExpectVal}
	%\begin{split}
		K \equiv {\mathcal H} - \mu &{\mathcal N} = \int \limits_\mathbf{k} \left(k^2 - \mu \right) 
		\left[ R(\mathbf{k}, \mathbf{k}, t) + \Psi^\ast (\mathbf{k}, t) \Psi (\mathbf{k}, t) \right] \nonumber \\
		&+ \left( \epsilon - 2\mu \right) \int \limits_\mathbf{k} \Phi^\ast (\mathbf{k}, t) \Phi (\mathbf{k}, t)
		+ \left( \varepsilon - 2 \mu \right) 
		\int \limits_\mathbf{k} X^\ast (\mathbf{k}, t) X (\mathbf{k}, t) \nonumber \\
		&+ \frac{1}{\sqrt{2}} \int \limits_{\mathbf{k}, \mathbf{k'}, \mathbf{k''}} \negthickspace
		\delta (\mathbf{k''} - \mathbf{k}
		+ \mathbf{k'}) f \left(\frac{\mathbf{k} + \mathbf{k'}}{2} \right) \nonumber \\
		& \hphantom{+  \int \limits_{\mathbf{k}, \mathbf{k'}, \mathbf{k''}}}
		\times \left\{ \Xi^\ast (\mathbf{k''}, t) 
		\left[ -D(\mathbf{k}, \mathbf{k'}, t) + \Psi(\mathbf{k}, t) \Psi(\mathbf{k'}, t) \right] \right. 
		\nonumber \\
		& \hphantom{+  \int \limits_{\mathbf{k}, \mathbf{k'}, \mathbf{k''}} \times}
		\left. + \, \Xi (\mathbf{k''}, t) \left[-D^\ast(\mathbf{k}, \mathbf{k'}, t) 
		+ \Psi^\ast(\mathbf{k}, t) \Psi^\ast(\mathbf{k'}, t) \right] \right\} .
	%\end{split}
\end{align}} \\
\noindent In the kinetic energy term, we have used the momentum space form $T(\mathbf{k}, \mathbf{k'}) = k^2 \delta(\mathbf{k} - \mathbf{k'})$, whereas the coupling term's form factor, $F$, has been chosen to be proportional to $f$ in the separable potential introduced in (\ref{TBSeparablePotential}).  Simplifying the notation, two molecular fields have been combined in the term $\Xi(\mathbf{k''},t) = \lambda \alpha_1 \Phi(\mathbf{k''},t) + g X(\mathbf{k''}, t)$.\footnote{$\Phi$ and $X$ are defined in analogy to (\ref{GVP_Meana}).}  Before trying to analyze the full time-dependent problem, it is beneficial to first examine the special case of the static, uniform medium.  Moreover, the uniform results may be applied to nonuniform trapping geometries by an application of the local density approximation.  
% Section: Static, Uniform Solution
\section{Static, Uniform Solution}\label{StaticUniformSolution}
The static, uniform case has zero momenta ($\Sigma= 0$, $\pi= \omega= \nu=0$) with constant mean fields in all of space, thus imparting a continuous translational symmetry to the system.  For a hard sphere bose gas, the uniform energy per particle (on the scale of $\hbar^2/2m$) is known to be
% Eq. 58
\begin{equation} \label{SUS_HardSphere}
	e = \frac{u}{\rho} = 4 \pi a \rho + \frac{512 \sqrt{\pi}}{15} a^{5/2} \rho^{3/2} + \ldots \, ,
\end{equation}
\noindent where $u$ is the energy density and $a$ is the s-wave scattering length \cite{Fetter}.  It has also been shown that the lowest order term $4 \pi a \rho$ is independent of the form of the two-body interaction \cite{Lieb}.  \\
\indent We compare this result with a variational many-body analysis of the Hamiltonian in (\ref{GVP_HExpectVal}) which is further simplified by noting that for a uniform system, the Gaussian trial functional's width is diagonal,
% Eq. 59
\begin{equation} \label{SUS_GDiag}
	G(\mathbf{k}, \mathbf{k'}) = G(\mathbf{k}) \delta(\mathbf{k} - \mathbf{k'}) .
\end{equation}
According to Eqs. \!(\ref{GVP_RD}), this yields diagonal forms for the fluctuations as well
% Eq. 60
\setlength{\jot}{.15in}
\begin{subequations} \label{SUS_RDDiag}
	\begin{align}
		\begin{split} \label{SUS_RDiag}
			R(\mathbf{k}, \mathbf{k'}) &= \frac{1}{2} \left[ \frac{1}{4} G(\mathbf{k})^{-1} 
			+ G(\mathbf{k}) - 1 \right] \delta(\mathbf{k} - \mathbf{k'}) \\
			&\equiv R(\mathbf{k}) \delta(\mathbf{k} - \mathbf{k'})
		\end{split} \\[.15in]
		\begin{split} \label{SUS_DDiag}
			D(\mathbf{k}, \mathbf{k'}) &= \frac{1}{2} \left[ \frac{1}{4} G(\mathbf{k})^{-1}
			- G(\mathbf{k}) \right] \delta (\mathbf{k} - \mathbf{k'}) \\
			&\equiv D(\mathbf{k}) \delta(\mathbf{k} - \mathbf{k'}) .
		\end{split}
	\end{align}
\end{subequations}
Likewise, in momentum space, the mean fields are simply constants multiplied by delta functions:
% Eqs. 61
\begin{subequations} \label{SUS_MeanFields}
	\begin{align}
		\psi(\mathbf{k}) &= \psi \delta(\mathbf{k}) \label{SUS_Meanpsi} \\
		\phi(\mathbf{k}) &= \phi \delta(\mathbf{k}) \label{SUS_Meanphi} \\
		\chi(\mathbf{k}) &= \chi \delta(\mathbf{k}) . \label{SUS_Meanchi}
	\end{align}
\end{subequations}
\indent To obtain finite quantities, the Hamiltonian is divided by the volume of space.  Applied to a uniform system, the First Law of thermodynamics, $dE = - P d{\mathcal V} + \mu d{\mathcal N}$, reveals the resulting quantity as the negative of the pressure,\footnote{The sign in front of $D(\mathbf{k})$ is unimportant since it can be absorbed into the as yet unknown parameter $\xi$.}
% Eq. 62
\begin{align} \label{SUS_Pressure1}
	\begin{split}
		-P = \frac{1}{\mathcal V} \left({\mathcal H} - \mu {\mathcal N} \right) 
		&= \int \limits_\mathbf{k} (k^2 - \mu) R(\mathbf{k}) 
		+ \xi \int \limits_\mathbf{k} f(\mathbf{k}) D(\mathbf{k}) + \left( \xi - \mu \right) \frac{1}{2} \psi^2 \\
		& + \left( \frac{1}{2} \epsilon - \mu \right) \left( \frac{\xi + \eta}{2 \lambda \alpha_1} \right)^2
		+ \left( \frac{1}{2} \varepsilon - \mu \right) \left( \frac{\xi - \eta}{2 g} \right)^2 .
	\end{split}
\end{align}
\noindent In addition to taking the mean fields as real quantities, we have defined
% Eqs. 63
\begin{subequations} \label{SUS_xieta}
	\begin{align}
		\xi &= \lambda \alpha_1 \phi + g \chi \label{SUS_xi} \\
		\eta &= \lambda \alpha_1 \phi - g \chi .
	\end{align}
\end{subequations}
Along with $G(\mathbf{k})$, the variational parameters include the mean fields $\psi$, $\xi$ and $\eta$.  Extremizing on each results in the following set:
% Eqs. 64
\begin{subequations} \label{SUS_Variations}
	\begin{align}
		\frac{\delta P}{\delta G} &= 0 \Rightarrow G(\mathbf{k}) = \frac{1}{2} 
		\sqrt{\frac{k^2 - \mu + \xi f(\mathbf{k})}{k^2 - \mu - \xi f(\mathbf{k})}} \label{SUS_GVariation} \\
		\frac{\delta P}{\delta \psi} &= \left( \xi - \mu \right) \psi = 0 \label{SUS_PsiVariation} \\
		\frac{\delta P}{\delta \xi} &= \frac{\delta P}{\delta \eta} = 0 \Rightarrow
		\frac{4 \tau \sigma}{\tau + \sigma} \xi + \int \limits_\mathbf{k} f(\mathbf{k}) D(\mathbf{k})
		+ \frac{1}{2} \psi^2 = 0, \label{SUS_MolecularVariations}
	\end{align}
\end{subequations}
 \noindent where $\sigma = (\epsilon - 2 \mu)/(2 \lambda \alpha_1)^2$ and $\tau = (\varepsilon - 2 \mu) / (2 g)^2$.  We use a step function form factor, $f(\mathbf{k}) = \theta(| \mathbf{k} | - 4 \pi^2 / b)$, so that both $R(\mathbf{k})$ and $D(\mathbf{k})$ vanish for $k > 4 \pi^2/b$ where $G=1/2$.  Thus, the upper limit in all radial integrals is cut off at $4 \pi^2 /b$ while taking $f(\mathbf{k}) = 1$ everywhere.  As a further reduction, the zero-range limit ($b \to 0$ and $\lambda \to 0^-$) is taken, whence we require the following quantities to the appropriate order in $\lambda$:\footnote{Recall that $g = \sqrt{-2/ \lambda}$ and $\varepsilon = 2/ \lambda^2$.}
% Eqs. 65
\begin{subequations} \label{SUS_LambdaExpansions}
	\begin{align}
		\frac{4 \tau \sigma}{\tau + \sigma} &\simeq - \frac{1}{\lambda} -  
		\frac{\alpha_1^2}{\epsilon - 2 \mu} \label{SUS_LambdaExpansion_a} \\
		\phi = \frac{\xi  + \eta}{2 \lambda \alpha_1} &\simeq - \frac{\alpha_1 \xi}{\epsilon - 2 \mu} 
		\label{SUS_LambdaExpansion_b} \\
		\chi^2 = \left( \frac{\xi - \eta}{2 g} \right)^2 &\simeq - \left( \frac{1}{2} \lambda +  
		\frac{\alpha_1^2}{\epsilon - 2 \mu} \lambda^2 \right)\xi^2. \label{SUS_LambdaExpansion_c}
	\end{align} \\
\end{subequations}
\indent After substitution of (\ref{SUS_GVariation}) into Eqs. \!(\ref{SUS_RDDiag}), the required integrals are written out explicitly, each being expressible in terms of elliptic integrals of the first and second kind, denoted as $F$ and $E$, respectively \cite{EllipInt}:
% Eqs. 66
\begin{subequations} \label{SUS_RDExpressions}
	{\allowdisplaybreaks \begin{align} 
		\begin{split} \label{SUS_k2RExpression}
			\int \limits_\mathbf{k} k^2 R(\mathbf{k}) &= \frac{1}{4 \pi^2} \int \limits_0^{4 \pi^2 /b}
			 k^4 \left[ \frac{k^2 + \gamma^2}{\sqrt{(k^2 + \gamma^2)^2 - \xi^2}} - 1 \right] dk \\
			&\xrightarrow[b \to 0]{} \frac{\sqrt{\gamma^2 + |\xi|}}{20 \pi^2} 
			\left[ - \left( 3 \xi^2 + \gamma^4 \right) E
			+ \gamma^2 \left( \gamma^2 - |\xi| \right) F \right] + \frac{\xi^2}{2 b} 
		\end{split} \\[.12in]
		\begin{split} \label{SUS_RExpression}
			\int \limits_\mathbf{k} R(\mathbf{k}) &= \frac{1}{4 \pi^2} \int \limits_0^{4 \pi^2 /b} k^2 \left[
			\frac{k^2 + \gamma^2}{\sqrt{(k^2 + \gamma^2)^2 - \xi^2}} - 1 \right] dk \\
			&\xrightarrow[b \to 0]{} \frac{\sqrt{\gamma^2 + |\xi|}}{12 \pi^2} 
			\left[ \gamma^2 E - \left( \gamma^2 
			- |\xi| \right) F \right] 
		\end{split} \\[.12in]
		\begin{split} \label{SUS_DExpression}
			\int \limits_\mathbf{k} D(\mathbf{k}) &= - \frac{\xi}{4 \pi^2} \int \limits_0^{4 \pi^2 / b} \left[
			\frac{k^2}{\sqrt{(k^2 + \gamma^2)^2 - \xi^2}} - 1 \right] dk - \frac{\xi}{b} \\
			&\xrightarrow[b \to 0]{} \frac{\xi}{4 \pi^2} \sqrt{\gamma^2 + |\xi|} E - \frac{\xi}{b}.
		\end{split}
	\end{align}} \\
\end{subequations}
\noindent The resulting expressions have been expanded in the $b \rightarrow 0$ limit, omitting terms of order $b$ and higher.  Identifying $\mu = -\gamma^2$, the elliptic integral arguments are implicitly understood to be $\sqrt{2 |\xi| / ( |\xi| + \gamma^2 )}$. \\
\indent Retaining only the nonvanishing terms in the $\lambda \rightarrow 0^-$ limit, the pressure and the density become
% Eq. 67
\begin{align} 
	-P &= \int \limits_\mathbf{k} (k^2 - \mu) R(\mathbf{k}) 
	+ \xi \int \limits_\mathbf{k} D(\mathbf{k}) 
	+ \left( \xi - \mu \right) \frac{1}{2} \psi^2 - \frac{1}{2} \frac{\alpha_1^2 \xi^2}{\epsilon - 2 \mu}
	- \frac{\xi^2}{2 \lambda} \nonumber \\ 
	\begin{split}
	&= \frac{1}{2} \int \limits_\mathbf{k} \left[ \sqrt{ \left( k^2 - \mu \right)^2 - \xi^2} 
	- \left( k^2 - \mu \right) + \frac{\xi^2}{2 k^2} \right]
	+ \left( \xi - \mu \right) \frac{1}{2} \psi^2 \\
	& \qquad - \frac{\xi^2}{16 \pi a(\epsilon - 2 \mu)}
	\end{split}
	 \label{SUS_Pressure}  \\
% Eq. 68
	\rho &= \int \limits_\mathbf{k} R(\mathbf{k}) + \frac{1}{2} \psi^2 
	+ \frac{\alpha_1^2 \xi^2}{\left( \epsilon - 2 \mu \right)^2}. \label{SUS_Density}
\end{align}
As in (\ref{SUS_Pressure}), we will frequently find it convenient to use the full scattering length (\ref{TBa_eff}), where $\epsilon$ is shifted by $- 2\mu$.  Both pressure and density depend on three parameters, $\psi$, $\gamma^2 = - \mu$ and $\xi$, which are constrained by the two variational Equations (\ref{SUS_PsiVariation}) and (\ref{SUS_MolecularVariations}).  Through the relation (\ref{SUS_Density}), the density is chosen as the free parameter.
% subsection: \psi = 0 Solution
\subsection{$\psi = 0$ Solution}
First, note that Eq. \!(\ref{SUS_PsiVariation}) admits two solutions, with one for $\psi =0$ and the other for $\xi = \mu = -\gamma^2$.  We begin with the former, describing a system composed entirely of a molecular condensate and correlated atom pairs, but with no atomic condensate component.  Substitution of (\ref{TBa_eff}), (\ref{SUS_LambdaExpansion_a}) and (\ref{SUS_DExpression}) into (\ref{SUS_MolecularVariations}) results in an expression that relates $\xi$ to $\gamma^2$:
% Eq. 69
%\begin{equation} \label{Psi0_Condition}
%	\frac{\xi}{8 \pi a(\epsilon + 2 \gamma^2)} - \frac{\xi}{4 \pi^2} \sqrt{ |\xi| + \gamma^2} E = 0 .
%\end{equation}
\begin{equation} \label{Psi0_Condition}
	\frac{1}{8 \pi a(\epsilon + 2 \gamma^2)} + \frac{1}{4 \pi^2} \int \limits_0^{4 \pi^2/b} 
	\left[ \frac{k^2}{\sqrt{\left(k^2 + \gamma^2 \right)^2 - \xi^2}} -1 \right] dk = 0,
\end{equation}
\noindent where, constrained by the form of the integral, $\xi$ ranges from $-\gamma^2$ to zero.  Differentiation of the density expansion (\ref{SUS_Density}) indicates that $d \rho / d \xi^2 > 0$, thus implying an increasing density with increasing $\xi^2$.  When $\xi = -\gamma^2$, a critical density is reached since (\ref{Psi0_Condition}) no longer admits a real solution for $\xi^2 > \gamma^4$.  According to (\ref{SUS_Density}), dilute gases  ($\rho a^3 \ll 1$) in the low-density regime correspond to small $\xi$.  In the limit $\xi \to 0$, (\ref{Psi0_Condition}) reduces to
% Eq. 70
\begin{equation} \label{Psi0_Binding}
	\gamma_0 \, a(\epsilon + 2 \gamma_0^2) = 1.
\end{equation}
\noindent When compared with (\ref{TBBE}), this is identified as the equation for the binding energy $2 \gamma_0^2$. \\
\indent In addition to (\ref{Psi0_Condition}), the equation of state for $\psi = 0$ is specified by the energy and number densities which, when cast in terms of the elliptic integrals become
% Eqs. 71, 72
\begin{align} 
	\begin{split} \label{Psi0_u} 
		u &= -P + \mu \rho = \frac{\sqrt{\gamma^2 + |\xi|}}{20 \pi^2} \left[ \left( 2 \xi^2 
		- \gamma^4 \right) E + \gamma^2 \left( \gamma^2 - |\xi| \right) F \right] \\
		& \hphantom{= -P + \mu \rho =} \quad -\frac{\xi^2}{16 \pi a (\epsilon + 2 \gamma^2)} 
		- \frac{\gamma^2 \alpha_1^2 \xi^2}{\left( \epsilon + 2 \gamma^2 \right)^2}
	\end{split} \\
	\rho &= \frac{\sqrt{\gamma^2 + |\xi|}}{12 \pi^2} \left[ \gamma^2 E - \left(\gamma^2 - |\xi| \right) F \right]
	+ \frac{\alpha_1^2 \xi^2}{\left( \epsilon + 2 \gamma^2 \right)^2}. \label{Psi_rho}
\end{align}
From the limit $\xi \to 0$, the energy per particle, $e = u/ \rho$, reduces to minus half the molecular binding energy, $- \gamma_0^2$.  This result makes physical sense because at zero density, only the molecular state remains.  Although the $\psi = 0$ solution terminates when $\xi = - \gamma^2$, the variational Equation (\ref{SUS_PsiVariation}) also permits an alternative solution given by $\xi = \mu = -\gamma^2$, which corresponds to a nonzero atomic field in general.
% subsection: \psi \neq 0 Solution
\subsection{$\psi \neq 0$ Solution}
\begin{figure} [h]
\begin{center}
\epsfxsize=2.75in\epsfbox{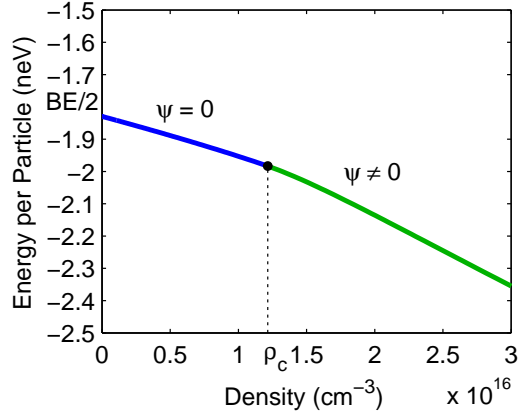}
\end{center}
\caption{Figure showing the two-piece collapsing ground state for the case of $^{85}$Rb at a magnetic field of $162.3$ G, corresponding to a critical density, $\rho_c$, of $1.22 \times 10^{16}$cm$^{-3}$ with a full scattering length of $193$ Bohr radii \cite{ParamValues}.  In the limit of zero density, the energy per particle simply reduces to half the molecular binding energy, BE/2 $\sim 1.85$neV. By tuning the magnetic field closer to resonance, the relatively high critical density may be brought within the regime of current experiments. \label{CollapsingGndSt}}
\end{figure}

The solution for $\xi = -\gamma^2$ permits an elementary evaluation of the integrals in (\ref{SUS_RDExpressions}):
% Eqs. 73
\begin{subequations} \label{Psin0_RDExpressions}
	{\allowdisplaybreaks \begin{align}
		\int \limits_\mathbf{k} k^2 R(\mathbf{k}) &\xrightarrow[b\to 0]{} \frac{\gamma^4}{2b} 
		- \frac{\sqrt{2}}{5 \pi^2} \gamma^5 \label{Psi0_k2R} \\
		\int \limits_\mathbf{k} R(\mathbf{k}) &\xrightarrow[b\to 0]{} 
		\frac{\sqrt{2}}{12 \pi^2} \gamma^3 \label{Psi0_R} \\
		\int \limits_\mathbf{k} D(\mathbf{k}) &\xrightarrow[b\to 0]{} \frac{\gamma^2}{b} 
		- \frac{\sqrt{2}}{4 \pi^2} \gamma^3. \label{Psi0_D}
	\end{align}} \\
\end{subequations}
Unlike the $\psi = 0$ case, (\ref{SUS_MolecularVariations}) is now solved for the atomic field:
% Eq. 74
\begin{equation} \label{Psin0_AtomicField}
	\frac{1}{2} \psi^2 \xrightarrow[b\to0]{} - \frac{\gamma^2}{8 \pi a(\epsilon + 2\gamma^2)}
	 + \frac{\sqrt{2}}{4 \pi^2} \gamma^3.
\end{equation}
With these expressions, both the energy and number densities are parametrized by $\gamma^2$:
% Eqs. 75, 76
\begin{align}
	u &= \left[ \frac{1}{16 \pi a_{bg}} + \frac{1}{2} \frac{\epsilon \alpha_1^2}
	{\left( \epsilon + 2 \gamma^2 \right)^2} \right] \gamma^4 - \frac{\sqrt{2}}{5 \pi^2} \gamma^5 
	\label{Psin0_u} \\
	\rho &= - \frac{1}{8 \pi a(\epsilon + 2 \gamma^2)} \gamma^2 
	+ \frac{\sqrt{2}}{3 \pi^2} \gamma^3 
	+ \frac{\alpha_1^2}{\left( \epsilon + 2 \gamma^2 \right)^2} \gamma^4. \label{Psin0_Rho}
\end{align}
\indent For the example of $^{85}$Rb, Figure \ref{CollapsingGndSt} shows the energy per particle curve formed by the merging of the $\psi = 0$ and $\psi \neq 0$ solutions at the critical point.  The fact that this point marks a quantum phase transition \cite{Rad1, Rad2, QPT} follows from a discontinuity in the density between the two cases.  That is, one piece has an atomic condensate density component and the other does not.  Due to the persistently negative slope of the energy per particle, the pressure, $P = \rho^2 de/ d\rho$, is always negative, by which we denote this two-piece solution as the ``collapsing ground state'' of the model Hamiltonian (\ref{H_2Res_Hamilt}).  Of particular importance is that this collapsing behavior persists even when the full scattering length is tuned to positive values, a regime thought to be stable against collapse.  Indeed, this result challenges the intuitive low-density scattering length behavior, $e \sim 4 \pi a \rho$, indicated in expansion (\ref{SUS_HardSphere}).
%Figure ******************
%Figure*******************
% Section: A Complex Chemical Potential
\section{A Complex Chemical Potential} \label{ComplexMu}
Due to an innately negative scattering length, $^{85}$Rb has two-body interactions that are attractive, resulting in an ensemble that tends toward collapse as quantum degeneracy is approached.  As such, this particular isotope cannot form a condensate containing more atoms than some critical number.  However, by exploiting a Feshbach resonance, the appearance of the molecular state allows the effective scattering length to be tuned to positive values, as indicated in Fig.\:\ref{Rba_vs_B}.  By tuning to positive values, a stable condensate has been experimentally observed in which the low-density energy per particle is found to have the usual behavior, $e \sim 4\pi a \rho$, where $a$ is the full scattering length \cite{Rb85BEC}. \\
\indent Despite the positive scattering length, our previous solutions predict that a uniform assembly should still tend toward collapse, confronting us with the dilemma that the separable potential somehow does not capture the correct many-body physics even though it does reproduce the correct two-body scattering (see Sec.\,\ref{Sec:SepPot}).  To resolve this issue, it is necessary to reexamine the solutions obtained in the collapsing case.  Since the experiments have observed the formation of an atomic condensate, we focus on the solution for $\psi \neq 0$, exploring the consequences of extending this solution to zero density.  To do so, the density Equation (\ref{Psin0_Rho}) is inverted, thereby obtaining $\mu$ as an expansion in $\sqrt{\rho}$:
% Eq. 77
\begin{equation} \label{CCP_MuExpansion}
	\begin{split}
		-\gamma^2 = \mu = 8 \pi a(\epsilon) \rho &- i\frac{\sqrt{\pi}}{3} \, 256 \, 
		a(\epsilon)^{5/2} \rho^{3/2} \\
		&-\Biggl[ \frac{64 \, a(\epsilon)^2}{3 \pi^2} + 8 \pi a(\epsilon) \frac{3\alpha_1^2}{\epsilon^2} 
		\Biggr] 64 \, \pi^2 a(\epsilon)^2 \rho^2 - \ldots \, .
	\end{split}
\end{equation}
\noindent When this expansion is substituted into (\ref{Psin0_u}), the usual low-density energy dependence, $e = u / \rho \sim 4 \pi a \rho$, is achieved.  Although this does give the appropriate form, the chemical potential becomes complex at higher order.  At first it may seem that this is simply an unphysical solution, but upon further consideration it is recognized as a signature of an instability that is inherent in the original system.  \\
\indent With the Hamiltonian given by (\ref{H_2Res_Hamilt}), consider the Heisenberg equation of motion for the atomic field operator $\hat{\Psi}^\dag_\zeta$: $i \hbar \, \partial_t \hat{\Psi}^\dag_\zeta = [\hat{\Psi}^\dag_\zeta, \hat{H}]$.  Employing the commutators $[\hat{\Psi}_\alpha, \hat{\Psi}^\dag_\beta] = \delta_{\alpha \beta}$, $[\hat{\Psi}^\dag_\zeta, \hat{\Psi}^\dag_\alpha \hat{\Psi}_\beta] = -\delta_{\zeta \beta} \, \hat{\Psi}^\dag_{\alpha}$ and $[\hat{\Psi}^\dag_\zeta, \hat{\Psi}_\gamma \hat{\Psi}_\beta] = -\delta_{\zeta \gamma} \, \hat{\Psi}_{\beta} -\delta_{\zeta \beta} \, \hat{\Psi}_\gamma$, the right-hand side can be readily evaluated using the expectation value of the mean atomic field (\ref{GVP_Meana}).  This leads to
% Eq. 78
\begin{align} \label{CCP_Heisenberg}
	i \hbar \, \frac{\partial \Psi^\ast}{\partial t} &= -\sqrt{2} \left( \lambda \alpha_1 \Phi^\ast
	+ g X^\ast \right) \Psi \nonumber \\
	&= -\sqrt{2} \, \Xi^\ast \Psi.
\end{align}
\indent We seek to add a phase to the mean fields such that (\ref{CCP_Heisenberg}) is consistent with the variational Equations (\ref{SUS_Variations}).  If we let
% Eqs. 79
\begin{subequations}
	\begin{align}
		\Psi &= \frac{1}{\sqrt{2}} \, \psi \, \mbox{e}^{-i \mu t /\hbar}, \label{CCP_PsiPhase} \\
		\Xi &= \frac{1}{\sqrt{2}} \, \xi \, \mbox{e}^{-2i \mu t /\hbar}, \label{CCP_XiPhase}
	\end{align}
\end{subequations}
\noindent then (\ref{CCP_Heisenberg}) gives back (\ref{SUS_PsiVariation}).  Thus, it is reasonable to interpret the chemical potential as the phase of the mean fields \cite{Leggett}, where the imaginary part of $\mu$ leads to a decay rate given by
% Eq. 80
\begin{equation} \label{CCP_Decayrate}
	\Gamma = \frac{\hbar}{2m} \frac{\sqrt{\pi}}{3} \, 512 \, a(\epsilon)^{5/2} \rho^{3/2} + \ldots .
\end{equation}
\noindent Displaying unique dependencies on both the scattering length [$\sim a(\epsilon)^{5/2}$] and the density ($\sim \rho^{3/2}$), this coherent rate can be tested by further experiments.  Under the conditions of the $^{85}$Rb experiment, there were $10^4$ atoms within a cloud of radius $25 \, \mu$m when the scattering length was tuned to $193 a_0$ ($a_0 =$ Bohr radius).  These parameters yield a decay time, $\tau \sim 1 / \Gamma$, of $14.3$ seconds,\footnote{For an attractive interaction, the longest decay time is attained in the limit $\lambda \to 0^-$ since the rate increases with decreasing $\lambda$ \cite{CoherentDecay}.}which is in qualitative agreement with the observed $10$ second lifetime \cite{Rb85BEC}.  \\
\indent Because all quantities depend on $\mu$, which is in general complex, it follows that the various other thermodynamic functions assume a complex character as well.  This is not unphysical since the imaginary parts should simply be regarded as signatures of the coherent decay with the real parts assuming their usual physical interpretations.  Although complex frequencies are well-known to signify the damping of collective modes \cite{Beliaev}, we emphasize that the distinctive effect here predicts a {\it decay of the condensate itself, in the absence of any excitations}.  Nonetheless, a physical analogy may be found in the context of quantum electrodynamics, where the instability of the electric field emerges through a complex action \cite{QEDSchwinger}. 
%Figure ******************
\begin{figure} [h]
\begin{center}
\epsfxsize=2.75in\epsfbox{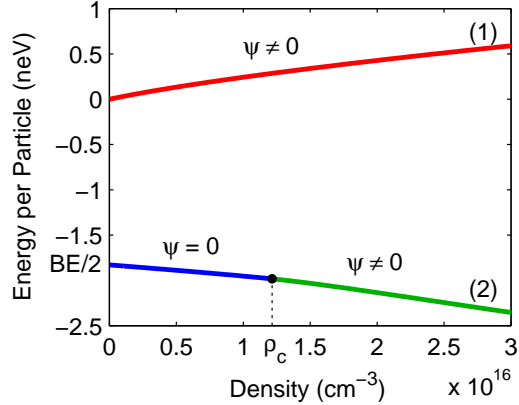}
\end{center}
\caption{For the case of $^{85}$Rb, we plot the real part of the energy per particle, $e = u/\rho$ [see Eqs. \!\!(\ref{Psin0_u}) and (\ref{CCP_MuExpansion})], corresponding to the decaying excited state (1).  Below this is the two-piece collapsing ground state (2) as shown in Fig.\:\ref{CollapsingGndSt}.  All results are for an applied magnetic field of $162.3$G.}  \label{AllSolns}
\end{figure}
%Figure*******************
% Here's where I stopped checking 8/9/06.  Need to check some two-body results too esp T matrix.
% Section: Decay in the Nonuniform Case
\section{Decay in the Nonuniform Case}\label{DecayinNonuniformCase}
\indent The complex chemical potential persists even in the static, but nonuniform case.  Thus far, only the uniform solutions have been discussed since it has been assumed that any nonuniformity can be accommodated using a local density approximation.  Therefore, one could conjecture that a chemical potential assuming a complex value is simply an artifact arising because the uniform solution is too restrictive to capture the physics observed in experiment.  \\
\indent To address this concern, the origin of the complex $\mu$ must be identified.  First, it is recognized that the expected solution has an equation of state associated with a positive chemical potential.  At low density, the energy per particle and its derivative satisfy $e \sim 4 \pi a \rho > 0$ and $de/ d\rho > 0$, respectively.  By the relationship\footnote{This can be seen by extremizing the pressure, $P = \mu \rho -u$, with respect to $\rho$, $\delta P / \delta \rho = 0$.}$\mu = du / d \rho$ and the definition of $e$ as $u / \rho$, it follows that $\mu = \rho(de/d\rho) +e$.  The positivity of $e$, $\rho$ and $de/d\rho$ thus implies the same for $\mu$.  From (\ref{SUS_GVariation}), it is seen that if $\mu$ is positive, then $G$ becomes imaginary for sufficiently small values of $k$.  By construction, $G^{-1}$ is the real part of the width, raising an apparent contradiction unless the chemical potential is allowed to be complex instead.  Therefore, an imaginary $G$ is also a harbinger of the same instability described in the previous section.  Most importantly, the ranging of $G$ into imaginary values provides a convenient test for this instability when discussing the nonuniform case where a full analytic expression for $\mu$ is not possible. \\
\indent For static nonuniformities, the grand canonical Hamiltonian can be written
% Eq. 81
\begin{align} \label{DNC_Ham}
	\begin{split}
		K &= \sum_{\alpha \beta} h_{\alpha \beta} \left( R_{\alpha \beta}
		+ \frac{1}{2} \, \psi_\alpha^\ast \, \psi_\beta \right) 
		+ \left( \frac{1}{2} \, \epsilon - \mu \right) \sum_\alpha \phi_\alpha^\ast \, \phi_\alpha \\ 
		& \hphantom{=} 
		+ \left( \frac{1}{2} \, \varepsilon - \mu \right) \sum_\alpha \chi_\alpha^\ast \, \chi_\alpha
		+ \frac{1}{2} \sum_{\alpha \beta \gamma} \xi^\ast_\alpha F_{\alpha; \beta \gamma} 
		\left( D_{\gamma \beta} + \frac{1}{2} \, \psi_\gamma \, \psi_\beta \right) + \mbox{H.c.},
	\end{split}
\end{align}
\noindent where $h_{\alpha  \beta} = T_{\alpha \beta} - \mu \, \delta_{\alpha \beta}$.  The fluctuation terms are given by the static form of (\ref{GVP_RD}):
% Eqs. 82
\begin{subequations} \label{DNC_RDStatic}
	\begin{align} 
		R_{\alpha \beta} &= \frac{1}{2} \left( \frac{1}{4} G_{\alpha \beta}^{-1} + G_{\alpha \beta}
		- \delta_{\alpha \beta} \right)\label{DNC_RStatic} \\
		D_{\alpha \beta} &= \frac{1}{2} \left( \frac{1}{4} G_{\alpha \beta}^{-1} - G_{\alpha \beta}
		\right)\! . \label{DNC_DStatic}
	\end{align}
\end{subequations}
\noindent Differentiating with respect to $G_{\rho \sigma}$ yields
% Eqs. 83
\begin{subequations} \label{DNC_RDDeriv}
	\begin{align} 
		\frac{\delta R_{\alpha \beta}}{\delta G_{\rho \sigma}} &= \frac{1}{2} \left( -\frac{1}{4} 
		G_{\alpha \rho}^{-1} G_{\beta \sigma}^{-1}
		+ \delta_{\alpha \rho} \, \delta_{\beta \sigma} \right) \label{DNC_RDeriv} \\
		\frac{\delta D_{\alpha \beta}}{\delta G_{\rho \sigma}} &= \frac{1}{2} \left( -\frac{1}{4} 
		G_{\alpha \rho}^{-1} G_{\beta \sigma}^{-1} 
		- \delta_{\alpha \rho} \, \delta_{\beta \sigma} \right)\!. \label{DNC_DDeriv}
	\end{align}
\end{subequations}
\noindent Using these derivatives, (\ref{DNC_Ham}) is extremized on the width
% Eq. 84
\begin{equation} \label{DNC_Ham_Deriv}
	\frac{\delta K}{\delta G_{\rho \sigma}} = -\frac{1}{8} \sum_{\alpha \beta}
	G_{\rho \alpha}^{-1} \, \bigl( h_{\alpha \beta} + \tilde{F}_{\alpha \beta} \bigr) \, 
	G_{\beta \sigma}^{-1} + \frac{1}{2}\sum_{\alpha \beta} \delta_{\rho \alpha} \bigl( h_{\alpha \beta} 
	- \tilde{F}_{\alpha \beta} \bigr) \, \delta_{\beta \sigma} = 0,
\end{equation}
\noindent with $\tilde{F}$ related to the form factor $F$ by\footnote{We have used the symmetry $F_{\tau; \alpha \beta} = F_{\tau; \beta \alpha}$.}
% Eq. 85
\begin{equation} \label{DNC_FTilde}
	\tilde{F}_{\alpha \beta} = \sum_\tau {\rm Re} \{ \xi_\tau \} \, 
	F_{\tau; \alpha \beta}.
\end{equation}
\indent After defining new matrices $Y$ and $Z$ as
% Eqs. 86
\begin{subequations} \label{DNC_ABhV}
	\begin{align}
		Y_{\alpha \beta} &= h_{\alpha \beta} + \tilde{F}_{\alpha \beta} \\
		Z_{\alpha \beta} &= h_{\alpha \beta} - \tilde{F}_{\alpha \beta},
	\end{align}
\end{subequations}
\noindent the variational Equation (\ref{DNC_Ham_Deriv}) can be put into the compact matrix form
% Eq. 87
\begin{equation} \label{DNC_ABMat_EqnforG}
	-\frac{1}{4} \, G^{-1} \, Y \, G^{-1} + Z = 0.
\end{equation}
\noindent To obtain a symmetric solution for $G^{-1}$, we first multiply by $\sqrt{Y}$ on the right and on the left,
% Eq. 88
\begin{equation}
	-\frac{1}{4} \sqrt{Y} \, G^{-1} \, \sqrt{Y} \, \sqrt{Y} \, G^{-1} \, \sqrt{Y} + \sqrt{Y} \, Z \, \sqrt{Y} = 0.
\end{equation}
\noindent Moving the first term to the right-hand side, then taking the square root, $G^{-1}$ is seen to be\footnote{This series of manipulations obtains the required symmetric form for $G^{-1}$.  In general, however, this is not a unique solution since multiplying (\ref{DNC_ABMat_EqnforG}) on the right by $Y$, then taking the square root results in $G^{-1} = 2 \, \sqrt{ZY} \, Y^{-1}$, which is different than the solution in (\ref{DNC_GinvNonUniformSoln}).}
% Eq. 89
\begin{equation} \label{DNC_GinvNonUniformSoln}
	G^{-1} = 2 \, \frac{1}{\sqrt{Y}} \, \sqrt{ \sqrt{Y} \, Z \, \sqrt{Y} } \, \frac{1}{\sqrt{Y}}.
\end{equation}
Likewise, the symmetric form of $G$ is most easily obtained by multiplying (\ref{DNC_ABMat_EqnforG}) on the left and right by $G$, then following the same steps to end up with
% Eq. 90
\begin{equation} \label{DNC_GNonUniformSoln}
	G = \frac{1}{2} \, \frac{1}{\sqrt{Z}} \, \sqrt{ \sqrt{Z} \, Y \, \sqrt{Z} } \, \frac{1}{\sqrt{Z}}.
\end{equation}
\indent In momentum space, the operator $h$ is given by $h(\mathbf{k}, \mathbf{k'}) = (k^2 - \mu) \delta (\mathbf{k} - \mathbf{k'})$.  For $k^2 = \mu$, $h = 0$.  At such values of $\mathbf{k}$, Eqs. \!(\ref{DNC_ABhV}) show that $Y = - Z = \tilde{F}$, thus giving
% Eq. 91
\begin{align} \label{DNC_ImagG}
	\begin{split}
		G \Big|_{k^2 = \mu} &= \frac{1}{2} \, \sqrt{ \frac{1}{\sqrt{-\tilde{F}}} \, \tilde{F} \,
		\frac{1}{\sqrt{-\tilde{F}}}} \\
		&= \pm \frac{i}{2},
	\end{split}
\end{align}
\noindent indicating that $G$ is in general complex.  By our earlier observation, it follows that the coherent decay is present even in the nonuniform case.
% Section: Small Oscillations: Generalized Random Phase Approximation
\section{Small Oscillations: Generalized Random Phase Approximation (RPA)}\label{RPA}
Evidence for the production of a coherent atom-molecule condensate has been demonstrated by the dynamical response of the $^{85}$Rb system near a Feshbach resonance.  In particular, oscillations in the atomic density were observed in trapped samples by Donley { \it et al.} \cite{AMCoherence}, indicating the coexistence of a molecular counterpart.  Although theoretical investigations have been conducted by Holland { \it et al.} \cite{PairField} and by Kokkelmans and Holland \cite{RamseyFringe}, it is perplexing that only the latter seems to predict a much longer damping rate in agreement with the experiment.  In any event, all of these investigations dealt with the dynamics of the expected solution which expands against the trap due to the positive scattering length.  Instead, our aim is to carry out a small oscillation expansion of the collapsing ground state since its spectrum completes the physical picture by providing an interpretation of the coherent decay uncovered in the previous two sections.\\
\indent To find the collapsing state's low-lying excitations, all variational quantities are expanded about their stationary values, resulting in an oscillator-like Hamiltonian expressed in terms of effective mass ($\mathfrak{A}$) and spring ($\mathfrak{B}$) matrices.  Solving the full problem is accomplished in a series of stages, with each including an added generality over the previous.  As the first and simplest step, we consider only the diagonal or noninteracting part of the Hamiltonian from which the corresponding ($\delta G$, $\delta \Sigma$) oscillations represent the energy of two free quasi-bosons.  Including both the diagonal and off-diagonal elements in $\mathfrak{A}$, a zero is obtained as a discrete point in its spectrum, thus verifying the presence of the Goldstone mode in general.  Nevertheless, a complete accounting of the excitations must be obtained from the product $\mathfrak{A} \cdot \mathfrak{B}$.  Correspondingly, the eigenvalue problem has the familiar Lippmann-Schwinger form, indicating the associated eigenfrequencies to be identified with the energy of two interacting quasi-bosons.  In addition to discrete eigenfrequencies, there exists a phonon continuum of the collapsing solution.  Since the expected energy per particle lies within the continuum, energy conservation gives a natural interpretation of the decay as a transition into the phonon excitations of the collapsing state.
% Subsection: General Expansion
\subsection{General Expansion}
All stationary quantities are expanded about their uniform equilibrium values:
% Eqs. 92
\begin{subequations} \label{SO_RPAexpansion}
	{\allowdisplaybreaks \begin{align}
		G(\mathbf{k},\mathbf{k'},t) & = G(\mathbf{k}) \delta(\mathbf{k} - \mathbf{k'}) + \delta 
		G(\mathbf{k},\mathbf{k'},t) \label{SO_RPA_G} \\
		\Sigma(\mathbf{k},\mathbf{k'},t) & = \delta \Sigma(\mathbf{k},\mathbf{k'},t) \label{SO_RPA_S} \\
		\psi(\mathbf{k},t) & = \psi \delta(\mathbf{k}) + \delta \psi(\mathbf{k},t) \label{SO_RPA_psi} \\
		\pi(\mathbf{k},t) & = \delta \pi(\mathbf{k},t)  \label{SO_RPA_pi} \\
		\phi(\mathbf{k},t) & = \phi \delta(\mathbf{k}) + \delta \phi(\mathbf{k},t) \label{SO_RPA_phi} \\
		\omega(\mathbf{k},t) & = \delta \omega(\mathbf{k},t)  \label{SO_RPA_omega} \\
		\chi(\mathbf{k},t) &= \chi \delta(\mathbf{k}) + \delta \chi(\mathbf{k},t) \label{SO_RPA_chi} \\
		\nu(\mathbf{k},t) &= \delta \nu(\mathbf{k},t) \label{SO_RPA_nu} \\
		\xi(\mathbf{k},t) &= \xi \delta(\mathbf{k}) + \delta \xi(\mathbf{k},t). \label{SO_RPA_xi}
	\end{align}}
\end{subequations}   
\noindent It will prove convenient to define new momenta
% Eqs. 93
\begin{subequations} \label{SO_NewMomentumCoords}
	{\allowdisplaybreaks \begin{align}
		\mathbf{P} & = \mathbf{k} - \mathbf{k'} \label{SO_NewMomentumCoordsP} \\
		\mathbf{q} & = \frac{1}{2} \left(\mathbf{k} + \mathbf{k'} \right)\!, 
		\label{SO_NewMomentumCoordsq}
	\end{align}}
\end{subequations}
\noindent \hspace{-7pt} having the interpretation as the respective total and relative momenta of a pair of quasiparticles.  \\
\indent To illustrate the relationship between the variations and their inverses, consider the expression for $G^{-1}(\mathbf{k},\mathbf{k'},t)$, which, to second-order in $\delta G$, we write as
% Eq. 94
\begin{equation} \label{SO_G_inv}
	G^{-1}(\mathbf{k},\mathbf{k'},t) = G(\mathbf{k})^{-1} \delta(\mathbf{k} - \mathbf{k'}) + \delta G^{-1}
	 (\mathbf{k}, \mathbf{k'},t) + \frac{1}{2} \delta G^{-2} (\mathbf{k}, \mathbf{k'},t).
\end{equation}
\noindent By using the identity $GG^{-1} = 1$, it follows that
% Eqs. 95
\begin{subequations} \label{SO_DeltaG12inv}
	\begin{align}
		\delta G^{-1} (\mathbf{k},\mathbf{k'},t) & = -G(\mathbf{k})^{-1} \delta G(\mathbf{k},\mathbf{k'},t) 
		G(\mathbf{k'})^{-1} \label{SO_DeltaG1inv} \\
		\delta G^{-2} (\mathbf{k},\mathbf{k'},t) & = 2G(\mathbf{k})^{-1} \int \limits_\mathbf{k''} 
		\delta G(\mathbf{k},\mathbf{k''},t) G(\mathbf{k''})^{-1} \delta G(\mathbf{k''},\mathbf{k'},t) 
		G(\mathbf{k'})^{-1}. \label{SO_DeltaG2inv}
	\end{align}
\end{subequations}
\noindent Upon promoting all quantities to the new variables $(\mathbf{P}, \mathbf{q})$, Hermitian conjugation, $\delta G (\mathbf{k}, \mathbf{k'},t)^\dag = \delta G^\ast (\mathbf{k'}, \mathbf{k}, t)$, obeys the mapping
% Eq. 96
\begin{equation} \label{SO_DeltaGtransf}
	\delta G(\mathbf{k},\mathbf{k'},t) \rightarrow \delta G(\mathbf{q}, \mathbf{P},t) \Leftrightarrow 
	\delta G^\ast (\mathbf{k'},\mathbf{k},t) \rightarrow \delta G^\ast (\mathbf{q}, -\mathbf{P},t),
\end{equation}
\noindent which, by the definition of $\mathbf{P}$ given in (\ref{SO_NewMomentumCoordsP}), follows from the interchange of $\mathbf{k}$ and $\mathbf{k'}$. \\
\indent Undertaking a harmonic analysis, the mean grand canonical Hamiltonian, $K = \langle \hat{H} - \mu \hat{N} \rangle$, must be expanded to second-order in the small deviations of (\ref{SO_RPAexpansion}), 
% Eq. 97
\begin{equation} \label{SO_Ham2ndorderexpnsn} 
	K = K^{(0)} + \delta K \delta {\mathcal P} + \frac{1}{2} \delta {\mathcal P}^\dag  \, \mathfrak{A} \, \delta {\mathcal P} + 
	\frac{1}{2} \delta {\mathcal Q}^\dag \, \mathfrak{B} \, \delta {\mathcal Q}, 
\end{equation}
\noindent where the vector $\delta {\mathcal Q}$ and its canonical conjugate $\delta {\mathcal P}$ are given as
% Eqs. 98
\begin{subequations} 
	\begin{align}
		\delta {\mathcal Q}^\dag (\mathbf{q},\mathbf{P},t) &= [ \delta \Sigma(\mathbf{q}, -\mathbf{P}, t), 
		\delta \pi(-\mathbf{P},t), \delta \omega(-\mathbf{P},t), \delta \nu(-\mathbf{P},t) ]
		\label{SO_dQ} \\
		\delta {\mathcal P}^\dag (\mathbf{q},\mathbf{P},t) &= [ \delta G(\mathbf{q}, -\mathbf{P}, t),
		\delta \psi(-\mathbf{P},t), \delta \phi(-\mathbf{P},t), \delta \chi(-\mathbf{P},t) ].
		\label{SO_dP}
	\end{align}
\end{subequations}
In expression (\ref{SO_Ham2ndorderexpnsn}), the zero-order constant term, $K^{(0)}$, is the static, uniform piece that can be dropped as it merely represents a constant energy shift, having no effect on any ensuing dynamics.  Also, the first variation, $\delta K$, is zero at the stationary points given by (\ref{SUS_Variations}), thus leaving a quadratic Hamiltonian in $\delta {\mathcal Q}$ and $\delta {\mathcal P}$.  Consequently, $\mathfrak{A}$ and $\mathfrak{B}$ can be interpreted as the mass and spring matrices whose product has eigenvalues that determine the low-lying excitations of the system. \\
\indent The second-order expansion of ${\mathcal H} - \mu {\mathcal N}$ has been carried out explicitly in the appendix, which finds the $\mathfrak{A}$ and $\mathfrak{B}$ matrices to be
% Eq. 99
%\begin{align} \label{SO_HamRPA}
%	\begin{split}
%		K_{RPA}(\mathbf{P},t) &= \frac{1}{2} \delta {\mathcal P}^\dag (\mathbf{q},\mathbf{P}, t) \, 
%		\mathfrak{A}(\mathbf{q},\mathbf{q'},\mathbf{P}) \delta {\mathcal P}(\mathbf{q'},\mathbf{P},t) \\
%		& + \frac{1}{2} \delta {\mathcal Q}^\dag (\mathbf{q},\mathbf{P},t) 
%		\mathfrak{B}(\mathbf{q},\mathbf{q'},\mathbf{P}) \delta {\mathcal Q} (\mathbf{q'}, \mathbf{P},t) ,
%	\end{split}
%\end{align}
%\noindent in which the $\mathbf{q}$ and $\mathbf{q'}$ integrations are understood.  Lastly, the $\mathfrak{A}$ and $\mathfrak{B}$ can be written as
% Eqs. 99
\begin{subequations} \label{SO_ABmatrices}
	\begin{align}
		\mathfrak{A}(\mathbf{q},\mathbf{q'},\mathbf{P},t) & = \begin{bmatrix} 
		s_M(\mathbf{q},\mathbf{P}) \delta(\mathbf{q} - \mathbf{q'})& c_M(\mathbf{q},\mathbf{P}) 
		\vspace{4pt} \\ 
		c_M^T(\mathbf{q'},\mathbf{P})&  
		A(\mathbf{P}) \end{bmatrix} \label{SO_Amatrix} \\
		\mathfrak{B}(\mathbf{q},\mathbf{q'},\mathbf{P},t) & = \begin{bmatrix}  
		s_K(\mathbf{q},\mathbf{P}) \delta(\mathbf{q} - \mathbf{q'})& c_K(\mathbf{q},\mathbf{P}) 
		\vspace{4pt} \\  
		c_K^T(\mathbf{q'},\mathbf{P})&  
		B(\mathbf{P}) \end{bmatrix} \label{SO_Bmatrix}\!,
	\end{align}
\end{subequations}
\noindent where $A(\mathbf{P})$ and $B(\mathbf{P})$ are $3 \times 3$ matrices [compare (\ref{AppbK_RPAmat})].  Finally, along with their corresponding transposes, $c_M^T$ and $c_K^T$, the off-diagonal terms are 
% Eqs. 100
\begin{subequations} \label{SO_cMcK}
	\begin{align}
		c_M (\mathbf{q}, \mathbf{P}) &= 
		\begin{bmatrix} \mathfrak{A}_{G \psi} (\mathbf{q}, \mathbf{P}) & 
		\mathfrak{A}_{G \phi}(\mathbf{q}, \mathbf{P}) & 
		\mathfrak{A}_{G \chi}(\mathbf{q},\mathbf{P}) \end{bmatrix} \label{SO_cM} \\ 
		c_K (\mathbf{q}, \mathbf{P}) &= 
		\begin{bmatrix} \mathfrak{B}_{\Sigma \pi} (\mathbf{q}, \mathbf{P}) & 
		\mathfrak{B}_{\Sigma \omega} (\mathbf{q}, \mathbf{P}) 
		& \mathfrak{B}_{\Sigma \nu} (\mathbf{q}, \mathbf{P}) \end{bmatrix}\!. \label{SO_cK}
	\end{align}
\end{subequations}
% Subsection: Quasi-Boson Interpretation
\subsection{Quasi-Boson Interpretation}
As suggested by (\ref{SO_ABmatrices}), it is natural to separate the diagonal or noninteracting part of $K_{RPA}$ from the off-diagonal piece so that\footnote{The subscript $0$ denotes the diagonal part of the matrix.}
% Eq. 101
\begin{equation} \label{QBI_KRPA_split}
	K_{RPA} = K_0 + K_{int},
\end{equation}
\noindent with
% Eqs. 102
\begin{subequations} \label{QBI_K0_Kint}
	\setlength{\jot}{.2in}
	\begin{align}
		K_0 = \frac{1}{2} \delta {\mathcal P}^\dag
		\begin{pmatrix} s_M & 0 \\ 0 & A_0 \end{pmatrix} 
		\delta {\mathcal P} & +
		\frac{1}{2} \delta {\mathcal Q}^\dag 
		\begin{pmatrix} s_K & 0 \\ 0 & B_0 \end{pmatrix}
		\delta {\mathcal Q} \label{QBI_K0} \\
		K_{int}  = \frac{1}{2} \delta {\mathcal P}^\dag 
		\begin{pmatrix} 0 & c_M \\ c_M^T & A-A_0 \end{pmatrix} 
		\delta {\mathcal P} & +
		\frac{1}{2} \delta {\mathcal Q}^\dag
		\begin{pmatrix} 0 & c_K \\ c_K^T & B-B_0 \end{pmatrix}
		\delta {\mathcal Q} \label{QBI_Kint}.
	\end{align} \\
\end{subequations}
\setlength{\jot}{.1in}
For the noninteracting case involving only $K_0$, it is simpler to introduce the multiplicative canonical transformation
% Eqs. 103
\begin{subequations} \label{QBI_CoordTransf}
	\begin{align}
		\begin{pmatrix} \delta \Sigma \\ \delta \pi \\ \delta \omega \\ \delta \nu 
		\end{pmatrix} & \rightarrow 
		\begin{pmatrix} \delta \Sigma \\ \delta \pi \\ \delta \omega \\ \delta \nu \end{pmatrix} '
		= \begin{pmatrix} \delta \Sigma \sqrt{s_M} \\ \delta \pi \sqrt{\mathfrak{A}_{\psi \psi}} \\
		\delta \omega \sqrt{\mathfrak{A}_{\phi \phi}} \\ \delta \nu \sqrt{\mathfrak{A}_{\chi \chi}} 	
		\end{pmatrix} \label{QBI_CoordTransfa} \\[.1 in]
		\begin{pmatrix} \delta G \\ \delta \psi \\ \delta \phi \\ \delta \chi \end{pmatrix} & \rightarrow 
		\begin{pmatrix} \delta G \\ \delta \psi \\ \delta \phi \\ \delta \chi \end{pmatrix} '
		= \begin{pmatrix} \delta G / \sqrt{s_M} \\ \delta \psi / \sqrt{\mathfrak{A}_{\psi \psi}} \\
		\delta \phi / \sqrt{\mathfrak{A}_{\phi \phi}} \\ 
		\delta \chi / \sqrt{\mathfrak{A}_{\chi \chi}} \end{pmatrix}\!, \label{QBI_CoordTransfb}
	\end{align}
\end{subequations}
\noindent resulting in
% Eq. 104
\begin{equation} \label{QBI_K0Transf}
	K_0 = \frac{1}{2} \begin{pmatrix} \delta \Sigma^\ast & \delta \pi^\ast 
	& \delta \omega^\ast & \delta \nu^\ast \end{pmatrix} '
	\begin{pmatrix} \Omega_2^2 & 0 & 0 & 0 \\ 
	0 & \omega_\Psi^2 & 0 & 0 \\
	0 & 0 & \omega_\Phi^2 & 0 \\
	0 & 0 & 0 & \omega_X^2 \end{pmatrix} 
	\begin{pmatrix} \delta \Sigma \\ \delta \pi \\ \delta \omega \\ \delta \nu \end{pmatrix} ' 
	+ \frac{1}{2} \left| \delta {\mathcal P}' \right|^2.
\end{equation}
\noindent Associated with each of the respective pairs, $(\delta \Sigma , \delta G)$, $(\delta \pi, \delta \psi)$, $(\delta \omega, \delta \phi)$, $(\delta \nu, \delta \chi)$, is a frequency given by
% Eqs. 105
\begin{subequations} \label{QBI_Frequencies}
	\begin{align}
		\Omega_2(\mathbf{q}, \mathbf{P}) &= \sqrt{s_M(\mathbf{q}, \mathbf{P}) 
		s_K(\mathbf{q}, \mathbf{P})} \label{QBI_Omega2} \\
		\omega_\Psi(\mathbf{P}) &= \sqrt{\left( \mathbf{P}^2 - \mu \right)^2 - \xi^2 } \label{QBI_omPsi} \\
		\omega_\Phi &= \epsilon - 2\mu \label{QBI_omPhi} \\
		\omega_X &= \varepsilon - 2\mu. \label{QBI_omXi}
	\end{align}
\end{subequations}
\indent An explicit calculation of $\Omega_2$ yields an interesting interpretation.  From (\ref{App_AGG}) and (\ref{App_BSigSig}) we have
% Eqs. 106
\begin{subequations} \label{QBI_sMsK}
	\setlength{\jot}{.1in}
	\begin{align}
		s_M(\mathbf{q},\mathbf{P}) & = \frac{1}{4} G_+^{-1} G_-^{-1} (\omega_+ + \omega_-)
		\label{QBI_sM} \\
		s_K(\mathbf{q}, \mathbf{P}) & = 4G_+G_- (\omega_+ + \omega_-), \label{QBI_sK} 
	\end{align}
\end{subequations}
\noindent where, in addition to the form of $G$ in (\ref{SUS_GVariation}), we have used the definition\footnote{The $\pm$ subscripts denote coordinate shifts of $\pm \mathbf{P} /2$ as in $f_{\pm} = f(\mathbf{q} \pm \mathbf{P}/2)$.}
% Eq. 107
\begin{equation} \label{4omega}
	\omega_{\pm} \equiv \omega_\Psi(\mathbf{q}_\pm) =  \sqrt{(q_{\pm}^2 - \mu)^2 - \xi^2}. \\ 
\end{equation}
\noindent Multiplication of $s_M$ and $s_K$ obtains
% Eq. 108
\begin{equation} \label{QBI_sqtsMsK}
	\sqrt{s_M s_K} = \omega_+ + \omega_- .
\end{equation}
\noindent Recalling the definitions of $\mathbf{q}$ and $\mathbf{P}$ in (\ref{SO_NewMomentumCoords}) leads to
% Eq. 109
\begin{equation} \label{QBI_Om2}
	\Omega_2(\mathbf{q},\mathbf{P}) = \omega(\mathbf{k}) + \omega(\mathbf{k'}),
\end{equation}
\noindent thus suggesting that $\omega$ and $\Omega_2$ are one and two free quasi-boson energies, respectively.  Although still incomplete, it is nonetheless possible to extrapolate these results to the more general case.  Because they arise from an effective Lippmann-Schwinger equation, the eigenfrequencies of the full problem are identified with the energy of two interacting quasi-bosons.  Accordingly, discrete, real-valued frequencies represent bound states whereas a continuous range represents the scattering continuum.
% Subsection: Eigenspectrum of A
\subsection{Eigenspectrum of $\mathfrak{A}$}
Before trying to solve for the spectrum of $\mathfrak{A} \cdot \mathfrak{B}$, it is helpful to first consider the solution for the eigenvalues of $\mathfrak{A}$ alone, for this simplified case serves to illustrate the approach in the more general problem.  Furthermore, a zero mode of either $\mathfrak{A}$ or $\mathfrak{B}$ implies its presence in their product.  For instance, consider a discrete basis where $\mathfrak{A}$ is diagonal.  If $\mathfrak{A}$ has a zero eigenvalue, then multiplication by $\mathfrak{B}$ reveals a product having an all-zero row, thus proving the persistence of the zero mode.  Therefore, the analysis of $\mathfrak{A}$ can also serve as a check on the results obtained thus far since the Goldstone mode is built into the equations from the particle number symmetry of the Hamiltonian.  \\ 
\indent Throughout, our attention is primarily concerned with the eigenspectrum for the $\psi = 0$ solution.  We begin with the eigenvalue equation\footnote{The subscript $a$ is used to avoid confusion with the eigenvalue label used for $\mathfrak{A} \cdot \mathfrak{B}$ (see Sec. \!\ref{SecABProd}).}
% Eq. 110
\begin{equation} \label{GM_EigA}
	\mathfrak{A} \delta P = \Omega_a \delta P,
\end{equation}
\noindent which, with the aid of (\ref{SO_dP}), (\ref{SO_Amatrix}) and (\ref{SO_cM}), can be written as four linearly independent equations in the variables $\delta G$, $\delta \psi$, $\delta \phi$, and $\delta \chi$:
% Eqs. 111
\begin{subequations} \label{GM_eigeqs}
	\begin{align}
		\int \limits_\mathbf{q'} s_M(\mathbf{q}) \delta(\mathbf{q} - \mathbf{q'}) 
		\delta G (\mathbf{q'}) + \mathfrak{A}_{G \phi}(\mathbf{q}) \delta \phi +
		\mathfrak{A}_{G \chi}(\mathbf{q}) \delta \chi &= \Omega_a \delta G(\mathbf{q}) 
		\label{GM_dG} \\ 
		\mathfrak{A}_{\psi \psi} \delta \psi &= \Omega_a \delta \psi \label{GM_dpsi} \\ 
		\int \limits_{\mathbf{q'}} \mathfrak{A}_{\phi G} (\mathbf{q'})
		\delta G (\mathbf{q'}) + \mathfrak{A}_{\phi \phi} \delta \phi &= 
		\Omega_a \delta \phi \label{GM_dphi} \\ 
		\int \limits_{\mathbf{q'}} \mathfrak{A}_{\chi G} (\mathbf{q'})
		\delta G (\mathbf{q'}) + \mathfrak{A}_{\chi \chi} \delta \chi &= 
		\Omega_a \delta \chi. \label{GM_dchi}
	\end{align}
\end{subequations}
\noindent In addition to the notational conciseness of dropping the $\mathbf{P}$ indices, we have used (\ref{App_Apsiphi}) and (\ref{App_Apsichi}) in setting $\mathfrak{A}_{\psi \phi}$ and $\mathfrak{A}_{\psi \chi}$ to zero for the $\psi = 0$ case.  Elimination of the fluctuations $\delta \phi$ and $\delta \chi$ leads to a single equation in $\delta G (\mathbf{q})$,
% Eq. 112
\begin{equation} \label{GM_dGeff}
	\int \limits_\mathbf{q'} \left\{ \left[ s_M(\mathbf{q}) - \Omega_a \right] \delta(\mathbf{q} - \mathbf{q'}) 
	+ \zeta(\Omega_a) V(\mathbf{q}, \mathbf{q'}) \right\} \delta G (\mathbf{q'}) = 0.
\end{equation}
% Eq. 121
%\begin{equation} \label{GM_Vdef}
%	V(\mathbf{q},\mathbf{q'}) = \frac{1}{\Omega^2 - \mathfrak{A}'_{\phi \phi}} 
%	\mathfrak{A}'_{G \phi}(\mathbf{q}) \mathfrak{A}'_{\phi G} (\mathbf{q'}) + 
%	\frac{1}{\Omega^2 - \mathfrak{A}'_{\chi \chi}} 
%	\mathfrak{A}'_{G \chi}(\mathbf{q}) \mathfrak{A}'_{\chi G}(\mathbf{q'}) .
%\end{equation}
\noindent Using the expressions for $\mathfrak{A}_{\phi \phi}$, $\mathfrak{A}_{\chi \chi}$, $\mathfrak{A}_{G \phi}$ and $\mathfrak{A}_{G \chi}$ in \ref{App_Amatrix_elmnts}, $V(\mathbf{q}, \mathbf{q'})$ simplifies to a single separable term,
% Eq. 113
\begin{equation} \label{GM_Vsep}
	V(\mathbf{q},\mathbf{q'}) = y(\mathbf{q}) y(\mathbf{q'}),
\end{equation}
\noindent where we have defined
{\allowdisplaybreaks \begin{align}
	% Eq. 114
	\zeta(\Omega_a) &= \frac{\lambda^2 \alpha_1^2}{\Omega_a - \epsilon + 2 \mu} 
	+ \frac{g^2}{\Omega_a - \varepsilon + 2 \mu} 
	\label{GM_sig} \\
	% Eq. 115 
	y(\mathbf{q}) &= \frac{1}{2}\left( \frac{1}{4} G_+^{-1} G_-^{-1} + 1\right)\! .
	\label{GM_v} 
\end{align}}
\indent \hspace{-6pt} Identifying the operator in (\ref{GM_dGeff}) as ${\mathcal O}(\mathbf{q},\mathbf{q'}) = [ s_M(\mathbf{q}) - \Omega_a] \delta (\mathbf{q} - \mathbf{q'}) + \zeta(\Omega_a) V(\mathbf{q},\mathbf{q'})$, its spectrum is found by solving for the values of $\Omega_a$ at which an inverse, ${\mathcal O}^{-1}$, does not exist.  Without loss of generality, it may be assumed that ${\mathcal O}^{-1}$ has the form 
% Eq. 116
\begin{equation} \label{GM_Oinv}
	{\mathcal O}^{-1}(\mathbf{q''}, \mathbf{q'}) = \frac{1}{s_M(\mathbf{q'})-\Omega_a} \, 
	\delta (\mathbf{q}'' - \mathbf{q'}) + {\mathcal S}(\mathbf{q''},\mathbf{q'}),
\end{equation}
\noindent where ${\mathcal S}(\mathbf{q''},\mathbf{q'})$ is obtained from the requirement $\int  {\mathcal O}(\mathbf{q}, \mathbf{q''}) {\mathcal O}^{-1}(\mathbf{q''}, \mathbf{q'}) = \delta (\mathbf{q} - \mathbf{q'})$.  Performing the necessary multiplications, this condition is met only if
% Eq. 117
\begin{equation} \label{GM_OinvEq}
\begin{split} 
	\left[ s_M(\mathbf{q}) - \Omega_a \right] \, {\mathcal S}(\mathbf{q}, \mathbf{q'}) & + 
	\zeta(\Omega_a) \int \limits_\mathbf{q''} V(\mathbf{q},\mathbf{q''}) 
	{\mathcal S} (\mathbf{q''}, \mathbf{q'}) \\
	&\qquad \qquad = -\zeta(\Omega_a)  V(\mathbf{q},\mathbf{q'}) \frac{1}{s_M(\mathbf{q'}) - 
	\Omega_a}.
\end{split}
\end{equation}
\noindent Defining a new matrix
% Eq. 118
\begin{equation} \label{GM_Tdef}
	T(\mathbf{q}, \mathbf{q'}) = \left[s_M(\mathbf{q}) - \Omega_a \right] 
	{\mathcal S}(\mathbf{q}, \mathbf{q'}) 
	 \left[s_M(\mathbf{q'}) - \Omega_a \right] \! ,
\end{equation}
\noindent gives the Lippmann-Schwinger equation used in scattering theory:
% Eq. 119
\begin{equation} \label{GM_LippSchw}
	T(\mathbf{q},\mathbf{q'}) + \zeta(\Omega_a) \int \limits_\mathbf{q''} 
	 V(\mathbf{q},\mathbf{q''}) \frac{1}{s_M(\mathbf{q''}) -\Omega_a} \, T(\mathbf{q''}, \mathbf{q'}) = 
	 -\zeta(\Omega_a) V(\mathbf{q}, \mathbf{q'}).
\end{equation}
\noindent Because this condition must be satisfied in order to invert the operator in (\ref{GM_dGeff}), values of $\Omega_a$ are sought for which (\ref{GM_LippSchw}) has no solution.  According to the Fredholm alternative \cite{IntegEq}, (\ref{GM_LippSchw}) cannot be solved if there exists a solution to its homogeneous counterpart,
% Eq. 120
\begin{equation} \label{GM_Fredholm}
	\frac{1}{\zeta(\Omega_a)} T(\mathbf{q},\mathbf{q'}) + \int \limits_\mathbf{q''}  
	V(\mathbf{q}, \mathbf{q''}) \frac{1}{s_M(\mathbf{q''}) - \Omega_a} \, T(\mathbf{q''},\mathbf{q'}) = 0.
\end{equation}
Thus, at least part of the spectrum is found by the $\{ \Omega_a \}$ that satisfy (\ref{GM_Fredholm}). \\
\indent As was done in the two-body case, we take a separable form for the $T$-matrix,
% Eq. 121
\begin{equation} \label{GM_Tsep}
	T(\mathbf{q},\mathbf{q'}) = t \, y(\mathbf{q}) \, y (\mathbf{q'}),
\end{equation}
\noindent which upon substitution into (\ref{GM_Fredholm}) gives
% Eq. 122
\begin{equation} \label{GM_Tsep_Fred}
	\frac{1}{\zeta(\Omega_a)} + \int \limits_{\mathbf{q}} 
	\frac{y(\mathbf{q})^2}{s_M(\mathbf{q}) - \Omega_a} = 0.
\end{equation}
\noindent An exhaustive solution for the $\Omega_a$'s is not attempted here as the non-zero elements of the spectrum reveal little in acquiring the eigenfrequencies of $\mathfrak{A} \cdot \mathfrak{B}$.  However, a zero mode of $\mathfrak{A}$ does prove its presence in the product, prompting us to check whether (\ref{GM_Tsep_Fred}) may be solved for $\Omega_a = 0$.  Since the Goldstone mode occurs at zero total momentum, the required expressions are obtained from the $\mathbf{P} = 0$ form of (\ref{GM_v}) and (\ref{App_AGG}), thus giving
% Eq. 123
\begin{align} \label{YS}
	\int \limits_\mathbf{q} \frac{y(\mathbf{q},\mathbf{P}=0)^2}{s_M(\mathbf{q}, \mathbf{P}=0)} &=
	\frac{1}{2} \int \limits_\mathbf{q} \frac{\left( q^2 - \mu \right)^2}{\left[ \left( q^2 - \mu \right)^2
	- \xi^2 \right]^{3/2}} \nonumber \\
	&= -\frac{\partial}{\partial \xi} \int \limits_\mathbf{q} D(\mathbf{q}) \nonumber \\
	&= - \frac{1}{\zeta(0)}.
\end{align}
\noindent Obtaining the second line requires the explicit form of $\int D(\mathbf{q})$ in (\ref{SUS_DExpression}), whereas the last line is derived from a combination of (\ref{SUS_MolecularVariations}) and (\ref{GM_sig}), thus verifying that the Goldstone mode is present in $\mathfrak{A}$. \\
\indent In addition to $\Omega_a = 0$, the inverse matrix, $\mathcal{O}^{-1}$, will not exist if 
% Eq. 124
\begin{align} \label{AEigs}
	\Omega_a &= s_M(\mathbf{q}, \mathbf{P}) \nonumber \\
	&= \sqrt{\frac{( q_+^2 - \mu - \xi )(q_-^2 - \mu - \xi)}{(q_+^2 - \mu + \xi)(q_-^2 - \mu + \xi)}}
	\left( \omega_+ + \omega_- \right).
\end{align}
Due to its dependence on $\mathbf{q}$, this solution represents the continuum.  More importantly, the system's excitations can be found from the eigenvalues of $\mathfrak{A} \cdot \mathfrak{B}$ using this same approach.
\subsection{The $\mathfrak{A} \cdot \mathfrak{B}$ Product} \label{SecABProd}
For the full excitation spectrum, it is necessary to first obtain the matrix product which, by using the definitions (\ref{SO_ABmatrices}), is\footnote{It is understood that all quantities within the matrix depend on the total momentum $\mathbf{P}=\mathbf{P}'$.}
% Eq. 125
\begin{multline} \label{MAB_Mmatrix}
	\mathfrak{A} \cdot \mathfrak{B} \, (\mathbf{q},\mathbf{q'},\mathbf{P})  = 
	M(\mathbf{q},\mathbf{q'},\mathbf{P}) \\ = \begin{bmatrix} 
	s_M(\mathbf{q}) s_K(\mathbf{q'}) \delta(\mathbf{q} - \mathbf{q'})
	+ c_M(\mathbf{q}) \, c_K^T(\mathbf{q'}) & 
	s_M(\mathbf{q}) \, c_K(\mathbf{q}) 
	+ c_M(\mathbf{q}) B \vspace{.1in} \\  
	 c_M^T(\mathbf{q'}) \, s_K(\mathbf{q'}) 
	 + A c_K^T(\mathbf{q'}) &  
	 {\displaystyle \int \limits_\mathbf{q''} c_M^T(\mathbf{q''}) c_K(\mathbf{q''})}
	+ A \cdot B \end{bmatrix} \\
	= \begin{bmatrix}
	\vspace{.05in}
	M_{G G}(\mathbf{q}, \mathbf{q'}, \mathbf{P}) & 
	M_{G \psi}(\mathbf{q}, \mathbf{P}) &
	M_{G \phi}(\mathbf{q}, \mathbf{P}) &
	M_{G \chi}(\mathbf{q}, \mathbf{P}) \\ 
	\vspace{.05in}
	M_{\psi G}(\mathbf{q'}, \mathbf{P}) &
	M_{\psi \psi}(\mathbf{P}) &
	M_{\psi \phi}(\mathbf{P}) &
	M_{\psi \chi}(\mathbf{P}) \\ 
	\vspace{.05in}
	M_{\phi G}(\mathbf{q'}, \mathbf{P}) &
	M_{\phi \psi}(\mathbf{P}) &
	M_{\phi \phi}(\mathbf{P}) &
	M_{\phi \chi}(\mathbf{P}) \\
	\vspace{.025in}
	M_{\chi G}(\mathbf{q'}, \mathbf{P}) &
	M_{\chi \psi}(\mathbf{P}) &
	M_{\chi \phi}(\mathbf{P}) &
	M_{\chi \chi}(\mathbf{P})
	\end{bmatrix}\! .
\end{multline}
\noindent Through (\ref{SUS_GVariation}), (\ref{SO_cMcK}), (\ref{QBI_sMsK}), (\ref{4omega}) and (\ref{GM_v}), the matrix elements are explicitly given by
% Eqs. 126
\begin{subequations} \label{MAB_Melements}
	{\allowdisplaybreaks \begin{align} 
		M_{G G}(\mathbf{q}, \mathbf{q'}) &= (\omega_+ + \omega_-)^2 
		\delta(\mathbf{q} - \mathbf{q'}) + (\lambda^2 \alpha_1^2 + g^2) y(\mathbf{q}) (G_+' + G_-') \\
		M_{G \psi}(\mathbf{q}) &= - (\lambda^2 \alpha_1^2 + g^2) y(\mathbf{q}) \psi \\
		M_{G \phi}(\mathbf{q}) &= -\frac{\lambda \alpha_1}{4} 
		(G_+^{-1} + G_-^{-1}) (\omega_+ + \omega_-)  
		- \lambda \alpha_1 \, y(\mathbf{q}) (\epsilon - 2\mu)\\
		M_{G \chi}(\mathbf{q}) &= -\frac{g}{4} 
		(G_+^{-1} + G_-^{-1}) (\omega_+ + \omega_-) 
		- g \, y(\mathbf{q}) (\varepsilon - 2\mu) \\
		M_{\psi G}(\mathbf{q'}) &= -(\lambda^2 \alpha_1^2 + g^2)(G_+' + G_-') \psi \\
		\begin{split}
			M_{\phi G}(\mathbf{q'}) &= -4 \, \lambda \alpha_1 \, y(\mathbf{q'})
			G_+'G_-'(\omega_+' + \omega_-') \\
			& \hphantom{= \, \,} - \lambda \alpha_1 \, (G_+' + G_-') ( \epsilon - 2\mu) 
		\end{split} \\
		\begin{split}
			M_{\chi G}(\mathbf{q'}) &= -4 \, g \, y(\mathbf{q'})
			 G_+'G_-' (\omega_+' + \omega_-') \\
			& \hphantom{= \, \,} - g \, (G_+' + G_-') ( \varepsilon - 2\mu) 
		\end{split} \\
		M_{\psi \psi} &= (P^2 - \mu)^2 - \xi^2 + (\lambda^2 \alpha_1^2 + g^2) \psi^2 \\
		M_{\psi \phi} &= \lambda \alpha_1 (P^2 - 3\mu + \xi + \epsilon) \psi \\
		M_{\psi \chi} &= g (P^2 - 3\mu + \xi + \varepsilon) \psi \\
		M_{\phi \psi} &= \lambda \alpha_1 (P^2 - 3\mu - \xi + \epsilon) \psi \\
		M_{\phi \phi} &= \lambda^2 \alpha_1^2 \, (2 I_1 + \psi^2 ) 
		+ (\epsilon - 2\mu)^2 \\
		M_{\phi \chi} &= M_{\chi \phi} = \lambda \alpha_1 g \, 
		( 2 I_1 +  \psi^2 \Bigr)  \\
		M_{\chi \psi} &= g (P^2 - 3\mu - \xi + \varepsilon) \psi \\
		M_{\chi \chi} &= g^2 ( 2 I_1 + \psi^2) + (\varepsilon - 2\mu)^2,
	\end{align}}
\end{subequations}
\noindent \hspace{-6pt} where we have conveniently defined
% Eq. 127
\begin{equation} \label{MAB_I}
	I_1 \equiv \frac{1}{2} \int \limits_\mathbf{q} y(\mathbf{q}) (G_+ + G_-).
\end{equation}
These elements are used in the set of equations corresponding to the eigenvalue problem, $M \, \delta P = \Omega^2 \delta P$:
% Eqs. 128
\setlength{\jot}{.025in}
\begin{subequations} \label{MAB_Eigset}
	{\allowdisplaybreaks \begin{align}
		\begin{split}
			\negthickspace \negthickspace \int \limits_\mathbf{q''} 
			M_{GG}(\mathbf{q},\mathbf{q''}) 
			\, \delta G(\mathbf{q''}) + M_{G \psi}(\mathbf{q}) \, \delta \psi
			+ M_{G \phi}(\mathbf{q}) \, \delta \phi
		 	+ M&_{G \chi}(\mathbf{q}) \, \delta \chi \\
			&= \Omega^2 \delta G(\mathbf{q}) 
		\end{split} \label{MAB_EigsetG} \\ 
		\int \limits_\mathbf{q''} M_{\psi G}(\mathbf{q''}) 
		\, \delta G(\mathbf{q''}) + M_{\psi \psi} \, \delta \psi 
		+ M_{\psi \phi} \, \delta \phi 
		+ M_{\psi \chi} \, \delta \chi 
		&= \Omega^2 \delta \psi \label{MAB_Eigsetpsi} \\
		\int \limits_\mathbf{q''} M_{\phi G}(\mathbf{q''}) 
		\, \delta G(\mathbf{q''}) + M_{\phi \psi} \, \delta \psi 
		+ M_{\phi \phi} \, \delta \phi 
		+ M_{\phi \chi} \, \delta \chi 
		& = \Omega^2 \delta \phi \label{MAB_Eigsetphi} \\
		\int \limits_\mathbf{q''} M_{\chi G}(\mathbf{q''}) 
		\, \delta G(\mathbf{q''}) + M_{\chi \psi} \, \delta \psi 
		+ M_{\chi \phi} \, \delta \phi 
		+ M_{\chi \chi} \, \delta \chi 
		&= \Omega^2 \delta \chi \label{MAB_Eigsetchi}.
	\end{align}} \\
\end{subequations}
\setlength{\jot}{.1in}
\indent Up to this point the equations are general, describing the excitation spectrum of any stationary solution.  For the $\psi = 0$ case, $M_{G \psi}$, $M_{\psi G}$, $ M_{\psi \phi}$, $M_{\phi \psi}$, $M_{\psi \chi}$ and $M_{\chi \psi}$ all vanish, reducing the above system to only three independent equations.  Further, we restrict the discussion to long wavelength excitations, $\mathbf{P} \rightarrow 0$, assuming the general case to have similar properties.  With these simplifications, the molecular field displacements are obtained in terms of $\delta G$:
% Eqs. 129
\begin{subequations} \label{MAB_phidisp_chidisp}
	\begin{align}
		\delta \phi &= \frac{1}{d} \int \limits_\mathbf{q''}  
		\left[ (\Omega^2 - M_{\chi \chi}) \,
		M_{\phi G}(\mathbf{q''}) + M_{\phi \chi} \, 
		M_{\chi G}(\mathbf{q''}) \right] \delta G(\mathbf{q''}) \label{MAB_phidisp} \\
		\delta \chi &= \frac{1}{d} \int \limits_\mathbf{q''} 
		\left[ (\Omega^2 - M_{\phi \phi}) \,
		M_{\chi G}(\mathbf{q''}) + M_{\chi \phi} \, 
		M_{\phi G}(\mathbf{q''}) \right] \delta G(\mathbf{q''}) \label{MAB_chidisp},
	\end{align}
\end{subequations}
\noindent where after defining
% Eqs. 130
\begin{subequations} \label{OmegaEpVarEp}
	\begin{align}
		\Omega_\epsilon^2 &= \Omega^2 - \left(\epsilon - 2 \mu \right)^2 \label{OmegaEp} \\
		\Omega_\varepsilon^2 &= \Omega^2 - \left(\varepsilon - 2 \mu \right)^2 \! , 
		\label{OmegaVarEp}
	\end{align}
\end{subequations} 
\noindent the denominator is expressed as
 % Eq. 131
 \begin{equation} \label{MAB_Mdet}
 	\begin{split}
 	d &= (\Omega^2 - M_{\phi \phi})(\Omega^2 - M_{\chi \chi})
	- M_{\phi \chi} M_{\chi \phi} \\
	&= \Omega_\epsilon^2 \, \Omega_\varepsilon^2 - 2g^2 I_1 \, \Omega_\epsilon^2
	-2 \lambda^2 \alpha_1^2 I_1 \, \Omega_\varepsilon^2 .
	\end{split}
\end{equation}
\indent Substitution of (\ref{MAB_phidisp_chidisp}) into (\ref{MAB_EigsetG}) obtains an effective matrix equation for $\delta G$,
% Eq. 132
\begin{equation} \label{MAB_effectiveGMat}
	\int \limits_\mathbf{q''} \left[ h(\mathbf{q}) \, 
	\delta (\mathbf{q} - \mathbf{q''})
	+ y(\mathbf{q}) \, r(\mathbf{q''}) + z(\mathbf{q}) \, s(\mathbf{q''}) \right] 
	\delta G (\mathbf{q''}) = 0.
\end{equation}
\noindent In addition to $y(\mathbf{q})$ given in (\ref{GM_v}), the terms appearing in this expression are identified by the following:
% Eqs. 133
\setlength{\jot}{.15in}
\begin{subequations} \label{MAB_hrs}
	\begin{align}
		h(\mathbf{q}) &= 4 \, \omega(\mathbf{q})^2 - \Omega^2 \label{MAB_h} \\
		z(\mathbf{q}) &= G(\mathbf{q})^{-1} \omega (\mathbf{q}) = 2 (q^2 - \mu - \xi) \label{MAB_z} \\
		\begin{split}
		r(\mathbf{q''}) &= 2 \left\{\lambda^2 \alpha_1^2 + g^2 
		+ \frac{1}{d} \left[ \lambda^2 \alpha_1^2 \left(\epsilon - 2 \mu \right)^2
		\Omega_\varepsilon^2 \right. \right. \\
		& \qquad \; \left. \left. + \, g^2 \! \left(\varepsilon - 2 \mu\right)^2 
		\Omega_\epsilon^2
		- 2 \lambda^2 \alpha_1^2 \,
		g^2 I_1 \left( \epsilon - \varepsilon \right)^2 
		\right] \vphantom{\frac{1}{d}} \right\} G(\mathbf{q''}) \\
		&\quad + \frac{8}{d} \left[ \lambda^2 \alpha_1^2 \left( \epsilon - 2\mu \right)
		\Omega_\varepsilon^2 
		+ \, g^2 \! \left( \varepsilon - 2\mu \right) 
		\Omega_\epsilon^2 \,
		\right]  y(\mathbf{q''}) G(\mathbf{q''})^2 \omega(\mathbf{q''}) 
		\end{split} \label{MAB_r} \\
		\begin{split}
		s(\mathbf{q''}) &= \frac{2}{d} \left[ \lambda^2 \alpha_1^2 \left(\epsilon - 2 \mu \right)
		\Omega_\varepsilon^2 
		+ \, g^2 \! \left(\varepsilon - 2 \mu\right) 
		\Omega_\epsilon^2 \, \right] G(\mathbf{q''})\\
		&\quad + \frac{8}{d} \left( \lambda^2 \alpha_1^2 \,
		\Omega_\varepsilon^2 
		+ \, g^2 \, \Omega_\epsilon^2 \right) 
		y(\mathbf{q''}) G(\mathbf{q''})^2 \omega(\mathbf{q''}) .
		\end{split} \label{MAB_s}
	\end{align}
\end{subequations}
% Eqs. 143
%\setlength{\jot}{.1in}
%\begin{subequations} \label{MAB_C1C2}
%	\begin{align}
%		C1 &= \frac{2 \lambda \alpha_1}{d} \, [ \lambda \alpha_1 \, 
%		(\Omega^2 - \mathfrak{M}_{\chi \chi})
%		+ g \, \mathfrak{M}_{\phi \chi} ] \label{MAB_C1} \\
%		C2 &=  \frac{2 g}{d} \, [ g \, (\Omega^2 - \mathfrak{M}_{\phi \phi})
%		+ \lambda \alpha_1 \, \mathfrak{M}_{\phi \chi} ] \label{MAB_C2} \\
%		C3 &= \frac{2 \lambda \alpha_1}{d} \, [ \lambda \alpha_1 \, 
%		(\Omega^2 - \mathfrak{M}_{\chi \chi}) (\epsilon - 2\mu)
%		+ g \, \mathfrak{M}_{\phi \chi} (\varepsilon - 2\mu) ] \label{MAB_C3} \\
%		C4 &= \frac{2 g}{d} \, [ g \, (\Omega^2 - \mathfrak{M}_{\phi \phi}) (\varepsilon - 2\mu)
%		+ \lambda \alpha_1 \, \mathfrak{M}_{\phi \chi} (\epsilon -2\mu) ] \label{MAB_C4} .
%	\end{align}
%\end{subequations}
The problem to be solved is to find all $\Omega$ such that Eq. \!(\ref{MAB_effectiveGMat}) has no solution except the trivial case $\delta G = 0$.  Along with a discrete set of eigenvalues, a continuum arises simply from the structure of the inverse matrix, as was indicated in the analogous calculation for $\mathfrak{A}$.  
% Subsection: {Point Spectrum at Long Wavelength}
\subsection{Point Spectrum at Long Wavelength $(\mathbf{P}=0)$}
Proceeding in analogy with the method used for $\mathfrak{A}$, the operator in (\ref{MAB_effectiveGMat}) can be written as
% Eq. 134
\begin{equation} \label{PS_GOp}
	{\mathcal O}(\mathbf{q}, \mathbf{q''}) = h(\mathbf{q}) \, \delta(\mathbf{q} - \mathbf{q''}) 
	+ \begin{bmatrix} 
		y(\mathbf{q}) & z(\mathbf{q}) 
	\end{bmatrix}
	\cdot \begin{bmatrix} 
		r(\mathbf{q''}) \vspace{.1in}\\
		s(\mathbf{q''})
	\end{bmatrix}\!.
\end{equation}
\noindent If it exists, the inverse, ${\mathcal O}^{-1}$, has the form
% Eq. 135
\begin{equation} \label{PS_GOinv}
	{\mathcal O}^{-1}(\mathbf{q''}, \mathbf{q'}) = \frac{1}{h(\mathbf{q''})} \, \delta(\mathbf{q''} - \mathbf{q'}) 
	+ \frac{1}{h(\mathbf{q''})} \begin{bmatrix} 
		y(\mathbf{q''}) & z(\mathbf{q''}) 
	\end{bmatrix}
	{\mathcal T} \begin{bmatrix} 
		r(\mathbf{q'}) \vspace{.1in}\\
		s(\mathbf{q'})
	\end{bmatrix} \frac{1}{h(\mathbf{q'})},
\end{equation}
\noindent where ${\mathcal T}$ is some $2 \times 2$ matrix, yet to be determined.  From the identity $\int_\mathbf{q''} {\mathcal O}(\mathbf{q},\mathbf{q''}) \, {\mathcal O}^{-1}(\mathbf{q''}, \mathbf{q'}) = \delta (\mathbf{q} - \mathbf{q'})$, we have
% Eq. 136
\begin{equation} \label{PS_noOinv}
	{\mathcal T} + 1 +
	\begin{bmatrix} 
		\, {\displaystyle \int \limits_\mathbf{q''} 
		\frac{r(\mathbf{q''}) y(\mathbf{q''})}{h(\mathbf{q''})}}
		& {\displaystyle \int \limits_\mathbf{q''}
		\frac{r(\mathbf{q''}) z(\mathbf{q''})}{h(\mathbf{q''})}} \vspace{.1in} \\
		{\displaystyle \int \limits_\mathbf{q''} 
		\frac{s(\mathbf{q''}) y(\mathbf{q''})}{h(\mathbf{q''})}}
		& {\displaystyle \int \limits_\mathbf{q''}
		\frac{s(\mathbf{q''}) z(\mathbf{q''})}{h(\mathbf{q''})}} \vspace{2pt}
	\end{bmatrix} {\mathcal T} 
	= 0.
\end{equation}
\noindent By the discussion following Eq. \!(\ref{GM_LippSchw}), there can be no solution if
% Eqs. 137
\begin{subequations} \label{PS_FredholmAB}
	\begin{align}
		\int \limits_\mathbf{q} 
		\frac{r(\mathbf{q}) y(\mathbf{q})}{h(\mathbf{q})} & = 
		\int \limits_\mathbf{q} 
		\frac{s(\mathbf{q}) z(\mathbf{q})}{h(\mathbf{q})}  =  -1 \\
		\int \limits_\mathbf{q} 
		\frac{r(\mathbf{q}) z(\mathbf{q})}{h(\mathbf{q})} &= 
		\int \limits_\mathbf{q}  
		\frac{s(\mathbf{q}) y(\mathbf{q})}{h(\mathbf{q})} = 0.
	\end{align}
\end{subequations}
\noindent Hence, the values of $\Omega$ that solve (\ref{PS_FredholmAB}) determine the point spectrum of $\mathfrak{A} \cdot \mathfrak{B}$.%are satisfied for any values of $\Omega$, the inverse, ${\mathcal O}^{-1}$ [given in (\ref{PS_GOinv})], cannot be constructed.
% Subsection: Continuous Spectrum of the Lower State
% Subsubsection {The Goldstone Mode}
\subsubsection{Verification of the Goldstone Mode}
\hspace{-3pt}Because of the particle number symmetry in the original Hamiltonian, a zero-frequency mode must emerge as a general feature of this description, independent of specific parameter values.  For a proof, it is easiest to take $\Omega = 0$ then see whether Eqs. \!(\ref{PS_FredholmAB}) are satisfied.  Utilizing the variational condition (\ref{SUS_MolecularVariations}) along with the appropriate partial derivatives of (\ref{SUS_DExpression}), all required integrals are expressed in terms of $\mu$, $\epsilon$, $\lambda^2 \alpha_1^2$ and $g^2$:
% Eqs. 138
\begin{subequations} \label{PSIntsABCD}
	{\allowdisplaybreaks \begin{align}
		%\begin{split} 
		\int \limits_\mathbf{q} \frac{G(\mathbf{q}) y(\mathbf{q})}{h(\mathbf{q}, \Omega=0)} 
		&= -\frac{1}{4 \xi}
		\frac{\partial}{\partial \mu} \int \limits_\mathbf{q} D(\mathbf{q}) \nonumber \\
		&= - \frac{1}{2} \frac{ g^2 \left(\epsilon - 2 \mu \right)^2
		+ \lambda^2 \alpha_1^2 \left( \varepsilon - 2 \mu \right)^2}
		{\left[ g^2 \left( \epsilon - 2 \mu \right) + \lambda^2 \alpha_1^2 
		\left( \varepsilon - 2 \mu \right) \right]^2} \label{PSIntsA} \\ \nonumber \\
		%\end{split} 
		%\begin{split}
		\int \limits_\mathbf{q} \frac{G(\mathbf{q}) z(\mathbf{q})}{h(\mathbf{q}, \Omega = 0)} 
		&= -\frac{1}{2 \xi}
		\int \limits_\mathbf{q} D(\mathbf{q}) \nonumber \\
		&= \frac{1}{2} \frac{\left(\epsilon - 2 \mu \right) \left( \varepsilon - 2 \mu \right)}
		{g^2 \left( \epsilon - 2 \mu \right) + \lambda^2 \alpha_1^2 \left( \varepsilon - 2 \mu \right)}
		\label{PSIntsB} \\
		%\end{split} 
		%\begin{split}
		\int \limits_\mathbf{q} \frac{y(\mathbf{q}) G(\mathbf{q})^2 \omega(\mathbf{q}) y(\mathbf{q})}
		{h(\mathbf{q}, \Omega=0)} &= - \frac{1}{8} \frac{\partial}{\partial \xi} \int \limits_\mathbf{q}
		D(\mathbf{q}) \nonumber \\
		&= \frac{1}{8} \frac{\left( \epsilon - 2 \mu \right) \left( \varepsilon - 2 \mu \right)}
		{g^2 \left( \epsilon - 2 \mu \right) + \lambda^2 \alpha_1^2 \left( \varepsilon - 2 \mu \right)}
		\label{PSIntsC} \\
		%\end{split} 
		\int \limits_\mathbf{q} \frac{y(\mathbf{q}) G(\mathbf{q})^2 \omega(\mathbf{q}) z(\mathbf{q})}
		{h(\mathbf{q}, \Omega=0)} &= \frac{1}{4}I_1. \label{PSIntsD} 
	\end{align}} \\
\end{subequations}
\noindent Also, the $\Omega = 0$ forms of $d$, $r$ and $s$ are
%{\allowdisplaybreaks \begin{align} 
%	d(\Omega = 0) &= \left( \epsilon - 2 \mu \right)^2 \left( \varepsilon - 2 \mu \right)^2
%	+ 2 g^2 I_1 \left( \epsilon - 2 \mu \right)^2 + 2 \lambda^2 \alpha_1^2 I_1
%	\left( \varepsilon - 2 \mu \right)^2 \\ \vspace{-6pt}
%	\begin{split}
%	r(\mathbf{q}, \Omega=0) &= \frac{4I_1}{d} \left[ g^2 \left( \epsilon - 2 \mu \right)
%	+ \lambda^2 \alpha_1^2 \left( \varepsilon - 2 \mu \right) \right]^2 G(\mathbf{q}) \\
%	& - \frac{8}{d} \left(\epsilon - 2 \mu \right) \left(\varepsilon - 2 \mu \right)
%	\left[ g^2 \left( \epsilon - 2 \mu \right) \right. \\
%	& \qquad \qquad \qquad \qquad \qquad \left. + \lambda^2 \alpha_1^2 
%	\left( \varepsilon - 2 \mu \right) \right] y(\mathbf{q}) G(\mathbf{q})^2 \omega(\mathbf{q})
%	\end{split} \\
%	\begin{split}
%	s(\mathbf{q}, \Omega=0) &= -\frac{2}{d} \left( \epsilon - 2 \mu \right) \left( \varepsilon - 2 \mu \right)
%	\left[ g^2 \left( \epsilon - 2 \mu \right)
%	+ \lambda^2 \alpha_1^2 \left( \varepsilon - 2 \mu \right) \right] G(\mathbf{q}) \\
%	& - \frac{8}{d} \left[ g^2 \left( \epsilon - 2 \mu \right)^2 + \lambda^2 \alpha_1^2 
%	\left( \varepsilon - 2 \mu \right)^2 \right] y(\mathbf{q}) G(\mathbf{q})^2 \omega(\mathbf{q})
%	\end{split}
%\end{align}}
\begin{align}
%\begin{equation} 
	% Eq. 139
	\begin{split}
	\negthickspace d(\Omega = 0) &= \left( \epsilon - 2 \mu \right)^2 \left( \varepsilon - 2 \mu \right)^2
	+ 2 g^2 I_1 \left( \epsilon - 2 \mu \right)^2 \\ 
	&\quad + 2 \lambda^2 \alpha_1^2 I_1
	\left( \varepsilon - 2 \mu \right)^2 
	\end{split} \label{DOm0} \\
%\end{equation} 
%\begin{equation} 
	% Eq. 140
	\begin{split}
	r(\mathbf{q}, \Omega=0) &= \frac{4I_1}{d} \left[ g^2 \left( \epsilon - 2 \mu \right)
	+ \lambda^2 \alpha_1^2 \left( \varepsilon - 2 \mu \right) \right]^2 G(\mathbf{q}) \\
	 & \quad - \frac{8}{d} \left(\epsilon - 2 \mu \right) \left(\varepsilon - 2 \mu \right)
	\left[ g^2 \left( \epsilon - 2 \mu \right) \right. \\
	& \qquad \qquad \qquad \qquad \left. + \, \lambda^2 \alpha_1^2 
	\left( \varepsilon - 2 \mu \right) \right] y(\mathbf{q}) G(\mathbf{q})^2 \omega(\mathbf{q})
	\end{split} \label{rOm0} \\
%\end{equation}
%\begin{equation} 
	% Eq. 141
	\begin{split}
	s(\mathbf{q}, \Omega=0) &= -\frac{2}{d} \left( \epsilon - 2 \mu \right) \left( \varepsilon - 2 \mu \right)
	\left[ g^2 \left( \epsilon - 2 \mu \right)
	+ \lambda^2 \alpha_1^2 \left( \varepsilon - 2 \mu \right) \right] G(\mathbf{q}) \\
	& \quad - \frac{8}{d} \left[ g^2 \left( \epsilon - 2 \mu \right)^2 + \lambda^2 \alpha_1^2 
	\left( \varepsilon - 2 \mu \right)^2 \right] y(\mathbf{q}) G(\mathbf{q})^2 \omega(\mathbf{q}) .
	\end{split} \label{sOm0}
%\end{equation}
\end{align}
Finally, a combination of (\ref{PSIntsABCD})-(\ref{sOm0}) satisfies the eigenvalue conditions (\ref{PS_FredholmAB}), confirming the presence of the Goldstone mode for arbitrary parameter values.
\subsubsection{Eigenfrequencies in the Zero-Range Limit}
\hspace{-3pt}We now solve for the nonzero solutions to the eigenvalue equations in the zero-range limit, $\lambda \rightarrow 0^-$.  First, $r$ and $s$ are expanded to the appropriate order in $\lambda$, then used in (\ref{PS_FredholmAB}).  Defining $I_1 = \kappa \lambda^{-3} + \mbox{const.}$, all subsequent expressions may be conveniently written in terms of $\kappa$.  Accordingly, the denominator has the limit
% Eq. 142
\begin{equation} \label{PS2d}
	\frac{2}{d} \xrightarrow[\lambda \rightarrow 0^-]{} \frac{\lambda^5}{4 \kappa \alpha_1^2} \left[
	1 - \frac{\kappa - 1}{2 \kappa \alpha_1^2} \Omega_\epsilon^2 \, \lambda + 2 \mu \, \lambda^2 
	+ \frac{(\kappa - 1)^2}{4 \kappa^2 \alpha_1^4} \Omega_\epsilon^4 \, \lambda^2 \right] \! .
\end{equation}
\noindent When substituted into (\ref{MAB_r}) and (\ref{MAB_s}) this leads to
% Eqs. 143
\begin{subequations} \label{PSrs}
	{\allowdisplaybreaks \begin{align}
	\begin{split}
		r(\mathbf{q}) &\xrightarrow[\lambda \rightarrow 0^-]{} \left[ - \frac{2}{\alpha_1^2} 
		\Omega_\epsilon^2 - 4 \left( \epsilon - 2 \mu \right) \lambda  
		+ \frac{\kappa - 1}{\kappa \alpha_1^4} \Omega_\epsilon^4 \,
		\lambda \right] G(\mathbf{q})\\
		& \qquad + \left[ -\frac{4}{\kappa \alpha_1^2} \Omega_\epsilon^2 \, \lambda^2 
		- \frac{4}{\kappa} \left( \epsilon - 2 \mu \right) \lambda^3 \right. \\
		& \qquad \qquad \qquad \qquad \qquad \quad \,
		\left. + \frac{2 \left( \kappa -1 \right)}{\kappa^2 \alpha_1^4} 
		\Omega_\epsilon^4 \, \lambda^3 \right]
		y(\mathbf{q}) G(\mathbf{q})^2 \omega(\mathbf{q}) 
	\end{split} \label{PSr} \\
		s(\mathbf{q}) &\xrightarrow[\lambda \rightarrow 0^-]{} - \frac{1}{\kappa \alpha_1^2}
		 \Omega_\epsilon^2 \, \lambda^2 G(\mathbf{q})
		-\frac{4 \lambda^3}{\kappa} y(\mathbf{q})G(\mathbf{q})^2 \omega(\mathbf{q}) .
		\label{PSs}
	\end{align}}\\
\end{subequations}
\indent Next, it is necessary to calculate the small $\lambda$ versions of (\ref{PSIntsABCD}).  However, in this case the variational condition cannot be utilized since $\Omega \neq 0$ in general.  It is easiest to write these expansions in terms of the following integrals\footnote{These expressions are obtained by first factorizing the integrand denominator as $(q +i \sqrt{-\mu - \Omega'})(q - i\sqrt{- \mu - \Omega'})(q + i\sqrt{-\mu + \Omega'})(q - i\sqrt{-\mu + \Omega'})$, then choosing either the upper or lower semicircular contour.}
% Eqs. 144
\begin{subequations} \label{PSJ1J2}
	{\allowdisplaybreaks \begin{align}
	J_1 &= \frac{1}{2} \int \limits_{-\infty}^{\infty} \frac{q^2 \left(q^2 - \mu \right)}
	{\sqrt{\left( q^2 - \mu \right)^2 - \xi^2}} \frac{1}{\left( q^2 - \mu \right)^2 - \Omega'^{\, 2}} \, dq 
	\nonumber \\
	&= -\frac{\pi}{4} \frac{1}{\sqrt{\Omega'^{\, 2} - \xi^2}} \left( \sqrt{-\mu - \Omega'} 
	+ \sqrt{-\mu + \Omega'} \right) \label{PSJ1} \\
	J_2 &= \frac{1}{2} \int \limits_{-\infty}^{\infty} \left[ \frac{q^2 \sqrt{\left( q^2 - \mu \right)^2
	- \xi^2}}{\left( q^2 - \mu \right)^2 - \Omega'^{\, 2}} - 1 \right] dq \nonumber \\
	&= - \frac{\pi}{4} \frac{\sqrt{\Omega'^{\, 2} - \xi^2}}{\Omega'} \left( \sqrt{-\mu - \Omega'}
	- \sqrt{-\mu + \Omega'} \right) \! , \label{PSJ2}
	\end{align}} \\
\end{subequations}
\noindent where $\Omega'^{\, 2} = \xi^2 + \Omega^2 / 4$.  To the required order in $\lambda$, the analogs of (\ref{PSIntsABCD}) may now be expressed as
% Eqs. 145
\begin{subequations} \label{PSZRIntsABCD}
	{\allowdisplaybreaks \begin{align}
		\int \limits_\mathbf{q} \frac{G(\mathbf{q}) y(\mathbf{q})}{h(\mathbf{q})} 
		& \xrightarrow[\lambda \rightarrow 0^-]{} \frac{1}{16 \pi^2} J_1 \label{PSZRIntsA} \\
		\int \limits_\mathbf{q} \frac{G(\mathbf{q}) z(\mathbf{q})}{h(\mathbf{q})} 
		& \xrightarrow[\lambda \rightarrow 0^-]{} - \frac{1}{2 \lambda} + \frac{1}{16 \pi a _{bg}}
		+ \frac{1}{8 \pi^2} J_2 \label{PSZRIntsB} \\
		\int \limits_\mathbf{q} \frac{y(\mathbf{q}) G(\mathbf{q})^2 \omega(\mathbf{q}) y(\mathbf{q})}
		{h(\mathbf{q})} 
		& \xrightarrow[\lambda \rightarrow 0^-]{} -\frac{1}{8 \lambda} + \mbox{const.}
		\label{PSZRIntsC} \\
		\int \limits_\mathbf{q} \frac{y(\mathbf{q}) G(\mathbf{q})^2 \omega(\mathbf{q}) z(\mathbf{q})}
		{h(\mathbf{q})} 
		& \xrightarrow[\lambda \rightarrow 0^-]{}  \frac{\kappa}{4 \lambda^3} 
		+ \mbox{const.}. \label{PSZRIntsD} 
	\end{align}} \\
\end{subequations}
\indent With Equations (\ref{PSrs}) and (\ref{PSZRIntsABCD}), the eigenvalue conditions (\ref{PS_FredholmAB}) become
% Eqs. 146
\begin{subequations} \label{PSZRconds}
	\begin{align}
	\int \limits_\mathbf{q} \frac{r(\mathbf{q}) y(\mathbf{q})}{h(\mathbf{q})} 
	& \xrightarrow[\lambda \rightarrow 0^-]{} -\frac{J_1}{8 \pi^2 \alpha_1^2} \,
	\Omega_\epsilon^2 = -1 \label{PSZRcondsA} \\
	 \int \limits_\mathbf{q} \frac{r(\mathbf{q}) z(\mathbf{q})}{h(\mathbf{q})} 
	& \xrightarrow[\lambda \rightarrow 0^-]{} \epsilon - 2\mu - \left( \frac{1}{8 \pi a_{bg}}
	+ \frac{J_2}{4 \pi^2} \right) \frac{1}{\alpha_1^2} \, \Omega_\epsilon^2 = 0 \label{PSZRcondsB} \\
	\int \limits_\mathbf{q} \frac{s(\mathbf{q}) y(\mathbf{q})}{h(\mathbf{q})} 
	& \xrightarrow[\lambda \rightarrow 0^-]{} \mathcal{O}(\lambda^2) = 0 \label{PSZRcondsC} \\
	\int \limits_\mathbf{q} \frac{s(\mathbf{q}) z(\mathbf{q})}{h(\mathbf{q})} 
	& \xrightarrow[\lambda \rightarrow 0^-]{} -1 + \mathcal{O}(\lambda), \label{PSZRcondsD}
	\end{align}
\end{subequations}
\noindent thus showing that the last two are automatically satisfied for zero range.  Aside from the Goldstone mode, the other eigenfrequencies are obtained most easily by solving (\ref{PSZRcondsA}) and (\ref{PSZRcondsB}) for $J_1$ and $J_2$ then equating their product to that given by a direct calculation using Eqs. \!(\ref{PSJ1J2}):
% Eq. 147
\begin{equation} \label{PSIntProd}
	J_1J_2 = - \frac{\pi^2}{8} = 32 \pi^4 \alpha_1^4 \left( \epsilon - 2 \mu \right)
	\frac{1}{\Omega_\epsilon^4}
	- \frac{4 \pi^3 \alpha_1^2}{a_{bg}} \frac{1}{\Omega_\epsilon^2}.
\end{equation}
\noindent From the definition of $\Omega_\epsilon^2$ (\ref{OmegaEp}) and the solution of (\ref{PSIntProd}), two distinct eigenfrequencies finally emerge as
% Eq. 148
\begin{equation} \label{PSModes}
	\Omega^2 = \left( \epsilon - 2 \mu \right)^2 + \frac{16 \pi \alpha_1^2}{a_{bg}}
	\left[ 1 \pm \sqrt{1 - a_{bg}^2 \left( \epsilon - 2 \mu \right)} \right] \! ,
\end{equation}
\noindent depending on the density through $\mu$ only.  In addition to a stable response frequency of the many-body collective, real values of $\Omega$ may be interpreted as the binding energies of two interacting quasi-bosons.  On the other hand, scattering states are implied by complex $\Omega$, with the imaginary part quantifying damping in the corresponding collective excitation.  It turns out that for the case depicted in Fig. \!\ref{Modes}, both roots of $\Omega$ are complex, having real parts that lie at $\sim 23$ neV, well within the scattering continuum.
\subsection{Continuous Spectrum of the Lower State}
\begin{figure} [h]
\begin{center}
\epsfxsize=2.75in\epsfbox{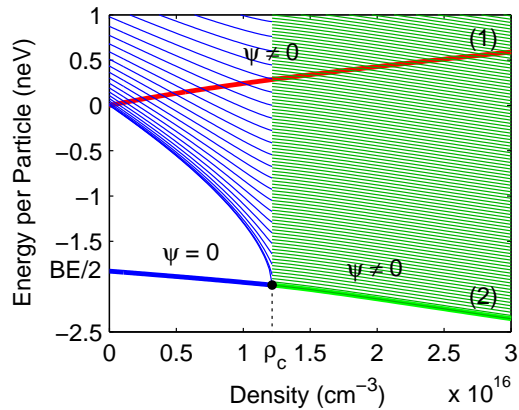}
\end{center}
\caption{A figure showing the same curves as in Fig.\:\ref{AllSolns}, where the hatched regions indicate the continuum of modes belonging to the collapsing lower state.  As discussed in the text, the excited ``false vacuum'' always lies in the continuum of the lower state, thus indicating that the decay represents a transition into collective phonon excitations of the collapsing solution.  Note that the density of states for $\psi \neq 0$ is greater than that for $\psi = 0$, as shown by the hatched lines above each region.  Additionally, the real parts of the discrete spectrum given by (\ref{CSLS_ContinuumBoundary}) lie relatively far up in the continuum at $\sim 23$ neV.} \label{Modes}
\end{figure}

Unlike the discrete elements, the continuum is obtained directly from the construction of the inverse.  Specifically, the condition $h(\mathbf{q}) = 0$ also precludes ${\mathcal O}^{-1}$ from existing, thus giving the continuous part of the spectrum,
% Eq. 149
\begin{align} \label{CSLS_Continuum}
	\begin{split}
		\Omega &= 2 \, \omega(\mathbf{q}) \\
		& = 2 \sqrt{(q^2 - \mu)^2 - \xi^2}.
	\end{split}
\end{align}
Since the form of $h(\mathbf{q})$ is the same regardless of whether $\psi = 0$ or $\psi \neq 0$, Eq. \!(\ref{CSLS_Continuum}) remains valid for the continuum of the entire lower state.  At long wavelength $(\mathbf{P} = 0)$, these excitations are parametrized only by $q^2$, with the lowest energy obtained for $\mathbf{q} = 0$.  Consequently, half of the lowest quasi-boson energy, $\Omega /2 |_{q=0}$, plus the energy per particle gives the continuum boundary:
% Eq. 150
\begin{equation} \label{CSLS_ContinuumBoundary}
	E_{exc} = e + \sqrt{\mu^2 - \xi^2}.
\end{equation}
\noindent As shown in Fig. \!\ref{Modes}, this threshold lies above the $\psi = 0$ solution ($\mu \neq \xi$), but intersects with the collapsing state at the critical point.  Thereafter, the boundary coincides with the $\psi \neq 0$ piece ($\mu = \xi$), indicating that the continuum contains the zero mode for the second piece of the collapsing solution.  \\
\indent It is now possible to give the decay found in Section \ref{ComplexMu} a more complete physical explanation.  Note that the real part of the energy per particle lies inside the continuum of the collapsing state.  Due to energy conservation, this upper solution can only evolve into a state at the same level, suggesting the coherent decay be interpreted as a transition into collective phonon excitations which inherit the same collapsing behavior as their associated ground state.  Hence, the condensate is initially lost through decay into the phonons of the lower level, as opposed to a physical loss of particles.  Nevertheless, as the system collapses, qusiparticles will emerge causing a kinematical atom loss akin to the semiclassical recombination models \cite{Dalibard, Shlya}.  However, the previous analysis is only valid for low-temperature systems, and therefore may not be applicable to these kinematic effects.
%Figure ******************
%Figure*******************
% Section Summary
\section{Summary}
We have considered the case of an atom-molecule condensate in which the interactions were attractive yet the effective scattering length was positive.  This situation raised the question of whether the condensate collapsed due to the mutual attractions or remained stable in accordance with the positive scattering length.  Starting with a two-body analysis, a separable potential was used to realistically model the interparticle interaction.  Due to its success, this separable form was implemented in the many-body Hamiltonian.  Equations of state were then obtained from the application of a variational principle that utilized a Gaussian trial wave functional for the many-body state.  Despite the positive scattering length, a collapsing solution was obtained, consisting of a relatively low-density piece having only a molecular component to its condensate.  At higher density, there occurred a quantum phase transition after which the solution comprised both atomic and molecular condensate components.  Only by allowing the chemical potential to assume complex values could the experimentally observed case be obtained.  As the phase of the order parameter, the chemical potential has an imaginary part quantifying a decay rate, assigned a physical meaning through a small oscillation analysis about the equilibrium solutions.  Expanding around the stationary points revealed two discrete eigenfrequencies associated with the low-density molecular condensate solution.  Moreover, the experimentally observed energy per particle lay within an excitation continuum of the collapsing two-piece lower state.  From energy conservation, the decay was interpreted as a coherent process corresponding to the evolution of the observed case into the excitations of the lower solution. 
%\subsection{<subsection title>}
%<subsection text>
%\subsection{<subsection title>}
%<subsection text>
\section*{Acknowledgement}
This work was primarily supported by the MIT-Los Alamos Collaborative Research Grant to Develop an Understanding of Bose-Einstein Condensates, contract number 19442-001-99-35.  Partial support also came from the LANL Laboratory Directed Research and Development (LDRD) Program.
\newpage
\begin{appendix}
%<insert any appendices here>
%\section{<appendix section title>}
%<insert appendix section text>
\setcounter{equation}{0}
\renewcommand{\theequation}{{\thesection}.\arabic{equation}}
\section{Derivation of the RPA Hamiltonian}
In the appendix, we expand the expectation value of the Hamiltonian (\ref{GVP_HExpectVal}) to second-order in the variations given by (\ref{SO_RPAexpansion}), thus deriving the $\mathfrak{A}$ and $\mathfrak{B}$ matrix elements used throughout Section \ref{RPA}.
% Subsection A1
\subsection{Expansion of R, D, $\Psi^2$, $\Phi^2$, $X^2$ and $\Xi$} \label{AppbexpRD}
As appears in (\ref{GVP_HExpectVal}), the momentum space forms of the quantities are given by
% Eqs. A1
\begin{subequations} \label{App_R_D_psi_phi_chi_xi}
\begin{align}
	\setlength{\jot}{.1in}
	%\begin{equation} 
		\begin{split}
			R(\mathbf{k},\mathbf{k'},t) &=  \frac{1}{2} \left[ \frac{1}{4} G^{-1}(\mathbf{k},\mathbf{k'},t) +
			G(\mathbf{k},\mathbf{k'},t) - \delta(\mathbf{k} - \mathbf{k'}) \right] \\
			& \quad + 2 \negthickspace \negthickspace 
			\int \limits_{\mathbf{k''},\mathbf{k'''}} \negthickspace \negthickspace
			\Sigma(\mathbf{k},\mathbf{k''},t) 
			G(\mathbf{k''},\mathbf{k'''},t) \Sigma(\mathbf{k'''},\mathbf{k'},t)
		\end{split} \label{App_R} \\
	%\end{equation}
	%\begin{equation}
		\begin{split}
			\hspace{-.25 in} D(\mathbf{k},\mathbf{k'},t) &= \frac{1}{2} \left[ \frac{1}{4} 
			G^{-1}(\mathbf{k},\mathbf{k'},t) -
			G(\mathbf{k},\mathbf{k'},t) \right] \\
			& \quad + 2 \negthickspace \negthickspace \int \limits_{\mathbf{k''},\mathbf{k'''}} 
			\negthickspace \negthickspace \Sigma(\mathbf{k},\mathbf{k''},t) 
			G(\mathbf{k''},\mathbf{k'''},t) \Sigma(\mathbf{k'''},\mathbf{k'},t) \\ 
			& \quad -i \int \limits_\mathbf{k''} \left[ \Sigma(\mathbf{k},\mathbf{k''},t) 
			G(\mathbf{k''},\mathbf{k'},t) + G(\mathbf{k},\mathbf{k''},t) \Sigma(\mathbf{k''},\mathbf{k'},t)
			\right]
		\end{split}  \label{App_D} \\
	%\end{equation}
	%\begin{equation} 
		\Psi(\mathbf{k},t) &= \frac{1}{\sqrt{2}}\Bigl[ \psi(\mathbf{k},t) + i \pi(\mathbf{k},t) \Bigr] .
		\label{App_psi}
	%\end{equation}
\end{align}
\end{subequations}
The molecular field expressions are straightforward, following from (\ref{App_psi}).  Expanding $K$ to second-order first requires an expansion of $R$, $D$, $\Psi$, $\Phi$, $X$ and $\Xi$ in terms of the variations (\ref{SO_RPAexpansion}).  Using (\ref{App_R_D_psi_phi_chi_xi}) obtains
% Eq. A2
\begin{equation} \label{App_Rexpansion}
	R(\mathbf{k},\mathbf{k'},t) = R^{(0)}(\mathbf{k},\mathbf{k'},t) + R^{(1)}(\mathbf{k},\mathbf{k'},t) + 
	R^{(2)}(\mathbf{k},\mathbf{k'},t),
\end{equation}
\noindent where
% Eqs. A3
\setlength{\jot}{.1in}
\begin{subequations} \label{App_R012}
	{\allowdisplaybreaks \begin{align}
		\begin{split}
			R^{(0)}(\mathbf{k},\mathbf{k'},t) & = \frac{1}{2}\left[\frac{1}{4} G(\mathbf{k})^{-1} + 
			G(\mathbf{k}) - 1 \right] \delta(\mathbf{k} - \mathbf{k'}) \\ 
			\vphantom{\frac{\Gamma}{\Gamma}} 
			& = R(\mathbf{k}) \delta(\mathbf{k} - \mathbf{k'}) \label{App_R0} 
		\end{split} \\
		R^{(1)}(\mathbf{k},\mathbf{k'},t) & = \frac{1}{2}\left[-\frac{1}{4}G(\mathbf{k})^{-1}
		G(\mathbf{k'})^{-1} + 1 \right] \delta G(\mathbf{k},\mathbf{k'},t) \label{App_R1} \\
		%\begin{split}
			R^{(2)}(\mathbf{k},\mathbf{k'},t) & = \frac{1}{8} G(\mathbf{k})^{-1} \int 
			\limits_\mathbf{k''} \delta G^\ast (\mathbf{k},\mathbf{k''},t) \, G(\mathbf{k''})^{-1} \,
			\delta G(\mathbf{k''},\mathbf{k'},t) G(\mathbf{k'})^{-1} \nonumber \\
			& \quad + 2 \int \limits_\mathbf{k''} \delta \Sigma^\ast (\mathbf{k},\mathbf{k''},t) \, 
			G(\mathbf{k''}) \, \delta \Sigma(\mathbf{k''},\mathbf{k'},t). \label{App_R2}
		%\end{split}
	\end{align}}
\end{subequations}
Similarly, the first few orders of the other fluctuations are
% Eqs. A4
\begin{subequations} \label{App_D012}
\setlength{\jot}{.15in}
	\begin{align}
		\begin{split}
			D^{(0)}(\mathbf{k},\mathbf{k'},t) & = \frac{1}{2}\left[\frac{1}{4} G(\mathbf{k})^{-1} - 
			G(\mathbf{k}) \right] \delta(\mathbf{k} - \mathbf{k'}) 
			\hphantom{************} \\
			%\vphantom{\frac{\Gamma}{\Gamma}}
			& = D(\mathbf{k}) \delta(\mathbf{k} - \mathbf{k'}) \label{App_D0}
		\end{split} \\
		\begin{split}
			D^{(1)}(\mathbf{k},\mathbf{k'},t) & = - \frac{1}{2}\left[\frac{1}{4}G(\mathbf{k})^{-1}
			G(\mathbf{k'})^{-1} + 1 \right] \delta G(\mathbf{k},\mathbf{k'},t) \\
			& \quad - i \left[G(\mathbf{k}) 
			+ G(\mathbf{k'})\right] \delta \Sigma(\mathbf{k},\mathbf{k'},t)
			\label{App_D1}  
		\end{split} \\
		%\begin{split}
			D^{(2)}(\mathbf{k},\mathbf{k'},t) & = \frac{1}{8} G(\mathbf{k})^{-1} \int 
			\limits_\mathbf{k''} \delta G^\ast (\mathbf{k},\mathbf{k''},t) \, G(\mathbf{k''})^{-1} \,
			\delta G(\mathbf{k''},\mathbf{k'},t) G(\mathbf{k'})^{-1}  \nonumber \\
			& \quad + 2 \int \limits_\mathbf{k''} \delta 
			\Sigma^\ast(\mathbf{k},\mathbf{k''},t) \, G(\mathbf{k''}) \,
			\delta \Sigma(\mathbf{k''},\mathbf{k'},t) \label{App_D2} \\
			& \hspace{-.5in} - i \int \limits_\mathbf{k''} \left[ \delta \Sigma^\ast(\mathbf{k},\mathbf{k''},t)
			\delta G(\mathbf{k''},\mathbf{k'},t) + \delta G^\ast(\mathbf{k},\mathbf{k''},t) 
			\delta \Sigma(\mathbf{k''},\mathbf{k'},t) \right] \! . \nonumber 
		%\end{split}
	\end{align}
\end{subequations}
Expanding the products of mean fields gives
% Eqs. A5
\begin{subequations} \label{App_psi*psi012}
	\begin{align}
		\Psi^\ast \Psi^{(0)}(\mathbf{k},\mathbf{k'},t) & = \frac{1}{2} \psi^2 \, \delta(\mathbf{k}) 
		\delta(\mathbf{k'}) \label{App_psi*psi0} \\
		\begin{split} 
			\Psi^\ast\Psi^{(1)}(\mathbf{k},\mathbf{k'},t) & = \frac{1}{2} \psi \, \delta(\mathbf{k'}) \left[ 
			\delta \psi^\ast(\mathbf{k},t) - i \delta \pi^\ast(\mathbf{k},t) \right] \\
			& + \frac{1}{2}  \psi \, \delta(\mathbf{k}) \left[ \delta \psi(\mathbf{k'},t) + i \delta 
			\pi(\mathbf{k'},t) \right] \label{App_psi*psi1}
		\end{split} \\
		\begin{split}
			\Psi^\ast\Psi^{(2)}(\mathbf{k},\mathbf{k'},t) & = \frac{1}{2} \left[ \delta \psi^\ast(\mathbf{k},t)
			\delta \psi(\mathbf{k'},t) + i \delta \psi^\ast(\mathbf{k},t) \delta \pi(\mathbf{k'},t) \right. \\
			& \qquad \left. - \, i \delta \psi(\mathbf{k'},t) \delta \pi^\ast(\mathbf{k},t) 
			+ \delta \pi^\ast(\mathbf{k},t) \delta \pi(\mathbf{k'},t) \right] \label{App_psi*psi2}
		\end{split}
	\end{align}
\end{subequations}
% Eqs. A6
\begin{subequations} \label{App_psipsi012}
	{\allowdisplaybreaks \begin{align}
		\Psi \Psi^{(0)}(\mathbf{k},\mathbf{k'},t) & = \frac{1}{2} \psi^2 \, \delta(\mathbf{k}) 
		\delta(\mathbf{k'}) \label{App_psipsi0} \\
		\begin{split} 
			\Psi\Psi^{(1)}(\mathbf{k},\mathbf{k'},t) & = \frac{1}{2} \psi \, \delta(\mathbf{k'}) \left[ 
			\delta \psi(\mathbf{k},t) + i \delta \pi(\mathbf{k},t) \right] \\
			& + \frac{1}{2}  \psi \, \delta(\mathbf{k}) \left[ \delta \psi(\mathbf{k'},t) + i \delta 
			\pi(\mathbf{k'},t) \right] \label{App_psipsi1}
		\end{split} \\
		\begin{split}
			\Psi\Psi^{(2)}(\mathbf{k},\mathbf{k'},t) 
			& = \frac{1}{2} \left[ \delta \psi(\mathbf{k},t)
			\delta \psi(\mathbf{k'},t) + i \delta \psi(\mathbf{k},t) \delta \pi(\mathbf{k'},t) \right. \\
			& \qquad \left. + \, i \delta \psi(\mathbf{k'},t) \delta \pi(\mathbf{k},t) 
			- \delta \pi(\mathbf{k},t) \delta \pi(\mathbf{k'},t) \right] \label{App_psipsi2}
		\end{split}
	\end{align}}
\end{subequations}
% Eqs. A7
\begin{subequations} \label{App_phi*phi012}
	\begin{align}
		\Phi^\ast \Phi^{(0)}(\mathbf{k},\mathbf{k'},t) & = \frac{1}{2} \phi^2 \, \delta(\mathbf{k}) 
		\delta(\mathbf{k'}) \label{App_phi*phi0} \\
		\begin{split} 
			\Phi^\ast\Phi^{(1)}(\mathbf{k},\mathbf{k'},t) & = \frac{1}{2} \phi \, \delta(\mathbf{k'}) \left[ 
			\delta \phi^\ast(\mathbf{k},t) - i \delta \omega^\ast(\mathbf{k},t) \right] \\
			& + \frac{1}{2}  \phi \, \delta(\mathbf{k}) \left[ \delta \phi(\mathbf{k'},t) + i \delta 
			\omega(\mathbf{k'},t) \right] \label{App_phi*phi1}
		\end{split} \\
		\begin{split}
			\Phi^\ast\Phi^{(2)}(\mathbf{k},\mathbf{k'},t) & = \frac{1}{2} \left[ \delta \phi^\ast(\mathbf{k},t)
			\delta \phi(\mathbf{k'},t) + i \delta \phi^\ast(\mathbf{k},t) \delta \omega(\mathbf{k'},t) \right. \\
			& \qquad \left. - \, i \delta \phi(\mathbf{k'},t) \delta \omega^\ast (\mathbf{k},t) 
			+ \delta \omega^\ast(\mathbf{k},t)
			\delta \omega (\mathbf{k'},t) \right] \label{App_phi*phi2}
		\end{split}
	\end{align}
\end{subequations}
% Eqs. A8
\begin{subequations} \label{App_chi*chi012}
	\begin{align}
		X^\ast X^{(0)}(\mathbf{k},\mathbf{k'},t) & = \frac{1}{2} \chi^2 \, \delta(\mathbf{k}) 
		\delta(\mathbf{k'}) \label{App_chi*chi0} \\
		\begin{split} 
			X^\ast X^{(1)}(\mathbf{k},\mathbf{k'},t) & = \frac{1}{2} \chi \, \delta(\mathbf{k'}) \left[ 
			\delta \chi^\ast(\mathbf{k},t) - i \delta \nu^\ast(\mathbf{k},t) \right] \\
			& + \frac{1}{2}  \chi \, \delta(\mathbf{k}) \left[ \delta \chi(\mathbf{k'},t) + i \delta 
			\nu(\mathbf{k'},t) \right] \label{App_chi*chi1}
		\end{split} \\
		\begin{split}
			X^\ast X^{(2)}(\mathbf{k},\mathbf{k'},t) & = \frac{1}{2} \left[ \delta \chi^\ast(\mathbf{k},t)
			\delta \chi(\mathbf{k'},t) + i \delta \chi^\ast(\mathbf{k},t) \delta \nu(\mathbf{k'},t) \right. \\
			&\qquad \left. - \, i \delta \chi(\mathbf{k'},t) \delta \nu^\ast(\mathbf{k},t) 
			+ \delta \nu^\ast(\mathbf{k},t) \delta \nu(\mathbf{k'},t) \right] \label{App_chi*chi2}
		\end{split}
	\end{align}
\end{subequations}
% Eqs. A9
\begin{subequations} \label{App_xi01}
	{\allowdisplaybreaks \begin{align}
		\begin{split}
			\Xi^{(0)}(\mathbf{k''},t)& = \frac{\xi}{\sqrt{2}} \delta (\mathbf{k''}) \\
			                                     & = \frac{1}{\sqrt{2}}
			                                     (\lambda \alpha_1 \phi + g \chi) \delta(\mathbf{k''}) 							\label{App_xi0}
		\end{split}\\
		\begin{split}
			\Xi^{(1)}(\mathbf{k''},t) & = \frac{1}{\sqrt{2}} \delta \xi (\mathbf{k''},t) \\
							& = \frac{1}{\sqrt{2}} 
							\left[ \lambda \alpha_1 \delta \phi(\mathbf{k''},t) + g \delta 										\chi(\mathbf{k''},t) \right] \\
							& + \frac{i}{\sqrt{2}} \left[ \lambda \alpha_1 \delta
							         \omega(\mathbf{k''},t) + g \delta \nu(\mathbf{k''},t) \right] \! .
							         %\vphantom{\int} 
							         \qquad \qquad
							         \label{App_xi1}
		\end{split}
	\end{align}}
\end{subequations}
\noindent \hspace{-8pt} Along with (\ref{App_R}) and (\ref{App_D}), these expansions are used to calculate the second-order contribution from each term in (\ref{GVP_HExpectVal}).
% Subsection A2
%\setcounter{equation}{0}
\subsection{Kinetic Contribution} \label{App_SecondOrderKin}
From Eqs. \!(\ref{App_R012}), the second-order part of $\int \limits_\mathbf{k} (k^2 - \mu) R(\mathbf{k},\mathbf{k},t)$ is
 % Eq. A10
 \begin{equation} \label{App_SOK}
	\begin{split}
		\int \limits_{\mathbf{k}} (k^2 - \mu) R^{(2)}(\mathbf{k},\mathbf{k},t)
		&= \int \limits_{\mathbf{q}, 
		\mathbf{P}} \left[ \delta G^{\ast} (\mathbf{q},\mathbf{P},t) (k^2 - \mu) 								R_{\, G^2}(\mathbf{q},\mathbf{P}) \delta G(\mathbf{q}, \mathbf{P},t) \right. \\
		& \left. \quad + \, \delta \Sigma^\ast(\mathbf{q},\mathbf{P},t) (k^2 - \mu) R_{\, \Sigma^2} 						(\mathbf{q},\mathbf{P}) \delta \Sigma(\mathbf{q},\mathbf{P},t) \right] \! , 
		\vphantom{\int}
	\end{split}
\end{equation}
where the total and relative momenta are identified in (\ref{SO_NewMomentumCoords}).
% Eq. A11
%\begin{subequations} \label{App_Pq}
%	\begin{align} 
%		\mathbf{P}& = \mathbf{k} - \mathbf{k'}&
%			\mathbf{P'}& = \mathbf{k''} - \mathbf{k'''}\\
%		\mathbf{q}& = \frac{1}{2}(\mathbf{k} + \mathbf{k'})&
%			\mathbf{q'}& = \frac{1}{2}(\mathbf{k''} + \mathbf{k'''}) .
%	\end{align}
%\end{subequations}
\noindent Furthermore, we have used the transformation $\delta G^\ast(\mathbf{k'},\mathbf{k},t) \rightarrow \delta G^\ast(\mathbf{q},\mathbf{P},t)$, with the symmetry under inversion $\mathbf{P} \rightarrow - \mathbf{P}$ incorporated by the following definitions
% Eqs. A11
\begin{subequations} \label{App_RG2_RSig2}
	\begin{align}
		\begin{split}
		(k^2 - \mu)R_{G^2}(\mathbf{q},\mathbf{q'},\mathbf{P}) & = \frac{1}{16} \Bigl[ (\mathbf{q}_+^2 - 
		\mu)G_+^{-2}G_-^{-1} \\
		& \qquad + (\mathbf{q}_-^2 - \mu)G_-^{-2}G_+^{-1} \Bigr] \delta(\mathbf{q} - 
		\mathbf{q'}) \label {App_RG2}
		\end{split} \\
		(k^2 - \mu)R_{\Sigma^2}(\mathbf{q},\mathbf{q'},\mathbf{P}) & = \Bigl[ (\mathbf{q}_+^2 - 
		\mu)G_- + (\mathbf{q}_-^2 - \mu)G_+ \Bigr] \delta(\mathbf{q} - \mathbf{q'}).
		\label{App_RSig2}
	\end{align}
\end{subequations}
\noindent For notational simplicity, the $\pm$ subscripts denote a shift of $\pm \mathbf{P}/2$ in the argument of any function $f$ such that
% Eq. A12
\begin{equation} \label{App_shift}
	f_{\pm} = f\Bigl( \mathbf{q} \pm \frac{1}{2}\mathbf{P} \Bigl) , \qquad  f'_{\pm} = f\Bigl( \mathbf{q'} \pm 
	\frac{1}{2}\mathbf{P'} \Bigl).
\end{equation}
\indent By (\ref{App_psi*psi2}), the second-order contribution from $\int \limits_\mathbf{k} (k^2 - \mu) \Psi^\ast(\mathbf{k},t)\Psi(\mathbf{k},t)$ is
% Eq. A13
\begin{align} \label{App_psi*psi2ndord}
	%\begin{split}
		\int \limits_\mathbf{k} (k^2 - \mu) \Psi^\ast \Psi^{(2)}(\mathbf{k},\mathbf{k},t) & = 
		\int \limits_\mathbf{k} (k^2 - \mu) \frac{1}{2} \left[ \delta \psi(\mathbf{k},t)^2 +
		\delta \pi(\mathbf{k},t)^2 \right] \nonumber \\
		& = \int \limits_\mathbf{P} \left[ \delta \psi^\ast(\mathbf{P},t) (k^2 - \mu) 
		\Psi^\ast\Psi_{\psi^2}(\mathbf{P}) \delta \psi(\mathbf{P},t) \right. \nonumber \\
		& \left. \quad + \, \delta \pi^\ast(\mathbf{P},t) (k^2 - \mu) 
		\Psi^\ast\Psi_{\pi^2}(\mathbf{P}) \delta \pi(\mathbf{P},t) \right] \vphantom{\int} \! .
	%\end{split}
\end{align}
\noindent In addition to using the fact that the fluctuations are real, we have defined
% Eq. A14
\begin{subequations} \label{App_psi*psi_phi^2_phi*phi_pi^2}
	\begin{align}
		(k^2 - \mu)\Psi^\ast\Psi_{\psi^2}(\mathbf{P}) &= \frac{1}{2}(\mathbf{P}^2 - \mu)
		\label{App_psi*psi_phi^2} \\
		(k^2 - \mu)\Psi^\ast\Psi_{\pi^2}(\mathbf{P}) & = \frac{1}{2}(\mathbf{P}^2 - \mu) 
		\label{App_psi*psi_pi^2}.
	\end{align}
\end{subequations}
% Subsection A3
%\setcounter{equation}{0}
\subsection{Molecular Contribution} \label{App_Sec_Ord_Molec}
The second-order contributions from the molecular detunings are given by
% Eq. A15
\begin{align} \label{App_SOMolec}
	%\begin{split}
		&\left(\epsilon - 2 \mu \right) \int \limits_{\mathbf{k}} \Phi^{\ast}
		\Phi^{(2)}(\mathbf{k},\mathbf{k},t) + \left(\varepsilon - 2\mu \right) \int 
		\limits_{\mathbf{k}} X^{\ast} X^{(2)} (\mathbf{k},\mathbf{k},t) 
		\hphantom{*************} 
		\nonumber \\
		&\quad = \int \limits_{\mathbf{P}} \Bigl[ \delta \phi^{\ast} (\mathbf{P},t) \Phi^{\ast}\Phi_{ 
		\phi^2} (\mathbf{P}) \delta \phi (\mathbf{P},t) + \delta \omega^{\ast} (\mathbf{P},t) \Phi^{\ast} 
		\Phi_{\omega^2} (\mathbf{P}) \delta \omega (\mathbf{P},t) \Bigr. \nonumber \\
		&\Bigl. \quad \quad + \delta \chi^{\ast} (\mathbf{P},t) X^{\ast} 
		X_{\chi^2} (\mathbf{P}) \delta \chi (\mathbf{P},t) + \delta \nu^{\ast} (\mathbf{P},t)
		X^{\ast}  X_{\nu^2} (\mathbf{P}) \delta \nu (\mathbf{P},t) \Bigr] .
	%\end{split}
\end{align}
\noindent Reading off the results from (\ref{App_phi*phi2}) and (\ref{App_chi*chi2}), we identify
% Eqs. A16
\begin{subequations} \label{App_phi2_chi2}
	\begin{align}
		\Phi^\ast \Phi_{\phi^2} (\mathbf{P}) &= \Phi^{\ast} \Phi_{\omega^2} (\mathbf{P}) = 		
		\frac{1}{2} \epsilon - \mu  \label{App_phi2} \\
		X^\ast X_{\chi^2} (\mathbf{P}) &= X^\ast  X_{\nu^2} (\mathbf{P}) =
		\frac{1}{2} \varepsilon - \mu . \label{App_chi2}
	\end{align}
\end{subequations}
% Subsection A4
%\setcounter{equation}{0}
\subsection{Contribution from the Coupling} \label{App_Sec_Ord_Coup}
The last integral of (\ref{GVP_HExpectVal}) gives the energy due to the coupling of atoms to both of the molecular states.  In this contribution, there is an $\Xi^{\ast} D$ term with a second-order part given by
% Eq. A17
{\allowdisplaybreaks \begin{align} 
	%\begin{split}
		& \frac{1}{\sqrt{2}} \int \limits_{\mathbf{k},\mathbf{k'},\mathbf{k''}} 
		\delta(\mathbf{k''} - \mathbf{k} + \mathbf{k'}) f\left( \frac{\mathbf{k}+\mathbf{k'}}{2}
		\right) \left[ \Xi^{\ast (0)} (\mathbf{k''},t) D^{(2)}(\mathbf{k},\mathbf{k'},t)
		\right. \nonumber \\
		& \left. \hphantom{\int \limits_{\mathbf{k},\mathbf{k'},\mathbf{k''}} 
		\delta(\mathbf{k''} - \mathbf{k} + \mathbf{k'}) f\left( \frac{\mathbf{k}+\mathbf{k'}}{2}
		\right)} 
		+ \Xi^{\ast (1)} (\mathbf{k''},t) D^{(1)}(\mathbf{k},\mathbf{k'},t) + \mbox{H. c.} 
		\right] \nonumber \\
		& \quad = \int \limits_{\mathbf{k}, \mathbf{k'''}} f(\mathbf{k}) \xi
		\left[ \delta G^\ast(\mathbf{k},\mathbf{k'''},t) \, \frac{1}{8} G(\mathbf{k})^{-2} G(\mathbf{k'''})^{-1} 
		\delta G (\mathbf{k'''}, \mathbf{k},t) \right. \nonumber \\
		& \hphantom{\int \limits_{\mathbf{k},\mathbf{k'},\mathbf{k''}} 
		\delta(\mathbf{k''} - \mathbf{k} + \mathbf{k'}) f\left( \frac{\mathbf{k}+\mathbf{k'}}{2}
		\right)} \left. \vphantom{\frac{1}{4}}
		+ \, \delta \Sigma^\ast(\mathbf{k},\mathbf{k'''},t) 2 G(\mathbf{k'''})
		\delta \Sigma(\mathbf{k'''}, \mathbf{k},t) \right] \nonumber \\
		&\qquad + \int \limits_{\mathbf{k}, \mathbf{k'}, \mathbf{k''}} \delta(\mathbf{k''} - \mathbf{k} + 
		\mathbf{k'}) f\left( \frac{\mathbf{k}+\mathbf{k'}}{2} \right) \nonumber \\
		& \qquad \qquad \qquad \times 
		\left\{ - \frac{\lambda \alpha_1}{2} \delta \phi^\ast (\mathbf{k''},t) 
		\left[ \frac{1}{4} G(\mathbf{k})^{-1} 
		G(\mathbf{k'})^{-1} + 1 \right] \delta G (\mathbf{k},\mathbf{k'},t) \right.  \nonumber \\
		& \qquad \qquad \qquad \qquad \; \;
		- \frac{g}{2} \delta \chi^\ast(\mathbf{k''},t) \left[ \frac{1}{4} G(\mathbf{k})^{-1} 
		G(\mathbf{k'})^{-1} + 1 \right]
		\delta G(\mathbf{k},\mathbf{k'},t) \nonumber \\
		& \qquad \qquad \qquad \qquad \; \;
		- \lambda \alpha_1 \delta \omega^\ast (\mathbf{k''},t) 
		\left[ G(\mathbf{k}) +G(\mathbf{k'}) \right] \delta \Sigma (\mathbf{k}, \mathbf{k'}, t) \nonumber \\
		& \left. \qquad \qquad \qquad \qquad \; \;
		- g \delta \nu^\ast(\mathbf{k''},t) \left[ G(\mathbf{k}) 
		+ G(\mathbf{k'}) \right] \delta \Sigma(\mathbf{k},\mathbf{k'},t) 
		\vphantom{\left[ \frac{1}{4} \right]} \right\} \! ,
	%\end{split}
\end{align}}
\noindent which can be more compactly written as
% Eq. A18
\begin{equation}  \label{App_SO_xiD}
	\begin{split}
		& \int \limits_{\mathbf{q}, \mathbf{q'}, \mathbf{P}} 
		\Bigl[ \delta G^{\ast} (\mathbf{q},\mathbf{P},t)
		D_{G^2}(\mathbf{q},\mathbf{q'},\mathbf{P}) \delta G(\mathbf{q},\mathbf{P},t) \Bigr. \\
		& \qquad \qquad \qquad \qquad \qquad \qquad 
		\Bigl. + \, \delta \Sigma^{\ast}(\mathbf{q},\mathbf{P},t) 
		D_{\Sigma^2} (\mathbf{q},\mathbf{q'},\mathbf{P}) 
		\delta \Sigma (\mathbf{q},\mathbf{P},t) \Bigr] \\
		& \negthickspace \negthickspace \negthickspace 
		 + \int \limits_{\mathbf{q}, \mathbf{P}} 
		\Bigl[ \delta \phi^{\ast}(\mathbf{P},t) D_{G \phi} (\mathbf{q}, \mathbf{P})
		\delta G (\mathbf{q},\mathbf{P},t)
		+ \delta \chi^{\ast}(\mathbf{P},t) D_{G \chi} (\mathbf{q},
		\mathbf{P},t) \delta G (\mathbf{q},\mathbf{P},t) \Bigr. \\
		&  \negthickspace \negthickspace \negthickspace \qquad 
		\Bigl. + \, \delta \omega^{\ast} (\mathbf{P},t) 
		D_{\Sigma \omega} (\mathbf{q}, \mathbf{P})
		\delta \Sigma(\mathbf{q}, \mathbf{P},t)
		+ \delta \nu^{\ast}(\mathbf{P},t) D_{\Sigma \nu} (\mathbf{q},\mathbf{P}) 
		\delta \Sigma(\mathbf{q},\mathbf{P},t) \Bigr] ,
	\end{split}
\end{equation}
where
% Eqs. A19
\begin{subequations} \label{App_xiD_terms}
	{\allowdisplaybreaks \begin{align}
	\begin{split} \label{App_xiDG2} 
		D_{G^2}(\mathbf{q},\mathbf{q'},\mathbf{P}) &= \xi \left[
		f\left(\mathbf{q} + \frac{1}{2} \mathbf{P} \right) \frac{1}{16} G_+^{-2} \, G_-^{-1} \right. \\
		& \qquad \qquad \qquad 
		\left. + f\left(\mathbf{q} - \frac{1}{2} \mathbf{P} \right) \frac{1}{16} G_-^{-2} \, G_+^{-1} \right] 
		\delta(\mathbf{q} - \mathbf{q'}) 
	\end{split} \\
	\begin{split} \label{App_DSig2}
		D_{\Sigma^2}(\mathbf{q},\mathbf{q'},\mathbf{P}) &= \xi \left[
		f\left(\mathbf{q} + \frac{1}{2} \mathbf{P} \right) G_- \right. \\
		& \qquad \qquad \qquad
		\left.+ f\left(\mathbf{q} - \frac{1}{2} \mathbf{P} \right) G_+ \right] 
		\delta(\mathbf{q} - \mathbf{q'})
	\end{split} \\
	%\begin{align} 
		D_{G \phi}(\mathbf{q},\mathbf{P}) &= - \frac{\lambda \alpha_1}{2} f(\mathbf{q}) 
		\left( \frac{1}{4} G_+^{-1} \, G_-^{-1} + 1 \right) \label{App_DGphi} \\
		D_{G \chi}(\mathbf{q},\mathbf{P}) &= - \frac{g}{2} f(\mathbf{q}) 
		\left( \frac{1}{4} G_+^{-1} \, G_-^{-1} + 1 \right) \label{App_DGchi} \\
		D_{\Sigma \omega}(\mathbf{q},\mathbf{P}) &= - \lambda \alpha_1 f(\mathbf{q}) 
		(G_+ + G_-) \label{App_DSigom} \\
		D_{\Sigma \nu}(\mathbf{q},\mathbf{P}) &= - g f(\mathbf{q}) 
		(G_+ + G_-) \label{App_DSignu}.
	\end{align}}
\end{subequations}
\negthickspace Likewise, the second-order part from the $\Xi^{\ast} \Psi \Psi$ term is
% Eq. A20
\begin{align} 
	\begin{split}
		& \frac{1}{\sqrt{2}} \int \limits_{\mathbf{k},\mathbf{k'},\mathbf{k''}} \negthickspace
		\delta(\mathbf{k''} - \mathbf{k} + \mathbf{k'}) f\left( \frac{\mathbf{k}+\mathbf{k'}}{2}
		\right) \left[ \Xi^{\ast (0)} (\mathbf{k''},t)
		\Psi \Psi^{(2)}(\mathbf{k},\mathbf{k'},t) \right. \\
		& \qquad \qquad \qquad \qquad \qquad \qquad \quad \left.
		+ \, \Xi^{\ast (1)}(\mathbf{k''},t) \Psi \Psi^{(1)}(\mathbf{k},\mathbf{k'},t) + \mbox{H. c.} \right] 
		\\
		& 
		= \int \limits_{\mathbf{k}} f(\mathbf{k}) \frac{\xi}{2}
		\left[ \delta \psi(\mathbf{k},t)^2 -  \delta \pi(\mathbf{k},t)^2 \right] \\
		& 
		+ \frac{1}{2} \int \limits_{\mathbf{k}, \mathbf{k'}, \mathbf{k''}} \negthickspace 
		\delta(\mathbf{k''} - \mathbf{k} + \mathbf{k'}) 
		f\left( \frac{\mathbf{k}+\mathbf{k'}}{2}\right) 
		\left\{ \left[ \lambda \alpha_1 \delta \phi (\mathbf{k''},t) 
		+ g \delta \chi(\mathbf{k''},t) \right]  \right. \\
		& \qquad \qquad \qquad \qquad \qquad \qquad \quad
		 \times \left[ \psi \delta(\mathbf{k'}) \delta \psi(\mathbf{k},t) 
		+ \psi \delta(\mathbf{k}) \delta \psi (\mathbf{k'},t) \right] \\
		& 
		\quad \left. 
		+ \left[ \lambda \alpha_1 \delta \omega(\mathbf{k''},t) + g \delta \nu(\mathbf{k''},t) \right]
		\left[ \psi \delta(\mathbf{k'}) \delta \pi(\mathbf{k},t) 
		+ \psi \delta (\mathbf{k}) \delta \pi(\mathbf{k'},t) \right] \right\} \! . 
	\end{split}
\end{align}
\noindent After multiplying out the integrand, this becomes
% Eq. A21
\begin{equation} \label{App_xipsi2}
	\begin{split}
		& \int \limits_\mathbf{P} \Bigl[ \delta \psi^\ast(\mathbf{P},t) 
		\Psi^2_{\, \psi^2}(\mathbf{P}) \delta \psi(\mathbf{P},t) 
		+ \delta \pi^\ast(\mathbf{P},t) \Psi^2_{\, \pi^2} (\mathbf{P}) \delta \pi (\mathbf{P},t)  \Bigr. \\
		& \hphantom{ \int \limits_\mathbf{P}} \vphantom{ \int \limits_\mathbf{P}}
		+ \delta \phi^\ast(\mathbf{P},t) \Psi^2_{\, \phi \psi} (\mathbf{P}) \delta \psi(\mathbf{P},t) 
		+ \delta \chi^\ast(\mathbf{P},t) \Bigl. \Psi^2_{\, \chi \psi} (\mathbf{P}) \delta \psi(\mathbf{P},t) \\
		& \hphantom{ \int \limits_\mathbf{P}} 
		\vphantom{ \int \limits_\mathbf{P}} \Bigl.
		+ \delta \omega^\ast(\mathbf{P},t) \Psi^2_{\, \omega \pi}(\mathbf{P}) \delta \pi (\mathbf{P},t)
		+ \delta \nu^\ast(\mathbf{P},t) \Psi^2_{\, \nu \pi} (\mathbf{P}) \delta \pi (\mathbf{P},t) \Bigr],
	\end{split}
\end{equation}
\noindent where
% Eqs. A22
\begin{subequations} \label{App_psi2_terms}
	{\allowdisplaybreaks \begin{align} 
		\Psi^2_{\, \psi^2}(\mathbf{P}) &= -\Psi^2_{\, \pi^2}(\mathbf{P}) 
		= \frac{\xi}{2} f(\mathbf{P}) \label{App_psi2_psi2} \\
		\Psi^2_{\, \phi \psi}(\mathbf{P}) &= \Psi^2_{\, \omega \pi}(\mathbf{P})
		= \lambda \alpha_1 \psi f\left(\frac{1}{2}\mathbf{P}\right)
		\label{App_psi2_phipsi} \\
		\Psi^2_{\, \chi \psi}(\mathbf{P}) &= \Psi^2_{\, \nu \pi}(\mathbf{P})
		= g \psi f\left(\frac{1}{2} \mathbf{P}\right) 
		\label{App_psi2_chipsi} \! . 
	\end{align}}
\end{subequations}
% Subsection A5
%\setcounter{equation}{0}
\subsection{Matrix Form of the RPA} \label{App_RPAmat}
Expressed in matrix form, the grand canonical RPA Hamiltonian corresponding to Eq. \!(\ref{GVP_HExpectVal}) can be written\footnote{The dependence on the total momentum $\mathbf{P}$ has been omitted for notational clarity.}
{\allowdisplaybreaks \begin{align} \label{AppbK_RPAmat}
	\begin{split}
		K_{RPA} &= \frac{1}{2} \delta \mathcal{P}^\dag(\mathbf{q}) \, 
		\begin{bmatrix}
		\mathfrak{A}_{GG}(\mathbf{q},\mathbf{q'}) & 
		\mathfrak{A}_{G \psi}(\mathbf{q}) &
		\mathfrak{A}_{G \phi}(\mathbf{q}) & 
		\mathfrak{A}_{G \chi}(\mathbf{q})\\
		\mathfrak{A}_{\psi G}(\mathbf{q'}) & 
		\mathfrak{A}_{\psi \psi} &
		\mathfrak{A}_{\psi \phi} &
		\mathfrak{A}_{\psi \chi} \\
		\mathfrak{A}_{\phi G}(\mathbf{q'}) & 
		\mathfrak{A}_{\phi \psi} &
		\mathfrak{A}_{\phi \phi} &
		\mathfrak{A}_{\phi \chi} \\
		\mathfrak{A}_{\chi G}(\mathbf{q'}) & 
		\mathfrak{A}_{\chi \psi} &
		\mathfrak{A}_{\chi \phi} &
		\mathfrak{A}_{\chi \chi} \\
		\end{bmatrix} \,
		\delta \mathcal{P}(\mathbf{q'}) \\[.1 in]
		& + \frac{1}{2} \delta \mathcal{Q}^\dag (\mathbf{q}) \, 
		 \begin{bmatrix}
		\mathfrak{B}_{\Sigma \Sigma}(\mathbf{q},\mathbf{q'}) & 
		\mathfrak{B}_{\Sigma \pi}(\mathbf{q}) &
		\mathfrak{B}_{\Sigma \omega}(\mathbf{q}) &
		\mathfrak{B}_{\Sigma \nu}(\mathbf{q}) \\
		\mathfrak{B}_{\pi \Sigma}(\mathbf{q'}) & 
		\mathfrak{B}_{\pi \pi} &
		\mathfrak{B}_{\pi \omega} &
		\mathfrak{B}_{\pi \nu} \\
		\mathfrak{B}_{\omega \Sigma}(\mathbf{q'}) &
		\mathfrak{B}_{\omega \pi} &
		\mathfrak{B}_{\omega \omega} &
		\mathfrak{B}_{\omega \nu} \\
		\mathfrak{B}_{\nu \Sigma}(\mathbf{q'}) &
		\mathfrak{B}_{\nu \pi} &
		\mathfrak{B}_{\nu \omega} &
		\mathfrak{B}_{\nu \nu}
		\end{bmatrix} \,
		\delta \mathcal{Q}(\mathbf{q'})
		\vphantom{\int \limits_\Sigma},
	\end{split}
\end{align}}
\noindent with the coordinates, $\delta \mathcal{Q}$, and momenta, $\delta \mathcal{P}$, defined as
% Eqs. A24
\begin{subequations} \label{App_QP}
	\begin{align}
		\delta \mathcal{Q}^\dag (\mathbf{q}, \mathbf{P}) &= 
		\begin{bmatrix} \delta \Sigma(\mathbf{q}, \mathbf{P}) & 
		\delta \pi(\mathbf{P}) & \delta \omega(\mathbf{P}) & 
		\delta \nu (\mathbf{P})\end{bmatrix} \, \\
		\delta \mathcal{P}^\dag (\mathbf{q}, \mathbf{P}) &= 
		\begin{bmatrix} \delta G(\mathbf{q}, \mathbf{P}) & 
		\delta \psi(\mathbf{P}) & \delta \phi (\mathbf{P}) & 
		\delta \chi (\mathbf{P})\end{bmatrix} \! .
	\end{align}
\end{subequations}
To find explicit expressions for the matrix elements, we combine the results of (\ref{App_RG2_RSig2}), (\ref{App_psi*psi_phi^2_phi*phi_pi^2}), (\ref{App_phi2_chi2}), (\ref{App_xiD_terms}) and (\ref{App_psi2_terms}).  
% Subsection A6
%\setcounter{equation}{0}
\subsection{$\mathfrak{A}$ Matrix Elements} \label{App_Amatrix_elmnts}
As in the discussion following (\ref{SUS_Variations}), we take $f(\mathbf{k}) = 1$ with a cutoff at $|\mathbf{k}| = 4 \pi^2 / b$.  Subsequently, the matrix elements are simply matched with the expansions, leading to the following identifications:
% Eq. A25
\begin{align} \label{App_AGG}
	\begin{split}
		\mathfrak{A}_{GG}(\mathbf{q},\mathbf{q'},\mathbf{P}) &=  
		2 R_{G^2}(\mathbf{q},\mathbf{q'},\mathbf{P}) + 
		2D_{G^2}(\mathbf{q},\mathbf{q'},\mathbf{P}) \\
		&= \frac{1}{8} \left[ (q_+^2 - \mu + \xi) G_+^{-2}G_-^{-1} + 
		(q_-^2 - \mu + \xi)G_-^{-2}G_+^{-1} \right]
		\delta(\mathbf{q} - \mathbf{q'}) \\
		&= s_M(\mathbf{q}, \mathbf{P}) \, \delta(\mathbf{q} - \mathbf{q'}).
	\end{split}
\end{align}
\noindent In the above case, for instance, the terms were obtained from (\ref{App_RG2}) and (\ref{App_xiDG2}).  Similarly, the remaining elements are:
% Eq. A26
\begin{equation} \label{App_AGpsi}
	\mathfrak{A}_{G \psi}(\mathbf{q},\mathbf{P}) = \mathfrak{A}_{\psi G}(\mathbf{q}, \mathbf{P}) = 0
\end{equation}
% Eq. A27
\begin{align} \label{App_AGphi}
	\begin{split}
		\mathfrak{A}_{G \phi}(\mathbf{q}, \mathbf{P}) = \mathfrak{A}_{\phi G}(\mathbf{q},\mathbf{P}) 
		&= D_{G \phi}(\mathbf{q}, \mathbf{P}) \\
		&= - \frac{\lambda \alpha_1}{2} \left( \frac{1}{4}G_+^{-1}G_-^{-1} + 1 \right)
	\end{split}
\end{align}
% Eq. A28
\begin{align} \label{App_AGchi}
	\begin{split}
		\mathfrak{A}_{G \chi}(\mathbf{q}, \mathbf{P}) = \mathfrak{A}_{\chi G}(\mathbf{q},\mathbf{P}) 
		&= D_{G \chi}(\mathbf{q}, \mathbf{P}) \\
		&= - \frac{g}{2} \left( \frac{1}{4}G_+^{-1}G_-^{-1} + 1 \right)
	\end{split}
\end{align}
% Eq. A29
\begin{align} \label{App_Apsipsi}
	\begin{split}
		\mathfrak{A}_{\psi \psi}(\mathbf{P}) &= 2(k^2 - \mu) \Psi^\ast \Psi_{\psi^2}(\mathbf{P}) 
		+ 2\Psi^2_{\, \psi^2}(\mathbf{P}) \\
		&= P^2 - \mu + \xi
	\end{split}
\end{align}
% Eq. A30
\begin{align} \label{App_Apsiphi}
	\begin{split}
		\mathfrak{A}_{\psi \phi}(\mathbf{P}) = \mathfrak{A}_{\phi \psi}(\mathbf{P}) &= 
		\Psi^2_{\, \phi \psi}(\mathbf{P}) \\ 
		&= \lambda \alpha_1 \psi
	\end{split}
\end{align}
% Eq. A31
\begin{align} \label{App_Apsichi}
	\begin{split}
		\mathfrak{A}_{\psi \chi}(\mathbf{P}) = \mathfrak{A}_{\chi \psi}(\mathbf{P}) &= 
		\Psi^2_{\, \chi \psi}(\mathbf{P}) \\ 
		&= g \psi
	\end{split}
\end{align}
% Eq. A32
\begin{align} \label{App_Aphiphi}
	\begin{split}
		\mathfrak{A}_{\phi \phi}(\mathbf{P}) &= 2 \Phi^\ast \Phi_{\phi^2}(\mathbf{P}) \\
		&= \epsilon - 2\mu
	\end{split}
\end{align}
% Eq. A33
\begin{align} \label{App_Aphichi}
	\mathfrak{A}_{\phi \chi}(\mathbf{P}) = \mathfrak{A}_{\chi \phi}(\mathbf{P}) = 0
\end{align}
% Eq. A34
\begin{align} \label{App_Achichi}
	\begin{split}
		\mathfrak{A}_{\chi \chi}(\mathbf{P}) &= 2 X^\ast X_{\chi^2}(\mathbf{P}) \\
		&= \varepsilon - 2\mu .
	\end{split}
\end{align}
% Subsection A7
%\setcounter{equation}{0}
\subsection{$\mathfrak{B}$ Matrix Elements}
% Eq. A35
\begin{align} \label{App_BSigSig}
	\begin{split}
		\mathfrak{B}_{\Sigma\Sigma}(\mathbf{q},\mathbf{q'},\mathbf{P}) &=  
		2 R_{\Sigma^2}(\mathbf{q},\mathbf{q'},\mathbf{P}) + 
		2D_{\Sigma^2}(\mathbf{q},\mathbf{q'},\mathbf{P}) \\
		&= 2 \left[ (q_+^2 - \mu + \xi) \, G_- + 
		(q_-^2 - \mu + \xi) \, G_+ \right]
		\delta(\mathbf{q} - \mathbf{q'}) \\
		&= s_K(\mathbf{q}, \mathbf{P}) \, \delta(\mathbf{q} - \mathbf{q'}) .
	\end{split}
\end{align}
% Eq. A36
\begin{equation} \label{App_BSigpi}
	\mathfrak{B}_{\Sigma \pi}(\mathbf{q},\mathbf{P}) = 
	\mathfrak{B}_{\pi \Sigma}(\mathbf{q}, \mathbf{P}) = 0
\end{equation}
% Eq. A37
\begin{align} \label{App_BSigomega}
	\begin{split}
		\mathfrak{B}_{\Sigma \omega}(\mathbf{q}, \mathbf{P}) = 
		\mathfrak{B}_{\omega \Sigma}(\mathbf{q},\mathbf{P}) 
		&= D_{\Sigma \omega}(\mathbf{q}, \mathbf{P}) \\
		&= - \lambda \alpha_1 \left( G_+ + G_- \right)
	\end{split}
\end{align}
% Eq. A38
\begin{align} \label{App_BSignu}
	\begin{split}
		\mathfrak{B}_{\Sigma \nu}(\mathbf{q}, \mathbf{P}) = 
		\mathfrak{B}_{\nu \Sigma}(\mathbf{q},\mathbf{P}) 
		&= D_{\Sigma \nu}(\mathbf{q}, \mathbf{P}) \\
		&= - g \left( G_+ + G_- \right)
	\end{split}
\end{align}
% Eq. A39
\begin{align} \label{App_Bpipi}
	\begin{split}
		\mathfrak{B}_{\pi \pi}(\mathbf{P}) &= 2(k^2 - \mu)\Psi^\ast \Psi_{\pi^2}(\mathbf{P}) 
		+ 2 \Psi^2_{\, \pi^2}(\mathbf{P}) \\
		&=  P^2 - \mu - \xi
	\end{split}
\end{align}
% Eq. A40
\begin{align} \label{App_Bpiomega}
	\begin{split}
		\mathfrak{B}_{\pi \omega}(\mathbf{P}) = \mathfrak{B}_{\omega \pi}(\mathbf{P}) &= 
		\Psi^2_{\, \omega \pi}(\mathbf{P}) \\ 
		&= \lambda \alpha_1 \psi
	\end{split}
\end{align}
% Eq. A41
\begin{align} \label{App_Bpinu}
	\begin{split}
		\mathfrak{B}_{\pi \nu}(\mathbf{P}) = \mathfrak{B}_{\nu \pi}(\mathbf{P}) &= 
		\Psi^2_{\, \nu \pi}(\mathbf{P}) \\ 
		&= g \psi
	\end{split}
\end{align}
% Eq. A42
\begin{align} \label{App_Bomegaomega}
	\begin{split}
		\mathfrak{B}_{\omega \omega}(\mathbf{P}) &= 2 \Phi^\ast \Phi_{\omega^2}(\mathbf{P}) \\
		&= \epsilon - 2\mu
	\end{split}
\end{align}
% Eq. A43
\begin{align} \label{App_Bomeganu}
	\mathfrak{B}_{\omega \nu}(\mathbf{P}) = \mathfrak{B}_{\nu \omega}(\mathbf{P}) = 0
\end{align}
% Eq. A44
\begin{align} \label{App_Bnunu}
	\begin{split}
		\mathfrak{B}_{\nu \nu}(\mathbf{P}) &= 2 X^\ast X_{\nu^2}(\mathbf{P}) \\
		&= \varepsilon - 2\mu .
	\end{split}
\end{align}

\end{appendix}

\bibliographystyle{my-h-elsevier}

\begin{thebibliography}{10}
\bibitem{ParamValues} The resonance parameters such as the width, $\Delta B$, resonance field, $B_0$, and the background scattering length, $a_{bg}$, were obtained in \cite{AMCoherence}, whereas the $\Delta g$ factor [see Eq. \!(\ref{TBdetuning})] was found in \cite{RamseyFringe}.
\end{thebibliography}

\end{document}